\documentclass[AMA,STIX1COL]{WileyNJD-v2}

\articletype{Research article}%

\received{xx April 2019}
\revised{xx xxx xxxx}
\accepted{xx xxx xxxx}

\usepackage{stmaryrd} 
\usepackage{geometry,graphicx}
\usepackage{color}
\usepackage{hyperref}
\usepackage[all]{hypcap}
\usepackage[T1]{fontenc} 							
\usepackage{amssymb,amsmath}
\usepackage{empheq}							
\usepackage{bm} 
\usepackage{thmbox}
\usepackage{shadethm}		
\usepackage{scalerel} 
\usepackage{nicefrac}								
\usepackage{siunitx}  		
\usepackage{booktabs} 
\usepackage{subfigure}		
\usepackage{mathtools}	
\usepackage{lipsum}
\usepackage{soul}

\newcommand\redsout{\bgroup\markoverwith{\textcolor{red}{\rule[0.5ex]{2pt}{0.5pt}}}\ULon}

\newlength\bshft
	\def\fakebold#1{\ThisStyle{\ooalign{$\SavedStyle#1$\cr%
  	\kern-\bshft$\SavedStyle#1$\cr%
  	\kern\bshft$\SavedStyle#1$}}}

\newcommand{\R}{\mathbb{R}}
\newcommand{\N}{\mathbb{N}}

\newcommand\Pk[2]{{ \mathbb{P}_{#1}{#2} }}

\newcommand\Qk[2]{{ \mathbb{Q}_{#1}{#2} }}
\newcommand\Qdk[2]{{ \mathbb{Q}_{#1}{#2} }}
\newcommand\QQk[2]{{ \fakebold{\mathbb{Q}}_{#1}{#2} }}
\newcommand\QQdk[2]{{ \fakebold{\mathbb{Q}}_{#1}{#2} }}

\newcommand\hK{\hat{K}}
\newcommand\hQk[2]{{ \hat{\mathbb{Q}}_{#1}{#2} }}

\newcommand\hQQk[2]{{ \hat{\fakebold{\mathbb{Q}}}_{#1}{#2} }}

\newcommand\RTk[2]{{ \fakebold{\mathbb{RT}}_{\sqb{#1}}{#2} }}

\newcommand\hRTk[2]{{ \hat{\fakebold{\mathbb{RT}}}_{\sqb{#1}}{#2} }}

\newcommand{\PIF}{ \boldsymbol{\Pi}_\F }

\newcommand{\dvg}{{ \mathrm{div} }}

\newcommand{\rms}{{ \mathrm{rms} }}
\newcommand{\OMEGA}{{ \rb{\Omega} }}

\newcommand\Rey{\mbox{\textit{Re}}}  

\makeatletter
\newcommand*\bigcdot{\mathpalette\bigcdot@{.635}}
\newcommand*\bigcdot@[2]{\mathbin{\vcenter{\hbox{\scalebox{#2}{$\m@th#1\bullet$}}}}}
\makeatother

\def\onedot{$\mathsurround0pt\ldotp$}
\def\cddot{
  \mathbin{\vcenter{\baselineskip.67ex
    \hbox{\onedot}\hbox{\onedot}}%
  }}%

\DeclareMathOperator{\ip}{{\bigcdot}}
\DeclareMathOperator{\Fip}{{\boldsymbol{\cddot}}}
\DeclareMathOperator{\DIV}{\nabla\ip}
\DeclareMathOperator{\DIVh}{\nabla_\textit{h}\ip}

\newcommand{\ff}{{ \boldsymbol{f} }}
\newcommand{\PHI}{{ \boldsymbol{\varphi} }}
\newcommand{\uu}{{ \boldsymbol{u} }}

\newcommand{\vv}{{ \boldsymbol{v} }}
\newcommand{\ww}{{ \boldsymbol{w} }}
\newcommand{\uuhat}{{ \boldsymbol{\widehat{u}} }}
\newcommand{\vvhat}{{ \boldsymbol{\widehat{v}} }}

\newcommand{\ssh}{{ \boldsymbol{s}_h }}

\newcommand{\xx}{{ \boldsymbol{x} }}
\newcommand{\sss}{{ \boldsymbol{s} }}

\newcommand{\nn}{{ \boldsymbol{n} }}
\newcommand{\tang}{{ \boldsymbol{t} }}

\newcommand{\zero}{{ \boldsymbol{0} }}
\newcommand{\tend}{{ T }}

\newcommand{\drm}{{ \mathrm{d} }}
\newcommand{\per}{{ \mathrm{per} }}
\newcommand{\dx}{{ \,\drm\xx }}

\newcommand{\ds}{{\,\drm\boldsymbol{s}}}

\newcommand{\ka}{{ \boldsymbol{\kappa} }}
\newcommand{\half}{{ \nicefrac{1}{2} }}

\newcommand{\lavg}{{ \big\{\hspace{-0.99ex}\big\{ }}						
\newcommand{\ravg}{{ \big\}\hspace{-0.99ex}\big\} }}		
\newcommand{\ljmp}{ \left\llbracket }	
\newcommand{\rjmp}{ \right\rrbracket }	
\newcommand\jmp[1]{{ \ljmp#1\rjmp }}							
\newcommand\avg[1]{{ \lavg#1\ravg }}

\newcommand{\nhphantom}[1]{\sbox0{#1}\hspace{-0.751\dimexpr\the\wd0 \relax}}
\newcommand\hjmp[1]{{ {\llbracket}\nhphantom{$\llbracket$}{\llbracket} #1 
	{\rrbracket}\nhphantom{$\rrbracket$}{\rrbracket} }}		

\newcommand{\Kin}{{ \mathcal{K} }}

\newcommand\Ltwo{{ L^{2} }}	
\newcommand\LTWO{{ \boldsymbol{L}^{2} }}	
\newcommand\Lp[2]{{ L^{#1}{#2} }} 
\newcommand\LP[2]{{ \boldsymbol{L}^{#1}{#2} }} 
\newcommand\Lpz[2]{{ L_0^{#1}{#2} }}
\newcommand\LPZ[2]{{ \boldsymbol{L}_0^{#1}{#2} }}

\newcommand\HM[2]{{ \boldsymbol{H}^{#1}{#2} }}

\newcommand\HDIV{{ \boldsymbol{H}{\rb{\dvg}} }}
\newcommand{\Hdiv}{{ \boldsymbol{H}_0{\rb{\dvg;\Omega}} }}		
			
\newcommand{\VV}{{ \boldsymbol{V} }}	
	
\newcommand{\VVhat}{{ \boldsymbol{\widehat{V}} }}

\newcommand{\Q}{{ Q }}

\newcommand{\T}{{ \mathcal{T}_h }} 
\newcommand{\F}{{ \mathcal{F}_h }}	
	
\newcommand{\Fi}{{ \mathcal{F}_h^i }}								
\newcommand{\Fb}{{ \mathcal{F}_h^\partial }}		

\newcommand\lift[1]{{ \mathcal{L}\rb{#1} }}	

\newcommand\rb[1]{{ \left(#1\right) }}
\newcommand\sqb[1]{{ \left[ #1 \right] }}
\newcommand\rsb[1]{{ \left(#1\right] }}
\newcommand\set[1]{{ \left\{ #1 \right\} }}
\newcommand\bra[1]{{ \langle #1 \rangle }}
\newcommand\abs[1]{{ \left\lvert#1\right\rvert }}
\newcommand\norm[1]{ \left\lVert#1\right\rVert }

\newcommand{\tripnorm}[1]{{\left\vert\kern-\nulldelimiterspace\left\vert\kern-\nulldelimiterspace\left\vert #1
	\right\vert\kern-\nulldelimiterspace\right\vert\kern-\nulldelimiterspace\right\vert}}

\newcommand\restr[2]{{												
	\left.\kern-\nulldelimiterspace									
	#1
	\vphantom{|}
	\right|_{#2}
	}}
	
\newcommand{\goodgap}{%
	\hspace{0.01\subfigtopskip}
	\hspace{0.01\subfigbottomskip}
	}	

\newtheorem[style=S,underline=true,bodystyle=\normalsize\noindent]{thmDef}{\textsc{Definition}}[section]
\newtheorem[style=S,cut=false]{thmCor}[thmDef]{\textsc{Corollary}}
\newtheorem[style=S,cut=false,headstyle=\normalsize\bfseries\boldmath####1~####2]{thmLem}[thmDef]{\textsc{Lemma}}

\newshadetheorem{thm}[thmDef]{\textsc{Theorem}}

\newenvironment{thmRem}
                [0]
                { \refstepcounter{thmDef} \begin{example}[\normalsize\textsc{Remark} \thesection.\arabic{thmDef}]  \normalsize}
                {  $\hfill\blacktriangle$ \end{example} }	
                
\newtheorem[style=S,underline=true,bodystyle=\noindent,cut=false]{thmAss}{\small\textsc{Assumption}}

\begin{document}

 
\title{High-order DG solvers for under-resolved turbulent incompressible flows: A comparison of $L^2$ and $H$(div) methods} 

\author[1]{Niklas Fehn}

\author[1]{Martin Kronbichler}

\author[2]{Christoph Lehrenfeld}

\author[2]{Gert Lube}

\author[2]{Philipp W.\ Schroeder}

\authormark{FEHN, KRONBICHLER, LEHRENFELD, LUBE, SCHROEDER}

\address[1]{
	Institute for Computational Mechanics, Technical University of Munich,
	85748 Garching, Germany.
	Email:
		fehn@lnm.mw.tum.de,
		kronbichler@lnm.mw.tum.de (\url{https://orcid.org/0000-0001-8406-835X}) \\
	}

\address[2]{
	Institute for Numerical and Applied Mathematics, Georg-August-Universit{\"a}t G{\"o}ttingen,
	37083 G{\"o}ttingen, Germany.
	Email:
		lehrenfeld@math.uni-goettingen.de (\url{https://orcid.org/0000-0003-0170-8468}),
		lube@math.uni-goettingen.de,
		p.schroeder@math.uni-goettingen.de (\url{https://orcid.org/0000-0001-7644-4693}) \\
	}

\corres{
	Christoph Lehrenfeld, lehrenfeld@math.uni-goettingen.de, Institute for Numerical and Applied Mathematics, Georg-August-Universit{\"a}t G{\"o}ttingen,
	37083 G{\"o}ttingen, Germany.
	}

\abstract[Abstract]{
The accurate numerical simulation of turbulent incompressible flows is a challenging topic in computational fluid dynamics.
For discretisation methods to be robust in the under-resolved regime, mass conservation as well as energy stability are key ingredients to obtain robust and accurate discretisations. 
Recently, two approaches have been proposed in the context of high-order discontinuous Galerkin (DG) discretisations that address these aspects differently. 
On the one hand, standard $L^2$-based DG discretisations enforce mass conservation and energy stability weakly by the use of additional stabilisation terms. 
On the other hand, pointwise divergence-free $H(\operatorname{div})$-conforming approaches ensure exact mass conservation and energy stability by the use of tailored finite element function spaces. 
The present work raises the question whether and to which extent these two approaches are equivalent when applied to under-resolved turbulent flows. 
This comparative study highlights similarities and differences of these two approaches. 
The numerical results emphasise that both discretisation strategies are promising for under-resolved simulations of turbulent flows due to their inherent dissipation mechanisms.
}



\maketitle


\section{Introduction}
\label{sec:Introduction}

\subsection{Motivation}
Simulating turbulent flows is still a challenging undertaking, even on today's high-performance computing architectures. 
Discontinuous Galerkin (DG) discretisations are currently investigated in order to develop new discretisation methods with inbuilt stabilisation mechanism rendering these methods robust and accurate when applied to turbulent flow problems.
In this contribution, we compare the accuracy of two high-order DG solvers for incompressible flows with a special emphasis on how they perform in the practically relevant situation of only being able to marginally resolve the occurring flow features.
While the two `extreme cases' of direct numerical simulation (DNS) on the one hand, and Reynolds-averaged Navier--Stokes (RANS) simulations on the other hand are relatively well-established and well-understood, being able to perform time-dependent turbulent flow simulations with only a limited amount of fine-scale accuracy usually goes by the name large-eddy simulation (LES).
This regime is exactly where we are interested in in this work.
~\\

LES simulations by now have a rich history in both the engineering and the mathematics literature and there are numerous promising approaches for different flow situations.
Our aim is to be able to compute (incompressible) flows without the need of choosing the `correct' turbulence model or set of parameters.
Therefore, we rely on what can be called a `no-model LES' in the sense that no explicit turbulence model shall be incorporated in the simulations in this work.
Such an approach, in turn, relies heavily on distinctive robustness as well as healthy and controllable dissipation properties of the numerical scheme to approximate the incompressible Navier--Stokes equations.
While DG methods have reached a mature state in the field of the compressible Navier--Stokes equations, see for example the works~\cite{Uranga2011,GassnerBeck12,WiartEtAl13,Beck2014,Moura2017,Fernandez2017,Winters2018} in the context of low- and moderate-Mach number turbulent flows, designing robust DG discretisations of the incompressible Navier--Stokes equations exhibits some subtleties in the context of under-resolved turbulence: applying well-known numerical flux formulations to the discretisation of convective and viscous terms only, see the works~\cite{hesthaven2007nodal,Shahbazi2007,Botti2011,Ferrer2011,Klein2015,Ferrer2017}, has been realised to be not robust enough in under-resolved scenarios, as pointed out in the recent works~\cite{Steinmoeller2013, Joshi2016, Krank2017,PiatkowskiEtAl18}. 
It is worth mentioning that this lack of robustness is not related to under-integration of nonlinear terms, commonly known as aliasing. 
Instead, additional techniques are required which are inherently linked to the nature of the incompressible Navier--Stokes equations and in particular the incompressibility constraint. 
Namely, effort has to be put into ensuring compliance with both the divergence-free and the normal-continuity constraints. 
~\\

Two promising methods have been proposed recently in~\cite{LehrenfeldSchoeberl16, Fehn2018a} which rely on different stabilisation concepts. 
The two candidates under investigation are intelligently stabilised~$\LTWO$-based DG methods~\cite{Fehn2018a} and exactly divergence-free~$\HDIV$-conforming methods~\cite{LehrenfeldSchoeberl16,Schroeder2019}. 
A first connection between both approaches has been pointed out in~\cite{Fehn2018a,Akbas2018}. 
However, it remains rather unexplored how these different stabilisation techniques behave regarding their dissipation mechanism in the under-resolved regime, which is particularly relevant for practical LES of incompressible turbulent flows. 
Therefore, the aim of this work is a comparison of the accuracy of both approaches in under-resolved turbulence simulation, which is examined numerically for the well-known 3D benchmark cases of freely decaying turbulence (Taylor--Green vortex), and wall-bounded turbulence in a channel.
The numerical results for the~$\HDIV$-conforming method have been obtained using NGSolve\cite{ngsolve}, the results for the~$\LTWO$-based DG using the open-source finite element library \texttt{deal.II}\cite{dealII90}. 

\subsection{Mathematical model}
The underlying mathematical model is the incompressible Navier--Stokes problem\cite{Tritton88,SchlichtingGersten00,Durst08} without any additional terms, i.e.\
\begin{subequations}\label{eq:TINS}
	\begin{empheq}[left=\empheqlbrace]{alignat=2} 
		\partial_t\uu - \nu\Delta \uu + \rb{\uu\ip\nabla}\uu +\nabla p &= \ff \qquad\quad 												&&\text{in }\rsb{0,\tend}\times\Omega, 			\\
		\DIV\uu &= 0 				&&\text{in }\rsb{0,\tend}\times\Omega, 			\\
		\uu\rb{0,\cdot} &=\uu_0\rb{\cdot} 	&&\text{in }\Omega.		
	\end{empheq} 
\end{subequations}
In \eqref{eq:TINS}, $\Omega\subset\R^3$ denotes a connected bounded Lipschitz domain.
Concerning boundary conditions (BCs) we assume that $\partial\Omega=\Gamma_0\,\dot{\cup}\,\Gamma_\per$, where the no-slip BC $\uu=\zero$ is prescribed on $\Gamma_0$ and periodic BCs are imposed on $\Gamma_\per$.
Moreover, $\uu \colon\rsb{0,\tend}\times\Omega\to\R^3$ indicates the velocity field, $p\colon\rsb{0,\tend}\times\Omega\to\R$ is the (zero-mean) kinematic pressure, $\ff\colon\rsb{0,\tend}\times\Omega\to\R^3$ represents external body forces and $\uu_0\colon \Omega\to\R^3$ stands for a suitable initial condition for the velocity. 
The underlying fluid is assumed to be Newtonian with constant (dimensionless) kinematic viscosity $0<\nu\ll 1$.

\subsection{Outline}
We introduce both DG discretisations under consideration with a focus on the mathematical formulation and discuss properties such as approximation abilities, mass conservation and dissipation mechanisms in Section~\ref{sec:DG-solvers}. 
The accuracy of both methods for the 3D Taylor-Green vortex problem as a prototype problem of decaying homogeneous isotropic turbulence is investigated in Section~\ref{sec:TaylorGreen}. 
Apart from a direct comparison in terms of the kinetic energy dissipation rate on different meshes, we also investigate the impact of low-order vs.\ high-order discretisations and the impact of variations of the DG formulation such as changes in the stabilisation terms, and the flux formulations for convective and viscous terms.
To include wall-bounded turbulent flows in our investigations, we consider the turbulent channel flow problem in Section~\ref{sec:Channel}. 
In that section, we compare the behaviour of both methods under mesh refinement and again discuss the influence of method variations including the handling of anisotropic refinements towards the wall boundary with affine and isoparametric finite element meshes. The robustness of the considered approaches and their reliability in correctly predicting turbulent flows is further assessed by parameter studies.
The final part, Section~\ref{sec:Conclusions}, summarises the observations made in this work and derives conclusions.

\section{Two DG-based solvers for turbulent incompressible flows}
\label{sec:DG-solvers}

\subsection{Notation}
Let $\T$ be a partition of $\Omega$ into hexahedra with mesh size $h=\max_{K\in\T} h_K$, where $h_K$ denotes the diameter of the particular element $K\in\T$. 
The skeleton $\F$ denotes the set of all facets and $\F=\Fi\cup\Fb$ where $\Fi$ is the subset of interior plus periodic facets and $\Fb$ collects all Dirichlet boundary facets $F\subset\Gamma_0$. 
To any $F\in\F$ we assign a unit normal vector $\nn_F$ where, for $F\subset\partial\Omega$, this is the outer unit normal vector $\nn$. 
If $F\in\Fi$, there are two adjacent elements $K^+$ and $K^-$ sharing the facet $F=\partial K^+\cap\partial K^-$ and $\nn_F$ points from $K^+$ to $K^-$. 
Let $\phi$ be any piecewise smooth (scalar-, vector- or matrix-valued) function with traces from within the interior of $K^\pm$ denoted by $\phi^\pm$, respectively. 
Then, we define the jump $\jmp{\cdot}_F$ and average $\avg{\cdot}_F$ operators across interior and periodic facets $F\in\Fi$ by
\begin{align}\label{eq:DefInteriorJumps}
	\jmp{\phi}_F= \phi^+-\phi^-	
	\quad \text{and}\quad
	\avg{\phi}_F=\frac{1}{2}\rb{\phi^+ + \phi^-}.
\end{align}
These operators act componentwise for vector- and matrix-valued functions. Frequently, the subscript indicating the particular facet is omitted. 
~\\

As is usual in DG methods, we make use of the broken vector-valued Sobolev space ($m\in\N_0$)
\begin{equation*}
	\HM{m}{\rb{\T}} = 
		\set{\vv\in\LP{2}{\OMEGA}\colon \restr{\vv}{K}\in\HM{m}{\rb{K}},~\forall\, K\in\T},
\end{equation*}
and define the (elementwise) broken Jacobian $\nabla_h\colon\HM{1}{\rb{\T}}\to\LP{2}{\OMEGA}$ by $\restr{\rb{\nabla_h\vv}}{K}=\nabla\rb{\restr{\vv}{K}}$ and the (elementwise) broken divergence $\DIVh\colon\HM{1}{\rb{\T}}\to\Lp{2}{\OMEGA}$ by $\restr{\rb{\DIVh\vv}}{K}=\DIV\rb{\restr{\vv}{K}}$ for all $K\in\T$. 
~\\

In the remainder of this subsection we introduce local and global finite element spaces. 
For ease of presentation we consider only hexahedral meshes which consist of straight hexahedra $K$ which are aligned with the coordinate axes and are the image of the reference hexahedron $\hat{K} = [0,1]^3$ under affine linear mappings. 
However, the general case of non-affinely mapped hexadra can also be dealt with easily, cf.\ the discussion in Section \ref{sec:AffineVsIso} below.
~\\

In the following, $\Qk{k}{\rb{K}}$ (vector-valued $\QQk{k}{\rb{K}}$) denotes the tensor-product space of all polynomials on $K$ with degree less or equal $k$ in each direction separately, $\Qk{k}{\rb{K}} = \Pk{k}{(K)} \otimes \Pk{k}{(K)} \otimes \Pk{k}{(K)}$.
The $\LTWO$-based method makes use of these local spaces by glueing them together discontinuously, resulting in the global spaces
\begin{equation}\label{eq:SpacesL2}
	\Qdk{k}{}	\coloneqq \bigoplus_{K \in \T} \Qk{k}{(K)}, \quad
	\QQdk{k}{}	\coloneqq \bigoplus_{K \in \T} \QQk{k}{(K)}.
\end{equation}
For the $\HDIV$-based method, let us first introduce
\begin{align*}
 	\Hdiv 
 		= \set{\vv\in\LP{2}{\OMEGA}\colon \DIV\vv\in\Lp{2}{\OMEGA};~\restr{\vv\ip\nn}{\Gamma_0}=0}.
\end{align*} 
Note that a function in $\Hdiv$ at least has to be normal-continuous.
Since we are working on hexahedral meshes in 3D in this work, the local velocity space is based on the Raviart--Thomas element \cite{BoffiEtAl13} 
\begin{equation*}
	\RTk{k}{\rb{K}}
		= \rb{\Pk{k+1,k,k}{\rb{K}},\Pk{k,k+1,k}{\rb{K}},\Pk{k,k,k+1}{\rb{K}}}^\dag, \quad \Pk{k_1,k_2,k_3}{\rb{K}} = \Pk{k_1}{(K)} \otimes \Pk{k_2}{(K)} \otimes \Pk{k_3}{(K)}.
\end{equation*}
$\RTk{k}(K)$ includes polynomials of order $k+1$ in certain directions (depending on the vector entry), but much less than $\QQk{k+1}(K)$. 
There holds the inclusion
$
\QQk{k}{(K)} \subset \RTk{k}{(K)} \subset \QQk{k+1}{(K)}.
$
We further note that the highest order polynomials in $\RTk{k}(K)$ do not improve the approximation order of the local polynomial space, but are added on top of $\QQk{k}{(K)}$ for the purpose of improving the approximation of the divergence operator as there holds 
$
\DIV \RTk{k}{(K)} = \Qk{k}{(K)} \supset \DIV \QQk{k}{(K)}.
$
We define the global finite element spaces for the $\HDIV$-based method as
\begin{equation}\label{eq:SpacesHdiv}
	\RTk{k}{}	\coloneqq \Hdiv \cap \bigoplus_{K \in \T} \RTk{k}{(K)} .
\end{equation}

\subsection{\emph{L}$\mathbf{^2}$-DG method with consistent stabilisation terms}
\label{sec:L2DG}
The first DG method investigated in this work is an $\LTWO$-based DG incompressible Navier--Stokes solver characterised by consistent divergence and continuity penalty terms acting as stabilisation terms for improved mass conservation and energy stability. 
This method is based on the formulation proposed in recent works of Fehn et al.~\cite{Fehn2018a,Fehn2017}.
Both velocity and pressure are approximated by discontinuous functions, where the polynomial space for the velocity is one degree higher than for the pressure for reasons of discrete inf-sup stability:
\begin{equation*}
  	\VV_h = \QQdk{k}{}, \quad
   	\Q_h = \Qdk{k-1}{} \cap \Lpz{2}{\OMEGA} .
\end{equation*}
To obtain the contribution of boundary face integrals from the weak forms given below, equation~\eqref{eq:DefInteriorJumps} holds where the exterior weighting function is simply set to zero on boundary facets ($\vv_h^+ = \bm{0}$,~$\nabla \vv_h^+ \ip \nn = \bm{0}, q_h^+=0$). Homogeneous Dirichlet boundary conditions are prescribed according to the mirror principle~$\uu_h^+ = - \uu_h^-$ (with~$\nabla \uu_h^+ \ip \nn= \nabla \uu_h^- \ip \nn$ for the velocity gradient and~$p_h^+ = p_h^-$ for the pressure). The space-semidiscrete weak formulation of the incompressible Navier--Stokes equations~\eqref{eq:TINS} reads as follows:
\begin{subequations} \label{eq:L2DGMethod}
	\begin{empheq}[left=\empheqlbrace]{align} 
	&\text{Find }\rb{\uu_h,p_h}\colon \rsb{0,\tend} \to \VV_h \times \Q_h
	\text{ with }\uu_h\rb{0} =\uu_{0h}\text{ s.t., }\forall\,\rb{\vv_h,q_h}\in\VV_h \times \Q_h,\\
		&\rb{\partial_t\uu_h,\vv_h} 
		+ \nu a_h\rb{\uu_h,\vv_h}
		+ c_h\rb{\uu_h;\uu_h,\vv_h}
		+ j_h\rb{\uu_h,\vv_h}		
		+ g_h\rb{p_h,\vv_h}
		+ d_h\rb{\uu_h,q_h}
		=\rb{\ff,\vv_h}.		
	\end{empheq} 
\end{subequations}
The DG discretisation of the viscous term is based on the symmetric interior penalty Galerkin (SIPG) method
\begin{equation*}
	a_h\rb{\uu_h,\vv_h} 
		=	\int_\Omega \nabla_h \uu_h \Fip \nabla_h \vv_h \dx 
		-\sum_{F\in\F} \int_F \left(\avg{\nabla \uu_h} \ip \nn \right) \ip \jmp{\vv_h} \ds
		-\sum_{F\in\F} \int_F \jmp{\uu_h} \ip \left( \avg{\nabla \vv_h} \ip \nn \right) \ds
		+\sum_{F\in\F} \int_F \tau \jmp{\uu_h} \ip \jmp{\vv_h} \ds,
\end{equation*}
where the penalty parameter~$\tau$ is
\begin{align}
\tau =
\begin{cases}
\max\left(\tau_{K^-},\tau_{K^+}\right) & \text{if facet } F \in \Fi\; ,\\
\tau_K & \text{if facet } F \in \Fb\; ,
\end{cases}\label{TauIP}
\end{align}
using the definition according to Hillewaert~\cite{Hillewaert2013} for hexahedral elements
\begin{align}
\tau_K = (k+1)^2 \frac{ \abs{\partial K \cap \Fi}_{d-1}/2 + \abs{\partial K \cap \Fb}_{d-1}}{ \abs{K}_{d}}\; .
\end{align}
Here,~$|\cdot|_d$ denotes the $d$-dimensional Lebesgue measure. For the convective term, the local Lax--Friedrichs flux is applied
\begin{equation} \label{eq:L2-convection}
	c_h\rb{\ww_h;\uu_h,\vv_h}	
		= -\int_\Omega \rb{\uu_h \otimes \ww_h} \Fip \nabla_h \vv_h \dx 
		+ \sum_{F\in\F} \int_F \rb{\avg{\uu_h \otimes \ww_h} \ip \nn + \frac{\Lambda(\ww_h)}{2}\jmp{\uu_n}} \ip \jmp{\vv_h} \ds,
\end{equation}
where~$\Lambda(\ww_h) = \max \rb{2 \vert\ww^-_h\ip \nn\vert, 2 \vert \ww^+_h \ip \nn \vert}$. 
Integration by parts of the velocity-pressure coupling terms along with central numerical flux functions yields
\begin{align}
	g_h\rb{p_h,\vv_h} 
		&= -\int_\Omega p_h\rb{\DIVh \vv_h} \dx
		+\sum_{F\in\F}\int_F \avg{p_h}\rb{\jmp{\vv_h}\ip\nn} \ds \ , \label{L2PressureGradientWeakForm}\\
	d_h\rb{\uu_h,q_h} 
		&= \int_\Omega \uu_h \ip \nabla_h q_h \dx
		-\sum_{F\in\F}\int_F \jmp{q_h}\rb{\avg{\uu_h}\ip\nn} \ds \ . \label{L2VelocityDivergenceWeakForm}
\end{align}
The pressure gradient term and velocity divergence term are implemented in the so-called weak formulation of DG methods, cf. Ref.~\cite{hesthaven2007nodal}, and it holds~$g_h\rb{p_h,\uu_h}  = d_h\rb{\uu_h,p_h} $ in case of exact numerical quadrature, cf.\ also the paragraph on numerical quadrature below. 
The consistent stabilisation term~$j_h=j_{\mathrm{div},h}+j_{\mathrm{conti},h}$ is composed of a divergence penalty term~$j_{\mathrm{div},h}$ and a continuity penalty term~$j_{\mathrm{conti},h}$, which are given as~\cite{Fehn2018a}
\begin{align}
	j_{\mathrm{div},h}\rb{\uu_h,\vv_h} &= \int_\Omega \tau_{\mathrm{D}} \rb{\DIVh \uu_h} \rb{\DIVh \vv_h} \dx \ , \label{DivergencePenalty} \\
	j_{\mathrm{conti},h}\rb{\uu_h,\vv_h} &= \sum_{F\in\Fi}  \int_F \avg{\tau_{\mathrm{C}}} \rb{\jmp{\uu_h}\ip\nn} \rb{\jmp{\vv_h}\ip\nn} \ds  \ , \label{ContinuityPenalty}
\end{align}
where the penalty parameters of the divergence and continuity penalty terms are defined as~$\tau_{\mathrm{D}} = \zeta \overline{\Vert \uu^{\mathrm{ex}}_h \Vert} \frac{h}{k+1}$ and~$\tau_{\mathrm{C}} = \zeta \overline{\Vert \uu^{\mathrm{ex}}_h \Vert}$, respectively. 
Here,~$\overline{\Vert \uu^{\mathrm{ex}}_h \Vert}$ is the magnitude of the velocity vector averaged over the element volume and~$\left( \cdot \right)^{\mathrm{ex}}$ indicates that an extrapolation of the velocity field from previous instants of time is used in the time-discrete setting, resulting in a weak formulation~$j_h$ that is linear in~$\uu_h$. 
This definition of the penalty parameters originates from a dimensional analysis ensuring that all terms in the weak formulation~\eqref{eq:L2DGMethod} have consistent physical units. 
It was shown by means of numerical investigation in~\cite{Fehn2018a} that with this definition of the penalty parameters, a default value of~$\zeta=1$ ensures robustness for under-resolved flows in the sense that stability has been demonstrated for a wide range of Reynolds numbers (and also in the inviscid limit), as well as for a wide range of the spatial resolution parameters~$h,k$.

\paragraph{Numerical quadrature} 
The terms in the weak formulation are integrated numerically with Gaussian quadrature and~$k+1$ quadrature points per coordinate direction, except for the convective term~$c_h$ where an over-integration strategy with~$\lceil \frac{3k+1}{2} \rceil$ quadrature points is used due to the quadratic nonlinearities of the convective term. 
On affine element geometries this setup ensures exact integration. 
Exact integration allows to apply integration by parts to the integrals in $c_h(\cdot,\cdot),~d_h(\cdot,\cdot),~g_h(\cdot,\cdot)$ which yields~$g_h\rb{p_h,\uu_h}  = d_h\rb{\uu_h,p_h} $ and~$c_h(\ww_h;\uu_h,\uu_h) \geq 0$ for divergence-free and normal-continuous velocities $\ww_h$ (and suitable boundary conditions) and arbitrary $\uu_h \in \VV_h$, $p_h \in \Q_h$.

\paragraph{Time integration and iterative solution of linear systems of equations} 
Discretisation in time is based on the backward differentiation formula (BDF) time integration method using a second order accurate formulation (BDF2). 
The convective term is formulated explicitly in time (using a second order accurate extrapolation scheme) which results in a CFL-type restriction of the time step size, see~\cite{Fehn2018a} for details on the time discretisation. For the fully discrete problem, an unsteady Stokes problem has to be solved in each time step. 
This coupled system of equations including the stabilisation terms is solved iteratively using the GMRES method with block preconditioning. 
Other solution strategies for time integration such as projection methods can be used as well to obtain computationally efficient solution algorithms and have been described in~\cite{Fehn2018a} in the context of the present $\LTWO$-based DG discretisation. 
The solver is stopped once the $l_2$ norm of the residual has been reduced by $10^{6}$ compared to the initial residual norm (where the initial solution is an extrapolation of the solution from previous time steps), or if the absolute value of the discrete $l_2$ norm of the residual goes below $10^{-12}$. 
All solver components rely on fast matrix-free operator evaluation (exploiting the so-called sum-factorisation technique) to achieve optimal computational complexity, and we refer to~\cite{Fehn2018b} for a documentation of the computational efficiency of this $\LTWO$-based DG solver.

\subsection{Divergence-free \emph{H}(div)-HDG method}
\label{sec:HdivDG}

For the $\HDIV$-based method we basically rely on the method in Ref. \cite{LehrenfeldSchoeberl16}; the underlying FE spaces are given by
\begin{equation}\label{eq:HDivSpaces}
  	\VV_h  = \RTk{k}{}, \quad \Q_h = \Qdk{k}{} \cap \Lpz{2}{\OMEGA}, \quad
   	\VVhat_h = \set{\vvhat_h\in\LP{2}{\rb{\F}}\colon \restr{\vvhat_h}{F}\in\QQk{k-1}{\rb{F}}; ~\restr{\vvhat_h\ip\nn}{F}=0,~\forall\, F\in\F; ~\restr{\vvhat_h}{\Gamma_0}=\zero},
\end{equation}
where $\VV_h$ and $\Q_h$ are the discrete velocity and pressure spaces, respectively, and $\VVhat_h$ is the hybrid facet space which contains discrete tangential velocities on the skeleton of $\T$.
Note that, thus, the global space $\VV_h$ has order $k$ approximation properties but in general, locally, the discrete velocity can even be a polynomial of order $k+1$. 
To obtain the contribution of boundary face integrals on Dirichlet boundary facets $F\in\Fb$ with homogeneous boundary values, we set $\avg{\phi}_F=\phi$ and $\jmp{\phi}_F=\phi$ for the average and jump operators introduced in equation~\eqref{eq:DefInteriorJumps}.

\begin{thmRem}
In order to remove some degrees of freedom (DOFs) for the velocity and all pressure unknowns except for piecewise constants, we exploit the \emph{a priori} knowledge that $\DIV\uu_h=0$; cf.\ {Remark 1}\cite{LehrenfeldSchoeberl16} and {Sec.~2.2.4.2}\cite{Lehrenfeld10}.
This can be achieved with a smart choice of basis functions for $\VV_h$ based on an exact sequence property (De Rham complex) on the discrete level; cf.\ Refs.~\cite{SchoeberlZaglmayr05,Zaglmayr06}.
\end{thmRem}

The space-semidiscrete weak formulation of the hybridised $\HDIV$-DG method for \eqref{eq:TINS} reads as follows:
\begin{subequations} \label{eq:HdivHDGMethod}
	\begin{empheq}[left=\empheqlbrace]{align} 
	&\text{Find }\rb{\uu_h,\uuhat_h,p_h}\colon \rsb{0,\tend} \to \VV_h \times \VVhat_h \times \Q_h
	\text{ with }\uu_h\rb{0} =\uu_{0h}\text{ s.t., }\forall\,\rb{\vv_h,\vvhat_h,q_h}\in\VV_h \times \VVhat_h \times \Q_h,\\
		&\rb{\partial_t\uu_h,\vv_h} 
		+ \nu a_h\rb{\rb{\uu_h,\uuhat_h},\rb{\vv_h,\vvhat_h}}
		+c_h\rb{\uu_h;\uu_h,\vv_h}
		+b_h\rb{\vv_h,p_h}
		+b_h\rb{\uu_h,q_h}
		=\rb{\ff,\vv_h}.		
	\end{empheq} 
\end{subequations}
Let us now specify the various terms in \eqref{eq:HdivHDGMethod}.
In order to discretise the viscous term, a hybridised variant of the symmetric interior penalty (SIP) method is used. 
The corresponding bilinear form, with $\sigma_K=6\rb{k+1}^2$, is given by
\begin{align*}
	a_h\rb{\!\rb{\uu_h,\uuhat_h},\!\rb{\vv_h,\vvhat_h}\!} 
		\!=\! \sum_{K\in\T} \sqb{
		\int_K \nabla\uu_h \Fip \nabla\vv_h \dx
		\!-\! \int_{\partial K} \rb{\nabla\uu_h \ip \nn} \ip \hjmp{\vv_h} \ds
		\!-\! \int_{\partial K} \hjmp{\uu_h} \ip \rb{\nabla\vv_h \ip \nn} \ds 
		\!+\! \int_{\partial K} \frac{\sigma_K}{h_K\rb{\sss}} \hjmp{\uu_h} \ip \hjmp{\vv_h} \ds
		},
\end{align*}
where $h_{K}\rb{\sss} = \abs{K}_{d}/\abs{F}_{d-1}$ with $\sss \in F \subset \partial K$.
Here, in contrast to $\jmp{\cdot}$, the operator $\hjmp{\cdot}$ represents the tangential jump between cell and facet velocity projected onto the space $\QQk{k-1}{(F)}$; that is, using the projection onto the tangent plane $\vv_\tang = \vv - \rb{\vv\ip\nn}\nn$,
\begin{equation*}
	\hjmp{\vv}
		= \rb{\PIF^{k-1}  \vv-\vvhat}_\tang
		= \PIF^{k-1} \vv_\tang - \vvhat,
\end{equation*}
where $\PIF^{k-1}$ is the facet-wise $\LTWO$-orthogonal projection onto $\QQk{k-1}{(F)}$.
The choice $\sigma_K=6\rb{k+1}^2$ for the SIP stabilisation parameter renders the SIP penalty scaling comparable to that of the $\LTWO$-based method.
Due to $\HDIV$-conformity, the pressure-velocity coupling is simply given by
\begin{align}
	b_h\rb{\uu_h,q_h} 
		= -\int_\Omega q_h\rb{\DIV \uu_h} \dx. \label{HdivContinuityEquationWeakForm}
\end{align} 
Here, the property $\DIV\VV_h=\Q_h$ allows to test with $q_h=\DIV\uu_h$ which results in the fact that we are obtaining an exactly divergence-free discrete velocity solution since discretely divergence-free velocities are actually exactly divergence-free; i.e.
\begin{equation} \label{HdivExactlyDivergenceFree}
	-\int_\Omega q_h\rb{\DIV \uu_h} \dx = 0,\quad \forall\,q_h\in\Q_h
		\quad \Longrightarrow \quad
	\int_\Omega \abs{\DIV\uu_h}^2 \dx = 0
		\quad \Longrightarrow \quad
	\DIV\uu_h=0. 
\end{equation}
Finally, for the nonlinear convection term, we use the form \\
\begin{equation} \label{eq:Hdiv-convection}
	c_h\rb{\ww_h;\uu_h,\vv_h}
		= \int_\Omega \rb{\ww_h\ip\nabla_h}\uu_h\ip\vv_h \dx 
		- \sum_{F\in\Fi} \int_F \rb{\ww_h\ip\nn} \jmp{\uu_h}\ip\avg{\vv_h} \ds
		+ \sum_{F\in\Fi} \int_F \frac{\theta}{2} \abs{\ww_h\ip\nn}\jmp{\uu_h}\ip\jmp{\vv_h} \ds,
\end{equation}
where the choice $\theta=1$ leads to an upwind-stabilised formulation and $\theta=0$ corresponds to a method without any kind of convection stabilisation.
Note that in the convective form \eqref{eq:Hdiv-convection}, instead of the hybrid jump $\hjmp{\cdot}$, the `normal' DG jump $\jmp\cdot{}$ between neighbouring cell velocities is used.

\paragraph{Numerical quadrature}
In the realisation of the bilinear forms, numerical quadrature is applied so that on affine linear elements, numerical integration is performed exactly.
Due to polynomials of degree $k+1$ in $\RTk{k}{}$ this requires $k+2$ Gaussian quadrature points in each direction for the integrals $a_h(\cdot,\cdot)$; for $b_h(\cdot,\cdot)$ $k+1$ Gaussian quadrature points suffice as $\DIV \RTk{k}{} = \Qk{k}{}$ while, due to the nonlinearity in $c_h(\cdot;\cdot,\cdot)$, $\lceil \frac{3k+3}{2} \rceil$ points are required.

\paragraph{Time integration and linear solvers}
Concerning time integration, we use the second-order Runge--Kutta variant ARS(2,2,2) of the implicit-explicit (IMEX) method introduced in Ref.~\cite{AscherEtAl97}.
Here, the pressure-velocity coupling $b_h\rb{\cdot}$ and the viscosity term $a_h\rb{\cdot}$ is always treated implicitly in order to maintain the exactly divergence-free property of the method, and the convection term $c_h\rb{\cdot}$ is always treated explicitly.
Therefore, only linear systems have to be solved in each time step. We use static condensation in order to eliminate element-local unknowns and solve the linear systems involving the Schur complement with a BDDC-preconditioned\cite{SchoeberlLehrenfeld13} CG solver.

\subsection{Comparison of approximation spaces under the divergence constraint}
\label{sec:DoFs}
For both the $\QQdk{k}{}$- and the $\RTk{k}{}$-based discretisations, the global number of velocity unknowns on a periodic Cartesian mesh with $N$ elements per coordinate direction in $d$ dimensions is given as 
\begin{equation} \label{eq:ndof}
N_{\mathrm{DOFs},\uu}(N,k) = N^d d \rb{k+1}^d \ .
\end{equation}
We note that the global number of unknowns for both velocity spaces is the same despite the fact that the local spaces are different according to~\eqref{eq:SpacesL2} and \eqref{eq:SpacesHdiv}. 
~\\

In order to characterise the approximation properties of the two methods, it is useful to incorporate the constraints acting on the velocity space. 
For the $\HDIV$ method, the divergence-free condition $b_h\rb{\uu_h,q_h}=0$ constrains $(k+1)^d$ out of the $d(k+1)^d$ velocity degrees of freedom per element. 
For the $\LTWO$-based method, the limit $\tau_{\mathrm{C}}, \tau_{\mathrm{D}}\to\infty$ implies continuity in the normal direction and pointwise divergence-free solutions. Both imply constraints on the velocity.
In Appendix~\ref{app:limitmethod} we give a derivation and characterisation of the limit method that the solution of the $\LTWO$-based method converges to as $\tau_{\mathrm{C}}, \tau_{\mathrm{D}}\to\infty$.
Normal-continuity ($\tau_{\mathrm{C}} \to \infty$) corresponds to $(k+1)^{d-1}$ DOFs per coordinate direction, i.e.\ $d (k+1)^{(d-1)}$ DOFs per element, whereas the pointwise divergence-free condition ($\tau_{\mathrm{D}} \to \infty$) corresponds to $(k+1)^d-1$ DOFs (cf.\ $\dim(Q_h^{\text{vol}})$ in Appendix~\ref{app:limitmethod}).
Further, it can be easily seen that pointwise divergence-free velocities imply the weak divergence-free condition, rendering the weak divergence-free condition $d_h\rb{\uu_h,q_h}=0$ in~\eqref{eq:L2DGMethod} superfluous. 
Taking these results together and denoting by $N_{\mathrm{DOFs},\uu}^\text{L2,c}$ and $N_{\mathrm{DOFs},\uu}^\text{Hdiv,c}$ the DOFs that remain after taking the constraints into account, we obtain
\begin{equation}
  N_{\mathrm{DOFs},\uu}^\text{L2,c}(N,k) = N^d \rb{(d-1) \rb{k+1}^d - d (k+1)^{d-1}+1},
\end{equation}
for the DG method with infinite stabilisation. Moreover, one obtains the relation
\begin{equation}
N_{\mathrm{DOFs},\uu}^\text{L2,c}(N,k) < N_{\mathrm{DOFs},\uu}^\text{Hdiv,c}(N,k) = N^d (d-1) \rb{k+1}^d < N_{\mathrm{DOFs},\uu}^\text{L2,c}(N,k+1).
\end{equation}
As a consequence, it makes sense to consider the bracket~$\{k, k+1\}$ as polynomial degrees for the $\LTWO$-DG method to compare against~$\RTk{k}{}$. 
In the setting~$\tau_{\mathrm{C}}<\infty$ and~$\tau_{\mathrm{D}}<\infty$, the velocity approximation has some freedom left that could in principle increase the solution quality and move the $\LTWO$-DG method of degree~$k$ closer to~$\RTk{k}{}$ (and beyond). 
This behaviour again motivates a closer study of the two adjacent polynomial degrees with respect to Raviart--Thomas. 
We argue that the added solution quality by~$\tau_{\mathrm{C}}<\infty$ and~$\tau_{\mathrm{D}}<\infty$ can be expected to be minor because the penalty terms are necessary to render the approximation stable, dominating over ``spurious'' contributions.

\subsection{Mass conservation}
\label{sec:MassConservation}
To highlight the similarities and differences of the two discretisation approaches in terms of discrete mass conservation, it is instructive to reformulate the discrete continuity equation~\eqref{L2VelocityDivergenceWeakForm} for the $\LTWO$-based approach
\begin{align*}
d_h\rb{\uu_h,q_h} 
		&= - \int_\Omega q_h \nabla_h \ip \uu_h \dx
		+ \sum_{F\in\F}\int_F \avg{q_h}\rb{\jmp{\uu_h}\ip\nn} \ds = 0 \ . \label{L2VelocityDivergenceStrongForm}
\end{align*}

For the $\HDIV$-conforming approach, the velocity is normal-continuous across interior facets so that the second term in the above equation vanishes, see equation~\eqref{HdivContinuityEquationWeakForm}. 
By using Raviart--Thomas function spaces the velocity is exactly divergence-free for the $\HDIV$-conforming approach, see equation~\eqref{HdivExactlyDivergenceFree}. 
Accordingly, it can be seen as a specialisation of the $\LTWO$-based approach with additional restrictions for the function spaces of velocity/pressure. 
In case of the $\LTWO$-based approach, these restrictions are not set explicitly; instead, these restrictions are enforced weakly by the use of standard function spaces along with additional stabilisation terms, see equations~\eqref{DivergencePenalty} and~\eqref{ContinuityPenalty}. 

\subsection{Energy balance and dissipation mechanisms}
\label{sec:EnergyBalanceDissipation}

For a unified presentation below one can formally write $\uuhat_h = \uuhat_h\rb{\uu_h}$ for the HDG method and obtain a corresponding DG bilinear form $a_h\rb{\rb{\uu_h,\uuhat_h},\rb{\uu_h,\uuhat_h}} \to a_h\rb{\uu_h,\uu_h}$ (with abuse of notation) which only depends on the usual velocity variable.
~\\

Assuming that $\ff\equiv\zero$, testing symmetrically in \eqref{eq:L2DGMethod} (respectively in \eqref{eq:HdivHDGMethod} with $\uuhat_h = \uuhat_h\rb{\uu_h}$) leads to the discrete energy balance
\begin{equation} \label{eq:DiscEnergyBalance}
	 \underbrace{\frac{1}{2}\frac{\drm}{\drm t} \norm{\uu_h}_\LP{2}{}^2}_{\texttt{dt\_ekin}}
	 + \underbrace{\nu a_h\rb{\uu_h,\uu_h}}_{\texttt{visc\_diss}}
	 + \underbrace{c_h\rb{\uu_h;\uu_h,\uu_h}}_{\texttt{num\_diss\_conv}}
	 + \underbrace{j_h\rb{\uu_h,\uu_h}}_{\texttt{num\_diss\_div}~+~\texttt{num\_diss\_conti}}
	 = 0.
\end{equation}

Whenever $c_h\rb{\uu_h;\uu_h,\uu_h} > 0$ or $j_h\rb{\uu_h,\uu_h} > 0$, the corresponding non-viscosity related mechanism can clearly be characterised as artificial or numerical dissipation since there is no counterpart in the continuous Navier--Stokes model \eqref{eq:TINS}.
~\\

It may occur for some discretisations that~$c_h\rb{\uu_h;\uu_h,\uu_h}$ is sign-indefinite, which may lead to an artificial increase of kinetic energy. 
This is often compensated by additional artificial or numerical dissipation in $a_h\rb{\uu_h,\uu_h}$ and/or $j_h\rb{\uu_h,\uu_h}$. 
To illustrate this, let us explicitly state the discrete energy balance for the $\LTWO$-based approach with local Lax--Friedrichs flux under the assumption of periodic boundary conditions and vanishing body forces (see the work of Fehn et al.~\cite{Fehn2018a} for details)
\begin{align}
\frac{1}{2}\frac{\drm}{\drm t} \norm{\uu_h}_\LP{2}{}^2 = & - \nu a_h\rb{\uu_h,\uu_h} \nonumber - \frac{1}{2} \int_\Omega \nabla_h \ip \uu_h \rb{\uu_h \ip \uu_h}\dx + \frac{1}{2} \sum_{F\in\F}\int_F \jmp{\uu_h} \ip \nn \rb{\uu_h^-\ip\uu_h^+} \ds \\
& -  \sum_{F\in\F}\int_F \jmp{\uu_h} \ip \frac{\Lambda}{2}\jmp{\uu_h} \ds - j_{\mathrm{div},h}\rb{\uu_h,\uu_h} - j_{\mathrm{conti},h}\rb{\uu_h,\uu_h} .
\label{eq:DiscEnergyBalanceL2}
\end{align}

The above equation highlights the second main difference between the $\LTWO$-based and $\HDIV$-based approaches considered in this work. 
The last two terms in the first row of the above equation (belonging to the convective term~$c_h\rb{\uu_h;\uu_h,\uu_h}$) are sign-indefinite for the $\LTWO$-based approach. 
Note that similar terms would also arise for other formulations of the convective term, i.e., the convective formulation with upwind flux discretisation. 
In case of $\HDIV$-conforming function spaces with exactly divergence-free velocity, these two terms add up to zero by definition and only the sign-definite stabilisation term of the convective term contributes to the discrete energy evolution. 
In case of the stabilised $\LTWO$-based approach, the aim of the divergence and continuity penalty terms (which are positive-semidefinite, i.e., they exhibit a purely dissipative character) is to compensate sign-indefinite terms of the convective term.
~\\

Let us also mention that in DG discretisations, $a_h\rb{\uu_h,\uu_h}$ consists of two types of viscous dissipation: physical and numerical dissipation. 
Despite its frequent use in the literature, the broken gradient norm $\norm{\nabla_h \uu_h}_\LP{2}{}^2$ is not a good measure for the physical dissipation in under-resolved situations as the corresponding numerical dissipation $a_h\rb{\uu_h,\uu_h} - \norm{\nabla_h \uu_h}_\LP{2}{}^2$ is in general not sign-definite (even if the viscosity bilinear form is coercive). 
We refer to Ref.~\cite{LehrenfeldLubeSchroeder18} for an alternative evaluation of physical and numerical dissipation for DG methods. 
In the remainder of this work, we circumvent this problem by only considering the sum of physical and numerical viscous dissipation.

\section{Decaying homogeneous isotropic turbulence: \\ The 3D Taylor--Green vortex problem}
\label{sec:TaylorGreen}

In the box $\Omega=\rb{0,2\pi}^3$, equipped with periodic boundary conditions on all faces, consider the case $\ff\equiv\zero$ and the space-periodic initial condition \cite{TaylorGreen37,Brachet91}
\begin{equation}
	\uu_0\rb{\xx}
	= \begin{pmatrix}
		\cos\rb{x_1} \sin\rb{x_2} \sin\rb{x_3} \\
		-\sin\rb{ x_1} \cos\rb{ x_2} \sin\rb{x_3} \\
		0
	\end{pmatrix}.
\end{equation}
~\\
This initial condition is imposed and the resulting flow is monitored over time.
The resulting Taylor--Green vortex (TGV) problem is possibly the easiest flow system for which one can observe the key physical mechanisms inherent to turbulence: transition, vortex roll-up, 3D vortex stretching and interaction, and finally, in the viscous case, molecular energy dissipation.
As done frequently in the literature, simulations are run for~$\Rey=\nu^{-1}=\num{1600}$ until~$\tend=20$; see Refs.~\cite{WiartEtAl13,GassnerBeck12,PiatkowskiEtAl18}. 
All simulations are performed on structured hexahedral meshes consisting of~$N^3$ cubes and polynomial order~$k$.
The accuracy of the under-resolved simulations performed here is assessed by comparing the results to accurate DNS reference data from Ref.~\cite{Fehn2018b} with~$k=7$ and~$N=128$.

\subsection{Comparison of methods}

\begin{figure}[t]
\centering
	\includegraphics[width=0.45\textwidth]
 		{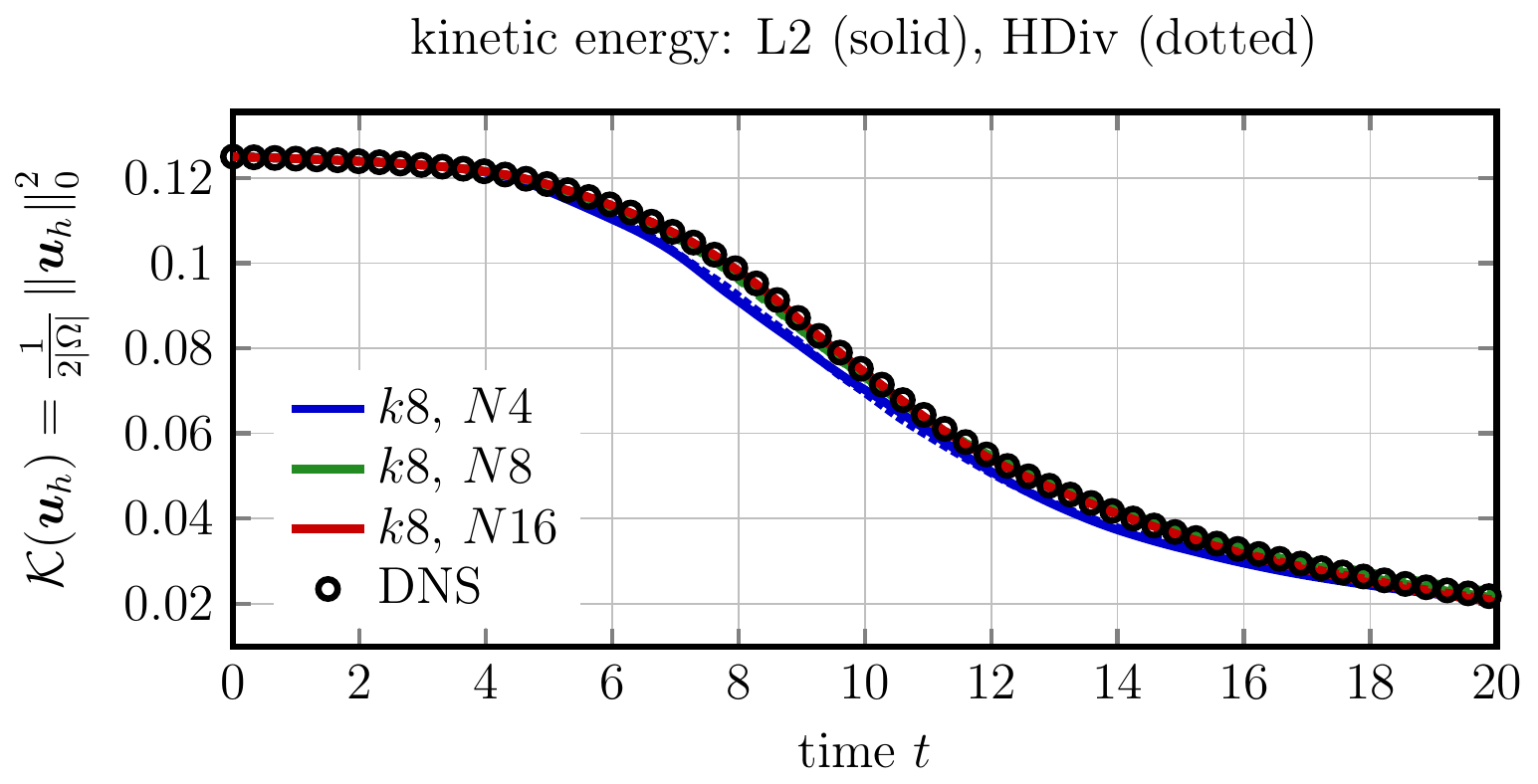} \goodgap
 	\includegraphics[width=0.45\textwidth]
		{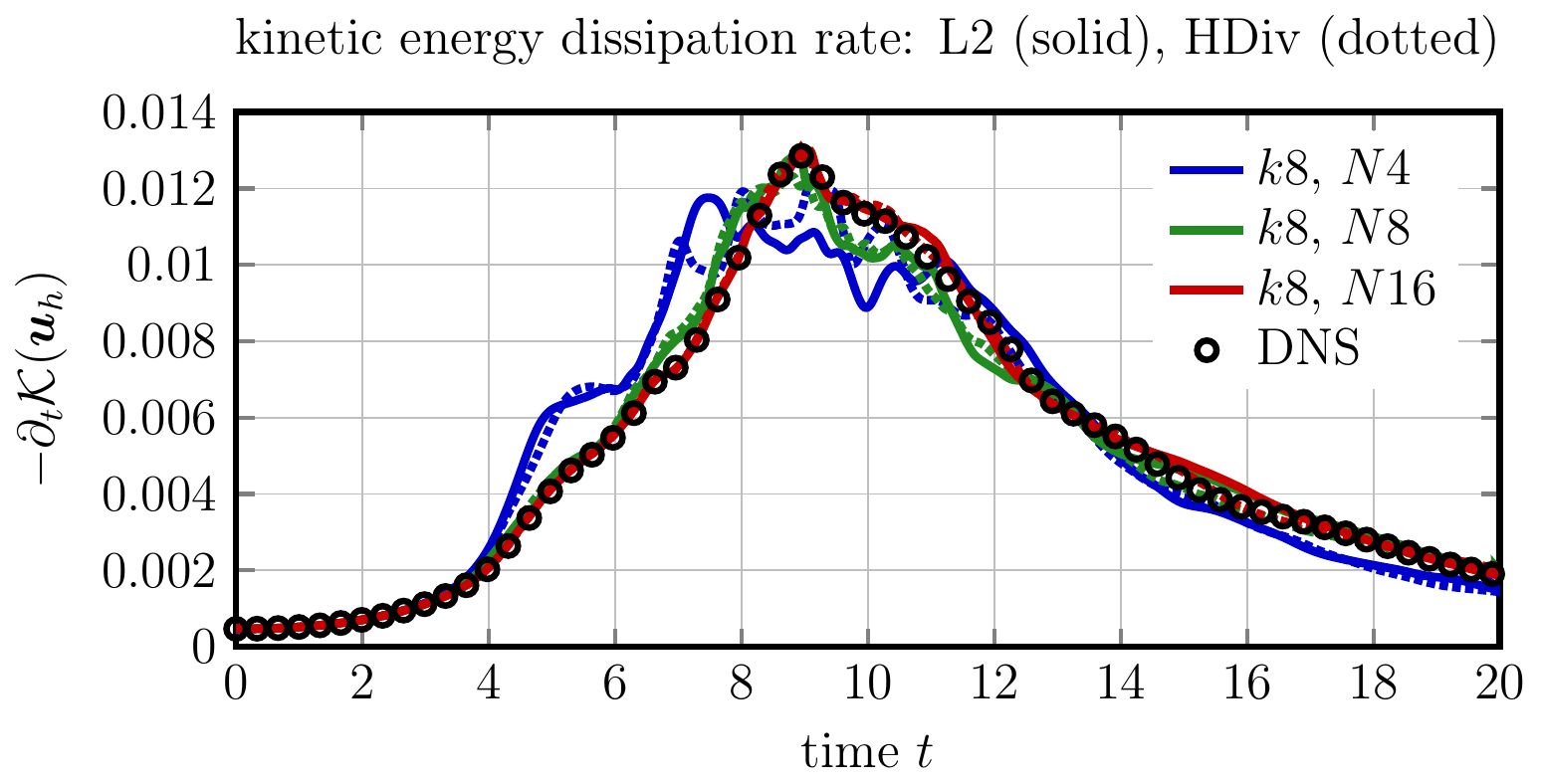} \\ \vspace{10pt}
 	\includegraphics[width=0.375\textwidth]
 		{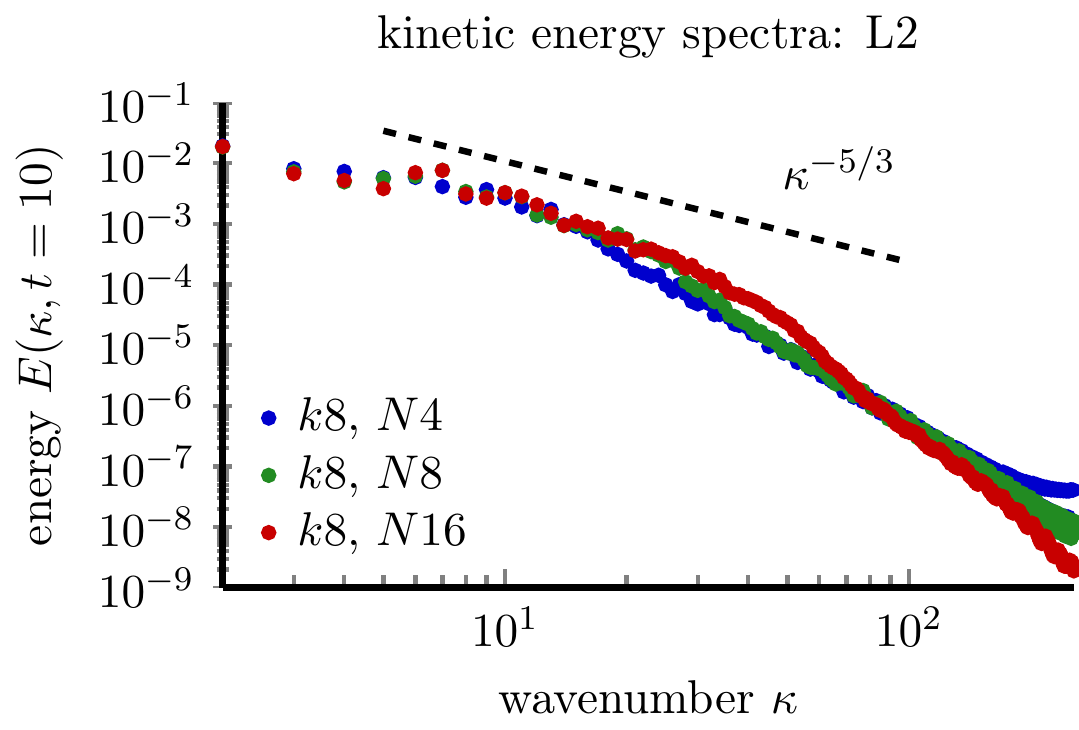} \goodgap \goodgap \goodgap 
 	\includegraphics[width=0.375\textwidth]
 		{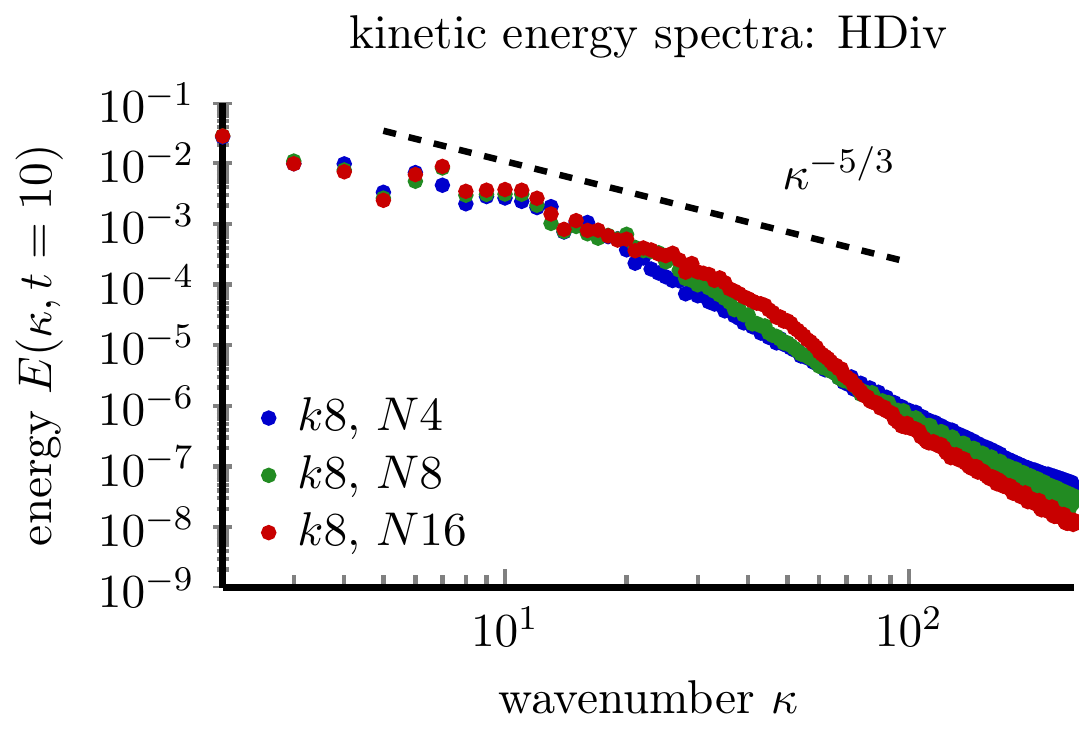} 
 	\caption{Comparisons of high-order ($k=8$) $\LTWO$- and $\HDIV$-based simulations for the TGV. Evolution of kinetic energy (top left), total kinetic energy dissipation rate (top right) and kinetic energy spectra (bottom) at $t=10$ on different meshes with $N\in\set{4,8,16}$. }
 	\label{fig:3DTGV-L2vsHdiv-EnergyDissRateSpectra}
 \end{figure}

At first, we want to compare the results obtained by the $\LTWO$ method \eqref{eq:L2DGMethod} and the $\HDIV$ method \eqref{eq:HdivHDGMethod} on the basis of the evolution of kinetic energy $\Kin\rb{\uu_h}=\frac{1}{2\abs{\Omega}}\norm{\uu_h}_\LP{2}{}^2$ and (negative) total kinetic energy dissipation rate $-\partial_t\Kin\rb{\uu_h}$ for the high-order choice $k=8$; see also \eqref{eq:DiscEnergyBalance}.
In Fig.~\ref{fig:3DTGV-L2vsHdiv-EnergyDissRateSpectra} (top) both quantities are displayed on a sequence of meshes $N\in\set{4,8,16}$, which correspond to strongly under-resolved, moderately resolved and essentially resolved simulations.
The most important observation is that the results are, in general, not significantly different with respect to the particular quantity.
While on the coarse meshes, slight differences between $\LTWO$ and $\HDIV$ results can be observed, they are basically identical on the finest mesh.
Moreover, it is comparably easy to have a good approximation for the kinetic energy, even in a strongly under-resolved situation, but in order to capture the total dissipation rate, significantly more resolution is necessary.
In the remainder of this section, we will thus not discuss the evolution of kinetic energy anymore, but focus on the more demanding energy dissipation rate.
~\\

Let $E\rb{\kappa}$ be defined as the amount of kinetic energy concentrated in the wavenumber vector $\ka\in\R^n$ with wavenumber $\kappa\coloneqq\abs{\ka}$, the energy spectrum.
Concerning the distribution of $E\rb{\kappa}$ over different wavenumbers $\kappa$, Fig~\ref{fig:3DTGV-L2vsHdiv-EnergyDissRateSpectra} (bottom) shows the kinetic energy spectrum at $t=10$ (shortly after the dissipation peak) for both methods under $h$-refinement.
The most important observation is that the $\LTWO$- and the $\HDIV$-based results are very similar to each other.
For the considered Reynolds number and spatial resolutions, the flow exhibits only a comparably small inertial range where the Kolmogorov $-5/3$ rule can be observed.
However, the inertial range becomes larger for finer resolutions for which viscous effects are better resolved and, hence, more energy gets dissipated in the fine scales. 
On the coarse mesh, the resolution is not sufficient to dissipate the fine-scale features on the molecular level.
~\\

\begin{figure}[t]
\centering
	\includegraphics[width=0.45\textwidth]
		{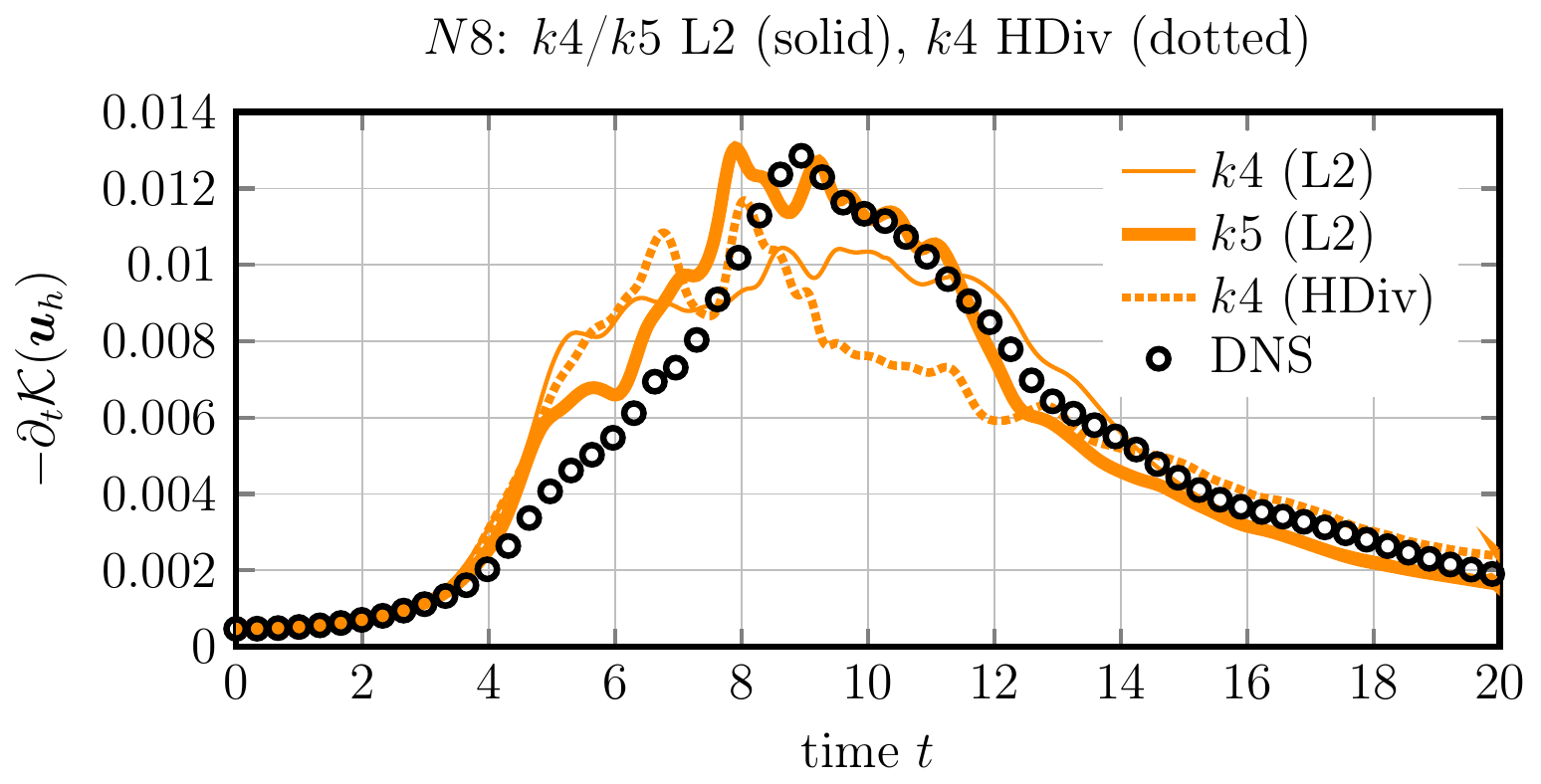} \goodgap
	\includegraphics[width=0.45\textwidth]
		{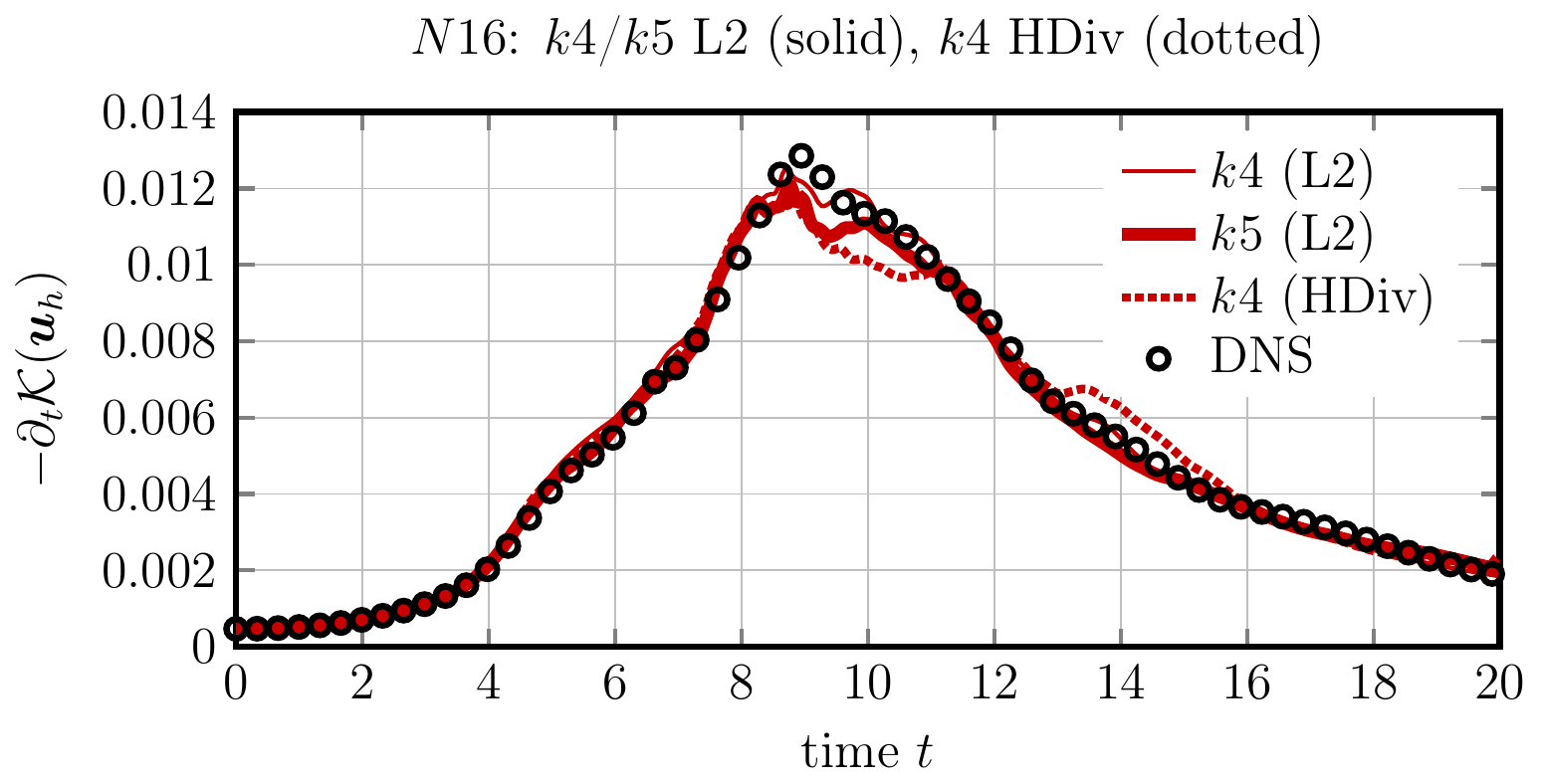} \\
	\includegraphics[width=0.45\textwidth]
		{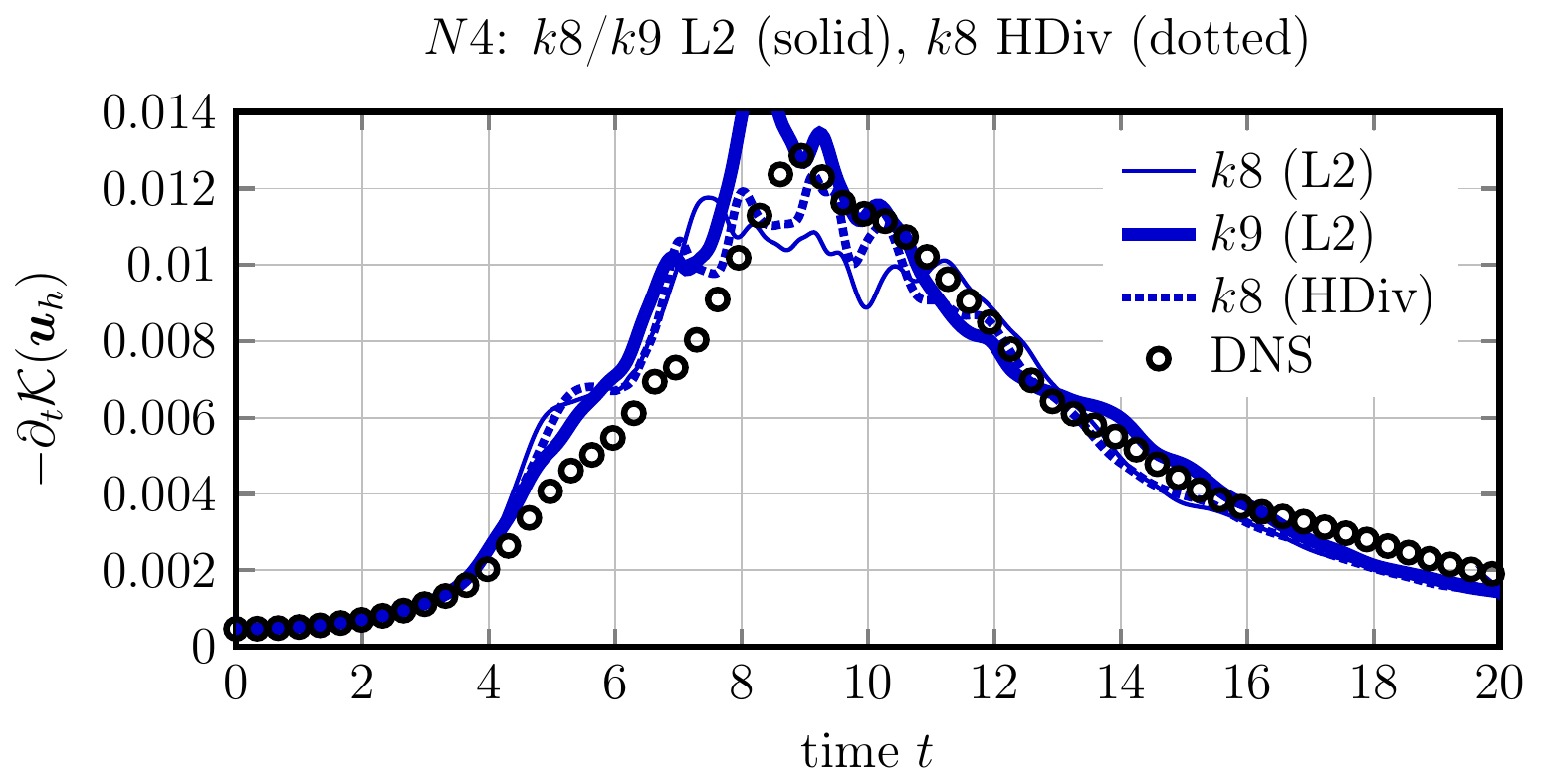} \goodgap
	\includegraphics[width=0.45\textwidth]
		{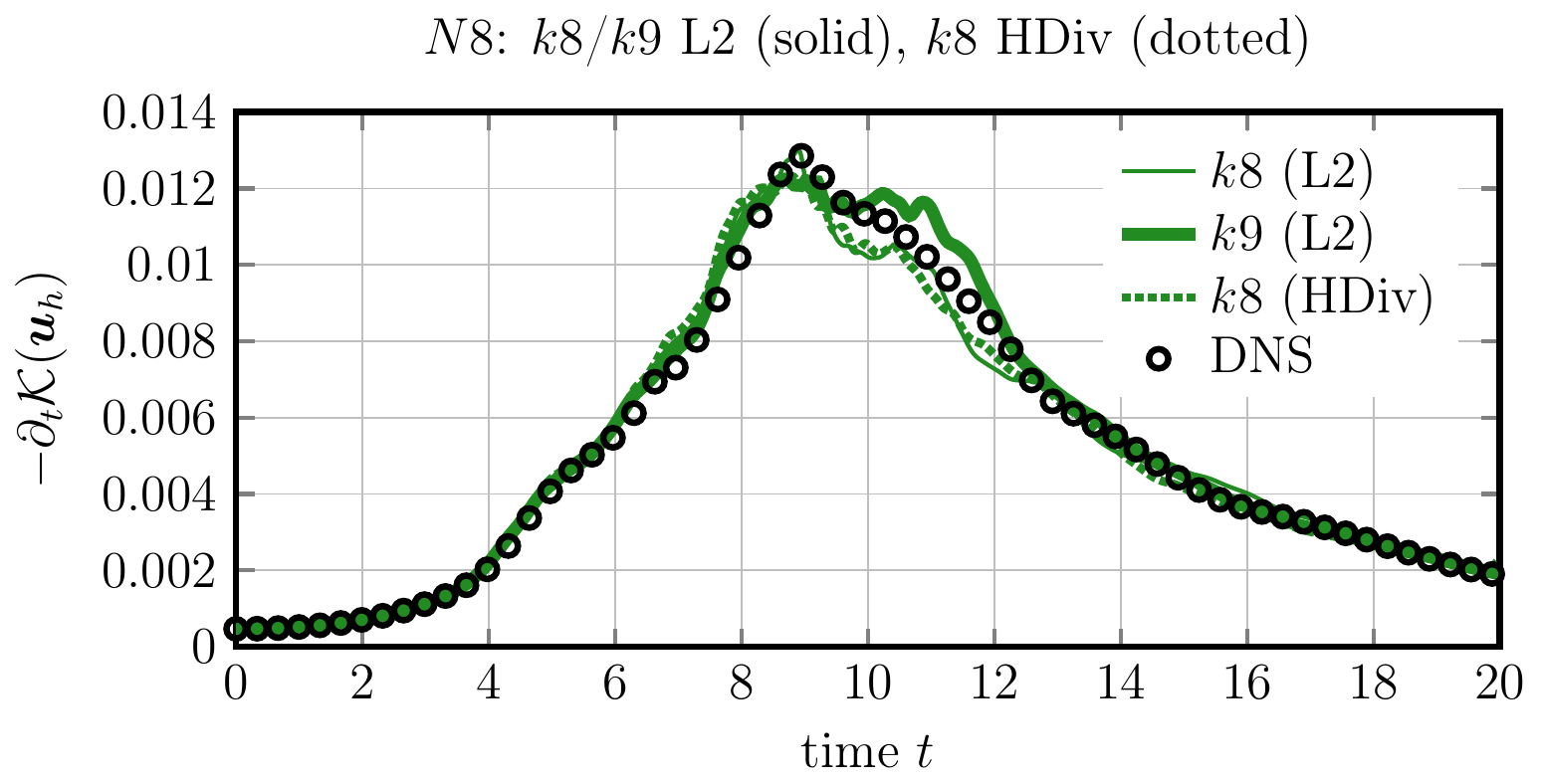} 
	\caption{Comparison of $\HDIV$-based method of order $k$ with $\LTWO$-based method of order $k$ and $k+1$. Shown are the respective total dissipation rates.}
	\label{fig:3DTGV-RTvsQkQkp1}
\end{figure}

Next, we refine the above analysis by comparing the $\HDIV$-conforming method of degree~$k$ with the $\LTWO$-conforming method of degrees~$k$ and~$k+1$, so that the $\HDIV$-conforming method is located in between the $\LTWO$-conforming method in terms of degrees of freedom, see Sec.~\ref{sec:DG-solvers}.
We show results of this comparison in Fig.~\ref{fig:3DTGV-RTvsQkQkp1}for~$k=4$ in the top row and for~$k=8$ in the bottom row. 
For~$k=4$, the $\LTWO$-conforming method of degree~$k$ appears to be of similar accuracy or slightly more accurate than the $\HDIV$-conforming method of degree~$k$. 
For degree~$k=8$, the $\LTWO$-method and $\HDIV$-method of degree~$k$ also show similar accuracy, where the $\HDIV$-method tends to be slightly more accurate for the~$N=4$ case. 
The $\LTWO$-conforming method of degree~$k+1$ seems to be the most accurate one in this comparison but we want to emphasise that it is difficult to draw precise conclusions due to the wriggling behaviour of the kinetic energy dissipation rate, especially for the coarser resolutions shown in the left part of the figure. 
Overall, these results suggest that both methods provide a similar level of accuracy for the Taylor--Green vortex problem investigated here, if the function spaces for the individual methods are chosen in a way that they offer comparable resolution capabilities in terms of DOFs. 
For all following considerations, comparisons between the two methods are made on the basis of the same polynomial degree~$k$, which also results in a fairly comparable number of degrees of freedom according to Sec.~\ref{sec:DG-solvers}.
 
\subsection{High-order dissipation mechanisms under $h$-refinement}

\begin{figure}[b]
 \centering
 	\includegraphics[width=0.275\textwidth]
		{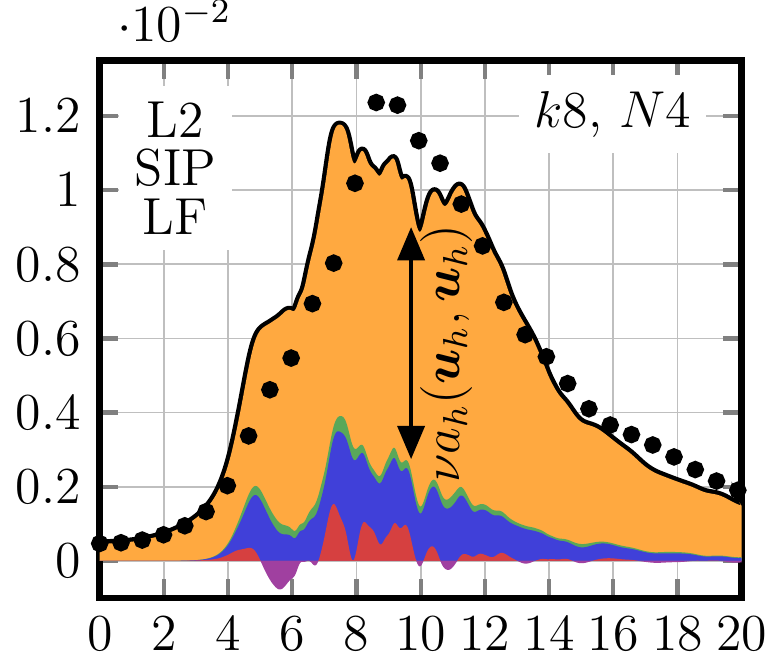}
 	\includegraphics[width=0.275\textwidth]
		{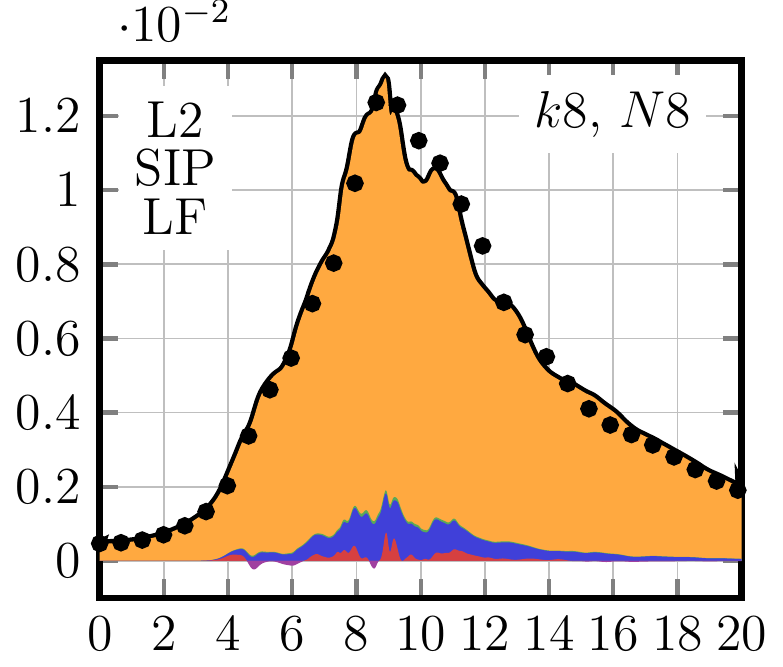}
 	\includegraphics[width=0.275\textwidth]
		{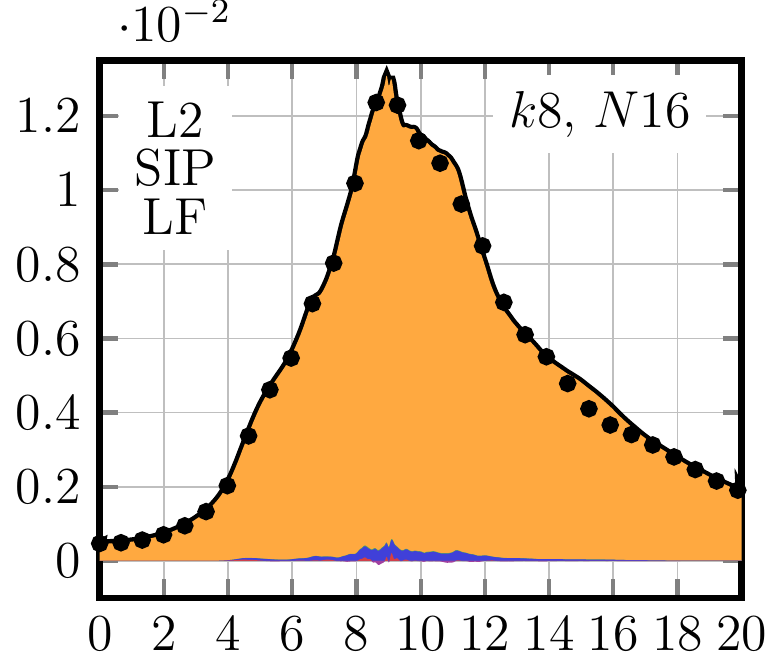}
 	\includegraphics[width=0.125\textwidth]
		{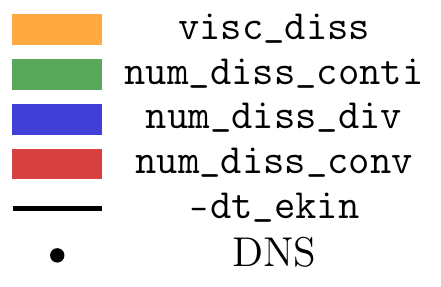}	\\
 	\includegraphics[width=0.275\textwidth]
		{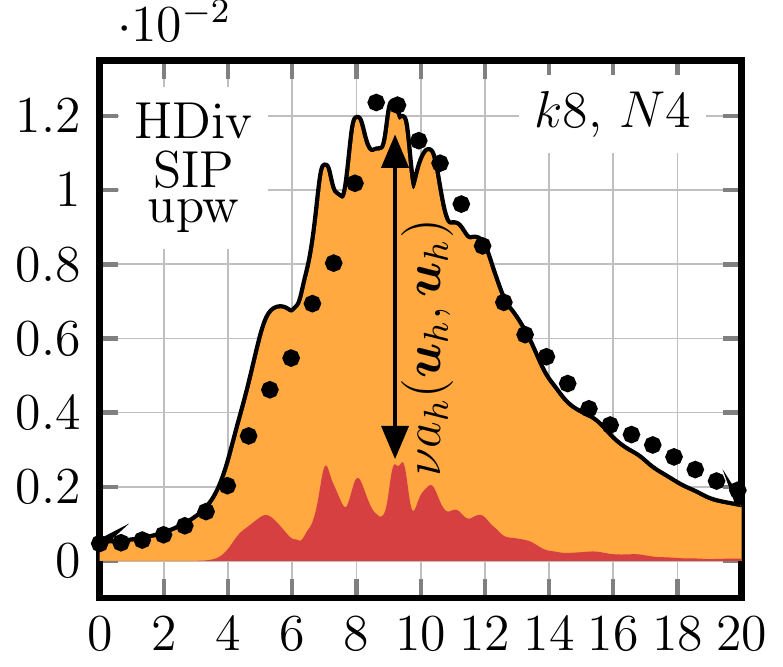}
 	\includegraphics[width=0.275\textwidth]
		{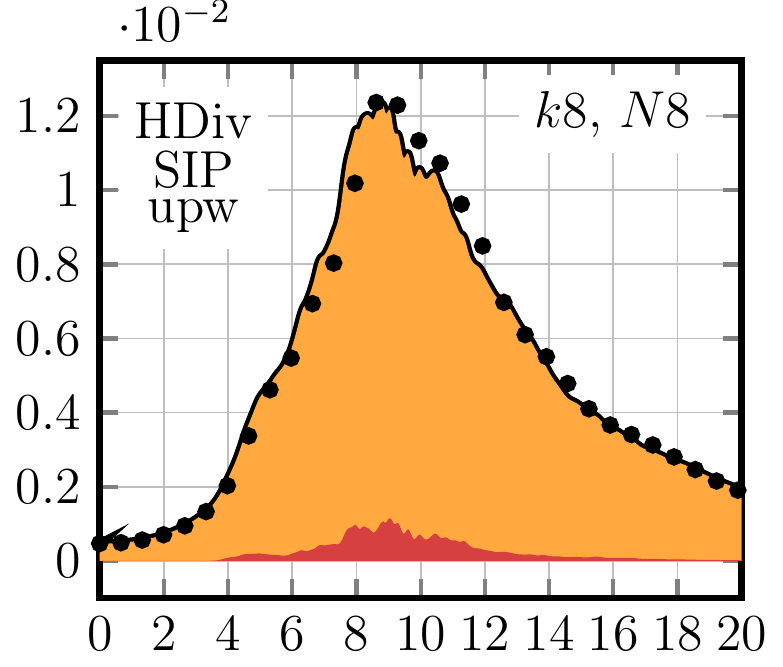}
 	\includegraphics[width=0.275\textwidth]
		{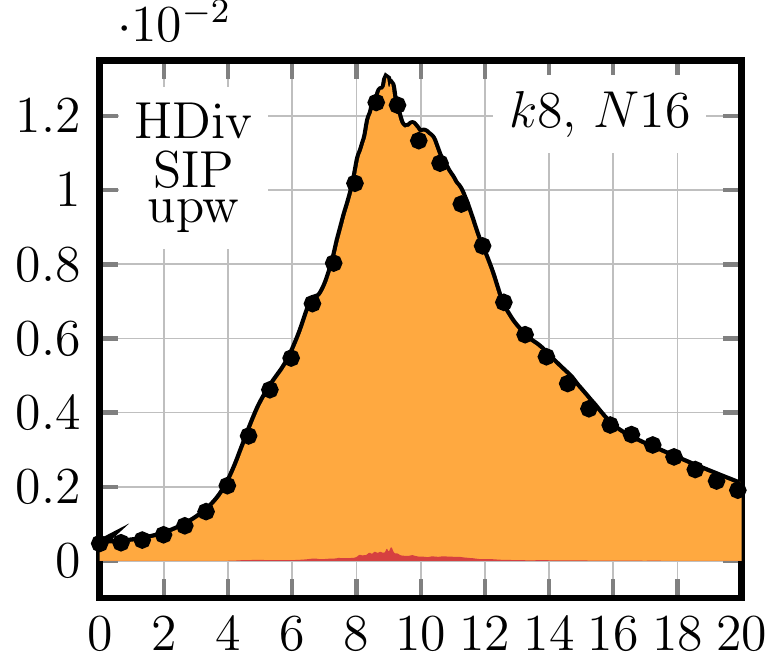}
 	\includegraphics[width=0.125\textwidth]
		{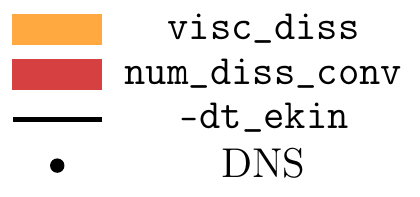}									
 	\caption{Dissipation mechanisms for $k=8$ under $h$-refinement (considering meshes with $N\in\set{4,8,16}$). The abscissa shows the time $t$. }
 	\label{fig:3DTGV-ILES-hRefine}
\end{figure}

In Fig.~\ref{fig:3DTGV-ILES-hRefine} we consider a fixed order ($k=8$) under $h$-refinement and display the sum of different contributions to the dissipation. 
Note that the plot is stacked in the sense that the solid line on the top of the orange area represents the sum of all the contributions below. 
For the $\LTWO$-based method (top row), there are four contributions, cf.\ \eqref{eq:DiscEnergyBalance}, while there are only two (non-zero) contributions for the $\HDIV$-based method. 
One observes that the sign-indefinite convection parts in the $\LTWO$-based methods (where the corresponding part is negative, the area is coloured purple) are compensated by the stabilisation terms. 
For both methods, one observes that the dominant contribution in the dissipation stems from the viscosity, which eventually is the only relevant dissipation mechanism once the flow is sufficiently resolved, e.g.\ $(k,N) = (8,16)$.
~\\

\subsection{High-order vs.\ low-order dissipation mechanisms for fixed under-resolution}
 
\begin{figure}[t]
\centering
 	\includegraphics[width=0.275\textwidth]
		{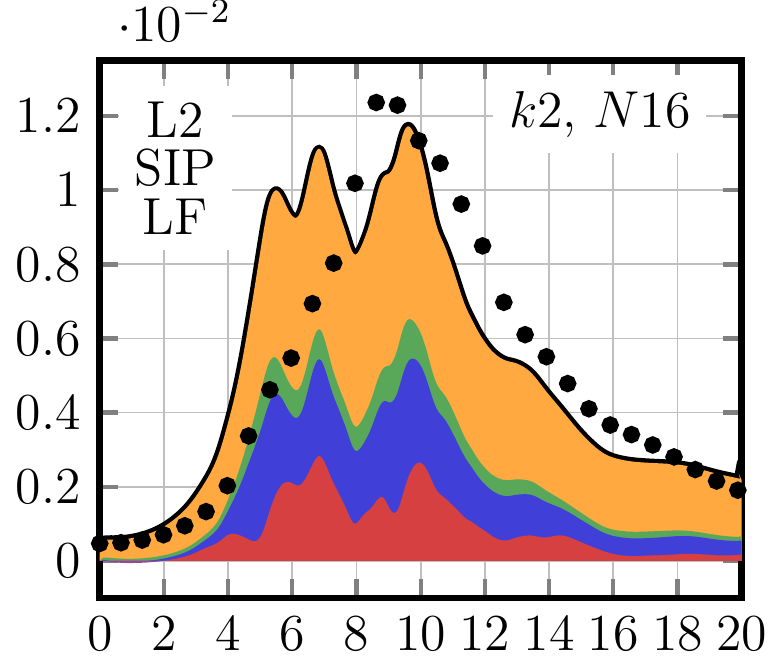}
 	\includegraphics[width=0.275\textwidth]
		{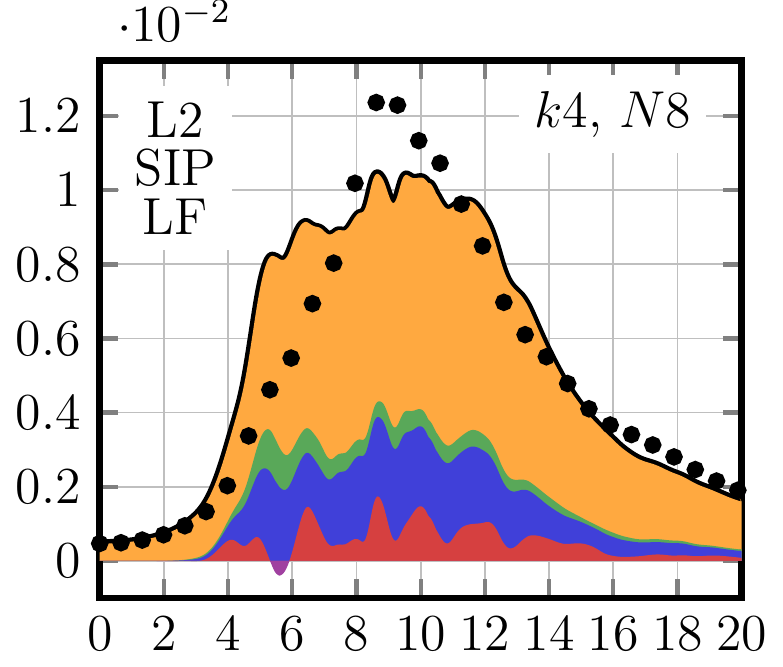}
 	\includegraphics[width=0.275\textwidth]
		{TaylorGreen/ILES-plot/tikz/3DTGV_ILES_L2_k8_N4.pdf}
 	\includegraphics[width=0.125\textwidth]
		{TaylorGreen/ILES-plot/tikz/3DTGV_ILES_L2_legend.pdf}	\\
 	\includegraphics[width=0.275\textwidth]
		{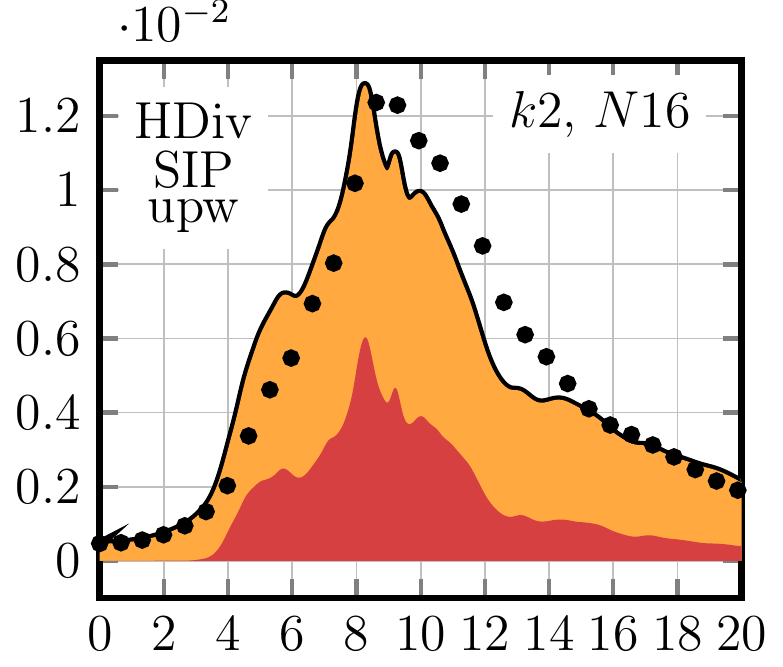}
 	\includegraphics[width=0.275\textwidth]
		{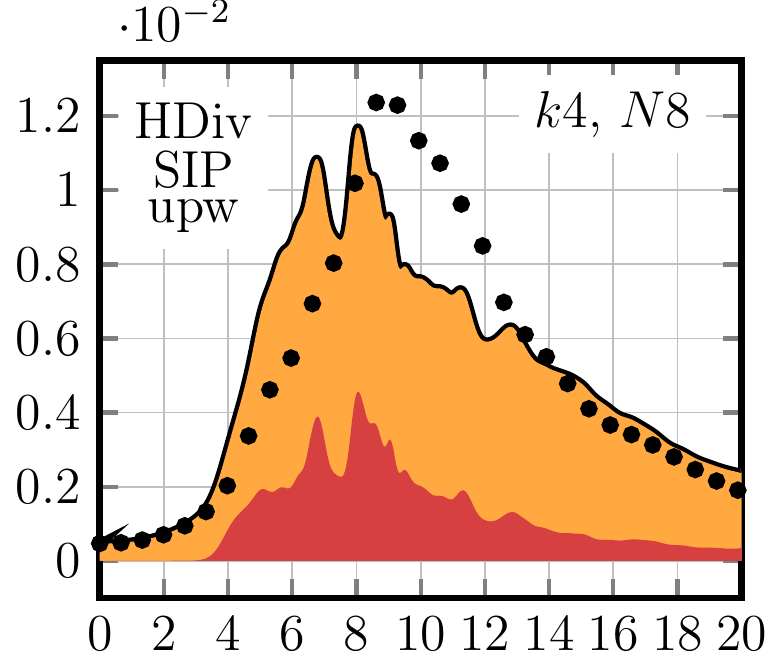}
 	\includegraphics[width=0.275\textwidth]
		{TaylorGreen/ILES-plot/tikz/3DTGV_ILES_HDiv_k8_N4.pdf}
 	\includegraphics[width=0.125\textwidth]
		{TaylorGreen/ILES-plot/tikz/3DTGV_ILES_HDiv_legend.pdf}									
 	\caption{Dissipation mechanisms for fixed strong under-resolution under $k$-refinement. The abscissa shows the time $t$.}
 	\label{fig:3DTGV-ILES-FixedRes}
\end{figure}

Up to now, we only considered a fixed polynomial order of~$k=8$ and did $h$-refinement procedures.
This section is concerned with the question whether high-order computations are actually superior to lower-order ones, given that a comparably fair environment is provided.
We already saw that whenever the resolution is sufficiently large, no differences between different methods can be observed anymore with the naked eye.
Therefore, here, we restrict ourselves to the strongly under-resolved situation~$(k,N)=(2,16),(4,8),(8,4)$ while we note that~$ 
N_{\mathrm{DOFs},\uu}(2,16) 
> N_{\mathrm{DOFs},\uu}(4,8) 
> N_{\mathrm{DOFs},\uu}(8,4) 
$
, cf.\ \eqref{eq:ndof}, i.e.\ the low-order methods have slightly more unknowns.
In Fig.~\ref{fig:3DTGV-ILES-FixedRes}, the evolution of the kinetic energy dissipation rate (\texttt{-dt\_ekin}) shows the differences between high-order or low-order methods.
We observe that the resolution improves when going to higher-order (although using slightly fewer unknowns).
However, one can also see that the total dissipation of high-order methods is concentrated in the viscous term (\texttt{visc\_diss}); cf.\ \eqref{eq:DiscEnergyBalance}.
Consequently, in the strongly under-resolved situation, low-order methods fundamentally rely on the numerical dissipation provided by stabilisation.
~\\
 
Summarising, we believe that high-order methods more accurate than low-order methods for freely decaying turbulence problems such as the Taylor--Green vortex.
Therefore, in the following, we will restrict ourselves to $k=8$ and investigate the influence of different treatments of convection, viscosity, and divergence-conformity.

\subsection{Influence of different treatments of convection, viscosity and divergence-conformity}

In this section, we will investigate the influence of small changes to the considered methods.
Namely, we compare different fluxes for the discrete convection term for both $\LTWO$ and $\HDIV$ method, investigate the impact of different viscosity treatments for the $\HDIV$ method and vary the penalisation parameter for the $\LTWO$ method.
As the particular differences decrease with an increasing resolution, we will only consider the strongly and mildly under-resolved cases $(k,N) = (8,4), (8,8)$.

\subsubsection{Different fluxes for the convection term}
\label{sec:TGVConvetiveTerm}

Let us briefly explain the different choices of the discrete convective term~$c_h\rb{\cdot;\cdot,\cdot}$ which are to be considered and compared here.
The basic variants (also used for all computations above) are the Lax--Friedrichs form~\eqref{eq:L2-convection} for $\LTWO$ and the upwind form~\eqref{eq:Hdiv-convection} with~$\theta=1$ for~$\HDIV$.
Now, we will also use the upwind form \eqref{eq:Hdiv-convection} with~$\theta=1$ and the central flux form~\eqref{eq:Hdiv-convection} with~$\theta=0$ for the~$\LTWO$ method, where a single-valued normal flux is ensured via~$\ww_h\ip\nn\to\avg{\ww_h}\ip\nn$.
For the~$\HDIV$ method, we additionally regard the performance of the central flux form~\eqref{eq:Hdiv-convection} with $\theta=0$, and the Lax--Friedrichs (LF) form~\eqref{eq:L2-convection}, where~$\Lambda(\ww_h) = 2\abs{\ww_h\ip\nn}$.
Note that for exactly divergence-free~$\HDIV$ methods (and under exact numerical integration), the LF form simply emerges from~\eqref{eq:Hdiv-convection} by choosing~$\theta=2$ and performing integration by parts of the corresponding volume term.
Put differently, the difference between LF and upwinding for the~$\HDIV$ method is solely a factor two in the positive semi-definite facet stabilisation term.
On the other hand, for the~$\LTWO$ method, integration by parts using a not exactly divergence-free convective velocity leads to a conceptually different discrete convection term.
~\\

\begin{figure}[t]
\centering
 	\includegraphics[width=0.275\textwidth]
		{TaylorGreen/ILES-plot/tikz/3DTGV_ILES_L2_k8_N4.pdf}
 	\includegraphics[width=0.275\textwidth]
		{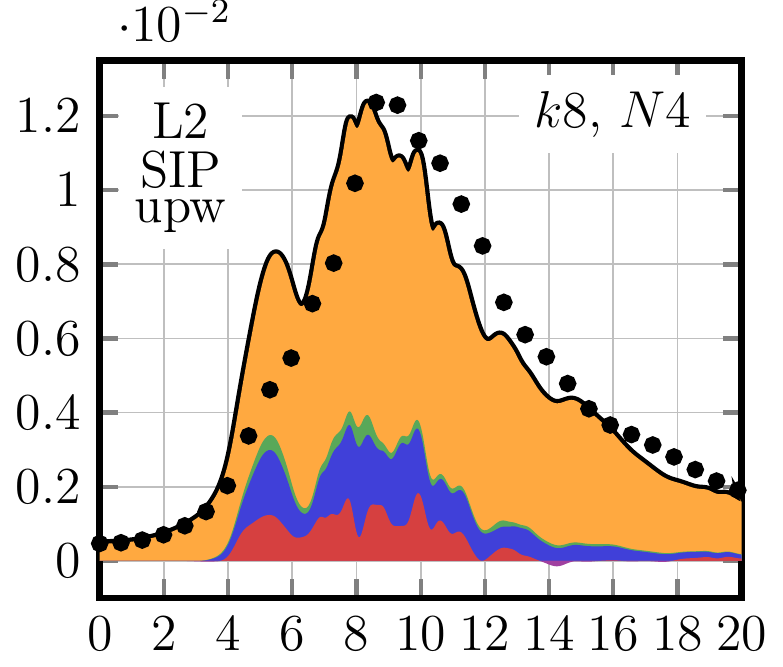} 
 	\includegraphics[width=0.275\textwidth]
		{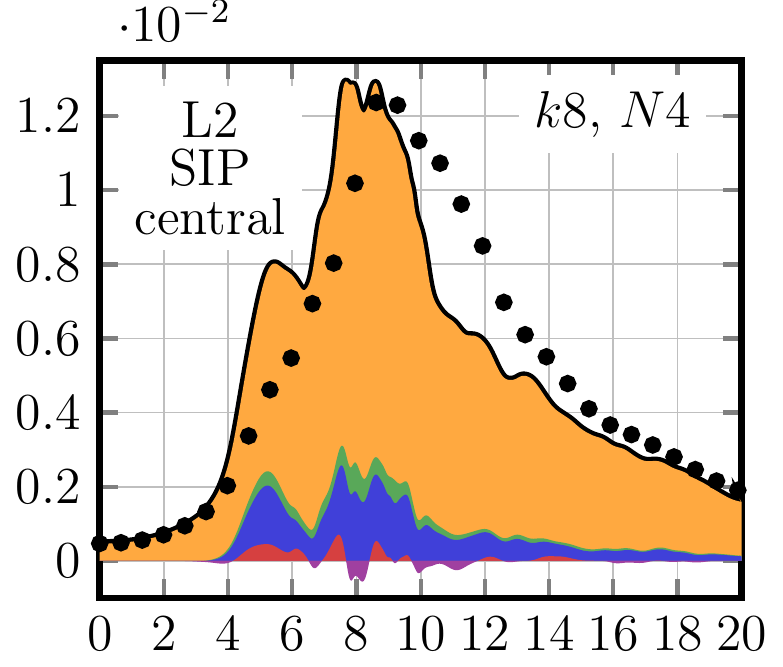} 
 	\includegraphics[width=0.125\textwidth]
		{TaylorGreen/ILES-plot/tikz/3DTGV_ILES_L2_legend.pdf} \\
 	\includegraphics[width=0.275\textwidth]
		{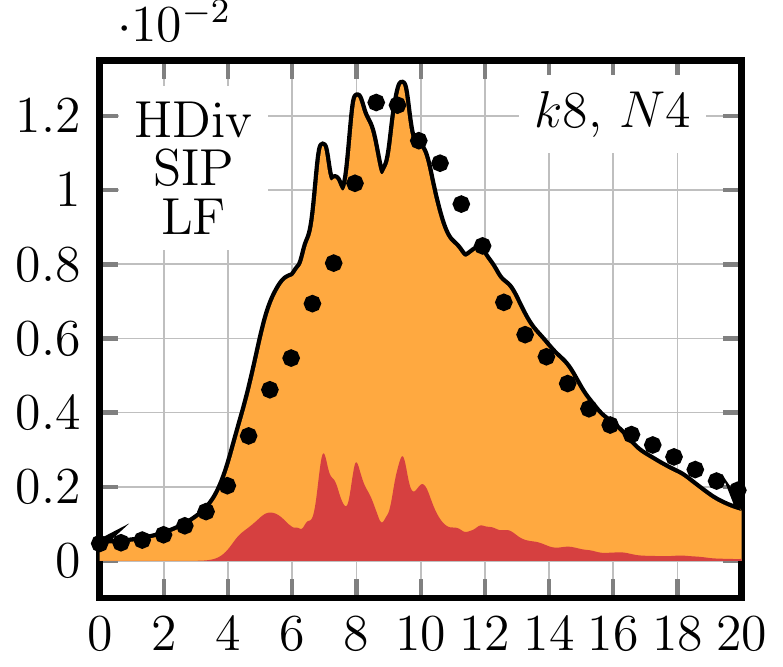}
 	\includegraphics[width=0.275\textwidth]
		{TaylorGreen/ILES-plot/tikz/3DTGV_ILES_HDiv_k8_N4.pdf}
 	\includegraphics[width=0.275\textwidth]
		{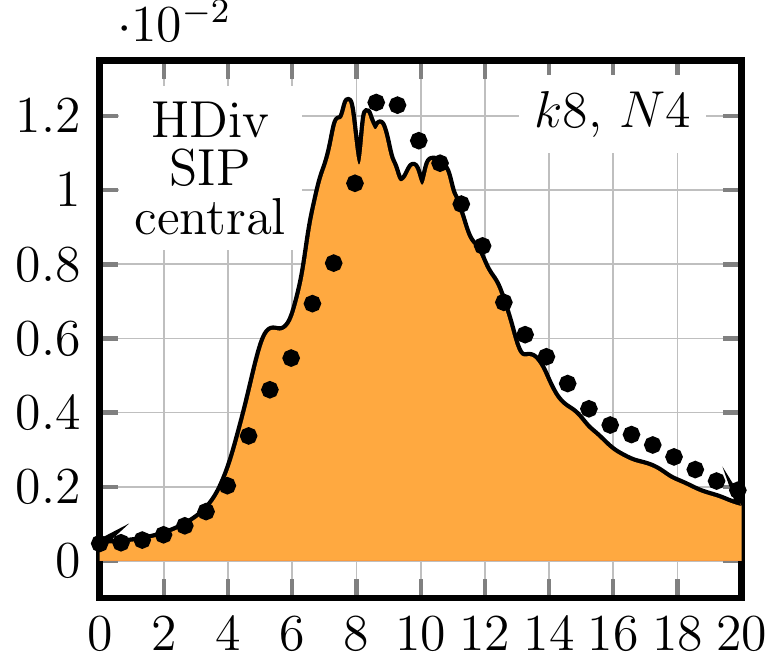}
 	\includegraphics[width=0.125\textwidth]
		{TaylorGreen/ILES-plot/tikz/3DTGV_ILES_HDiv_legend.pdf}
 	\caption{Dissipation mechanisms for different convection fluxes for $k=8$, $N=4$. Stacked dissipation rates for $\LTWO$ results (top) and $\HDIV$ results (bottom). Lax--Friedrichs form (left column), upwind form (middle column) and central flux/no stabilisation (right column). The abscissa shows the time $t$.}
 	\label{fig:3DTGV-ILES-convection}
\end{figure}

The corresponding results for the stacked dissipation rates for~$k=4$, $N=8$ simulations can be seen in Fig.~\ref{fig:3DTGV-ILES-convection}.
Regarding the total dissipation rate (\texttt{-dt\_ekin}), it does not seem to be possible to draw a conclusion as to whether one particular convective term is superior to the others.
In fact, all variants result in a comparably accurate solution even in this strongly under-resolved situation.
Moreover, regarding the LF and upwinding~$\HDIV$ results, one can observe that the amount of convective dissipation remains approximately the same regardless of which convection stabilisation is used.
In particular, the additional factor two in the LF form does not lead to more numerical dissipation.
Surprisingly, the third column shows that both the~$\LTWO$- and the~$\HDIV$-based method can even manage the case without convection stabilisation in form of upwinding or LF for the TGV problem.
We conjecture that choosing a seemingly less dissipative convection term is simply counterbalanced by other (numerical) dissipation mechanisms.
Summarising, the particular choice of the discretisation of the convection term does not have a large impact on the results for this TGV problem.
~\\

\subsubsection{Different viscosity treatment for the \emph{H}(div)-HDG method}

In addition to considering variations in the convection term in the last subsection, let us now take a look at what happens when we do not use the popular SIP method for the treatment of the viscosity effects.
The motivation for investigating a different method for Laplacian-like terms originates in the frequently heard criticism that the SIP parameter choice can lead to overly large penalties and, therefore, a seemingly large amount of numerical viscous dissipation.
To this end, we use a more subtle mechanism to ensure discrete coercivity of the viscosity bilinear form through a so-called lifting technique.
The resulting idea/concept goes back to Bassi \& Rebay~\cite{BassiRebay97} and we will use it only in the $\HDIV$-HDG context; see also Ref.~\cite{Oikawa10,SchoeberlLehrenfeld13}.
~\\

The usual penalty term in the SIP formulation is responsible for controlling the skeleton integrals involving non-quadratic forms. 
To obtain this control with the penalty integral, the penalty parameter has to be chosen sufficiently large, depending on a constant which, in turn, depends on polynomial trace inverse inequalities. 
Lifting methods tackle the problem differently by reformulating the problematic skeleton integrals as volume integrals. 
From this a stabilisation bilinear form can be characterised that guarantees non-negativity of the resulting bilinear form without the usual SIP jump penalty. 
This additional stabilisation bilinear form relies on a new variable, the discrete lifting, whose DOFs can all be eliminated locally by static condensation. 
Therefore, their efficiency is comparable to the corresponding method without lifting.
Similarly, in the context of hybrid mixed DG methods, the additional (dual) diffusive flux variable for approximating~$\nabla\uu$ can also be eliminated locally by static condensation, thereby resulting in a comparable situation in terms of global free DOFs. 
In fact, many mixed formulations can be expressed as primal formulations involving lifting operations \cite{CockburnEtAl09}.
~\\

\begin{figure}[b]
\centering
 	\includegraphics[width=0.275\textwidth]
		{TaylorGreen/ILES-plot/tikz/3DTGV_ILES_HDiv_k8_N4.pdf}
 	\includegraphics[width=0.275\textwidth]
		{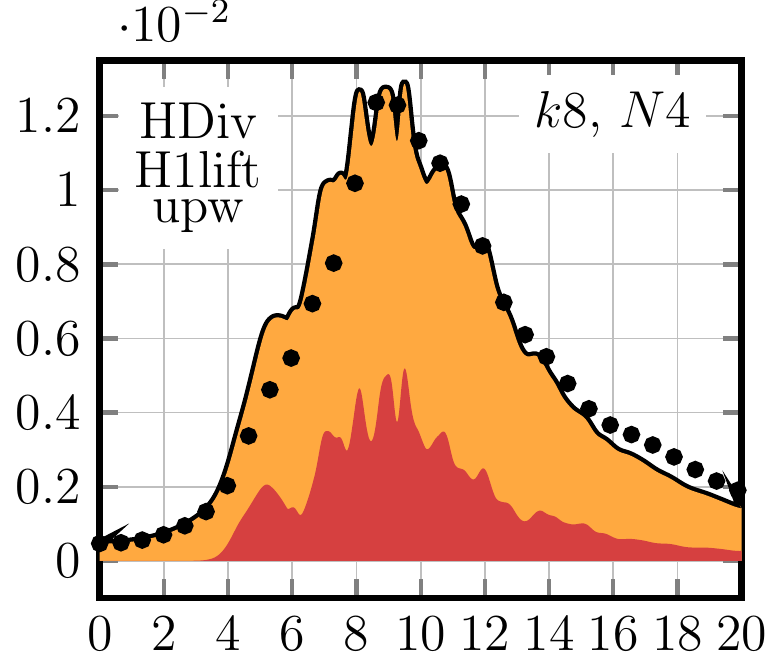} 
 	\includegraphics[width=0.125\textwidth]
		{TaylorGreen/ILES-plot/tikz/3DTGV_ILES_HDiv_legend.pdf}	\\
 	\includegraphics[width=0.275\textwidth]
		{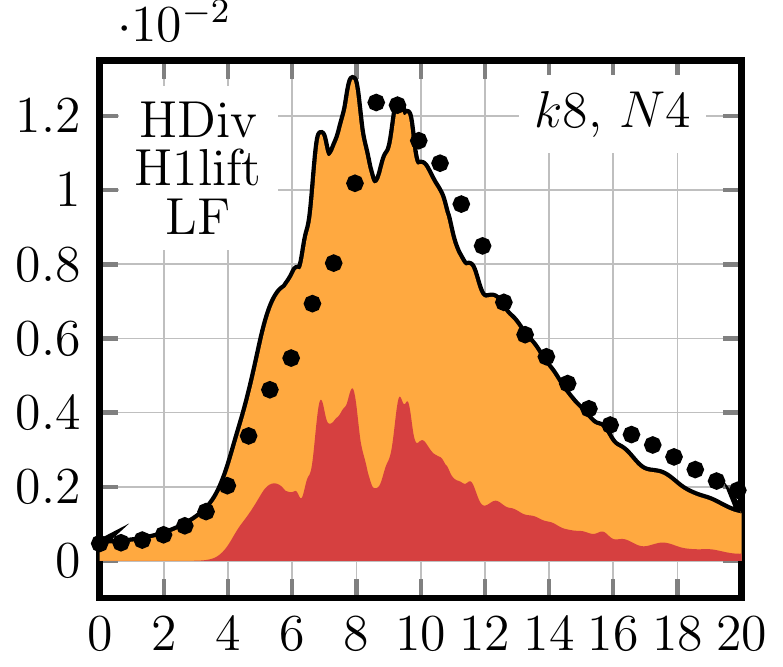}
 	\includegraphics[width=0.275\textwidth]
		{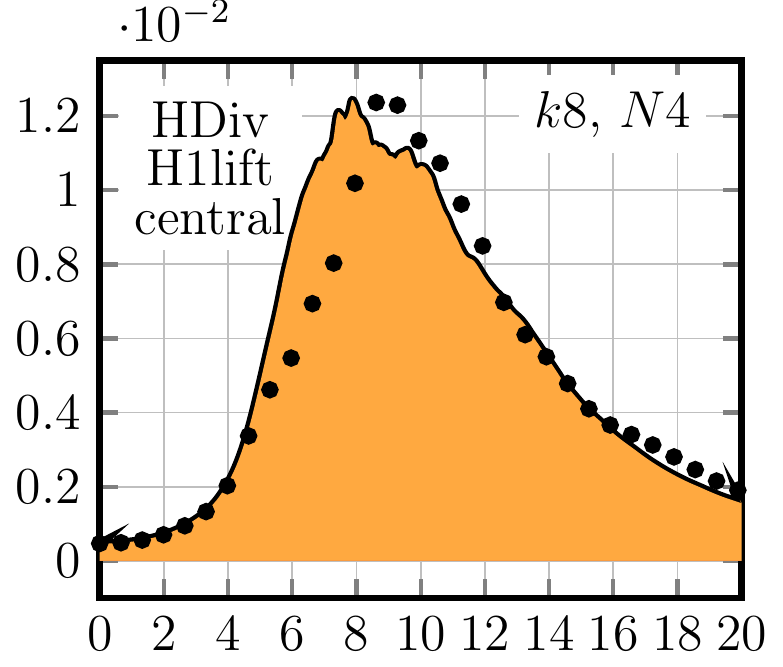}		
 	\includegraphics[width=0.125\textwidth]
		{TaylorGreen/ILES-plot/tikz/3DTGV_ILES_HDiv_legend.pdf}
 	\caption{$\HDIV$ dissipation mechanisms for different viscosity and convection treatment for $(k,N)=(8,4)$. The ($\HM{1}{}$) lifting results are the subject of discussion here; the SIP plot is repeated for a clearer presentation. The abscissa shows the time $t$.}
 	\label{fig:3DTGV-ILES-viscosity}
\end{figure}

We define the discrete ($\HM{1}{}$) lifting of a vector-valued function living on the boundary of each element. 
\begin{thmDef} \label{def:VectorH1Lifting}%
For all $K\in\T$ and $\PHI\in \LP{2}{\rb{ \partial K}}$, the discrete ($\HM{1}{}$) lifting $\lift{\PHI}\in \restr{\VV_h}{K}\cap\LPZ{2}{\rb{K}}$ is defined as
\begin{equation*}
	\int_K \nabla \lift{\PHI} \Fip \nabla \ssh \dx
		= - \int_{\partial K} \varphi \ip \rb{\nabla \ssh} \nn_K \ds,
		\quad \forall\,\ssh\in \restr{\VV_h}{K}.
\end{equation*}	
\end{thmDef}

Note that the space for the discrete lifting is chosen such that $\lift{\cdot}$ is unique and allows to rewrite 
$
- \int_{\partial K} \hjmp{\ww_h} \ip \rb{\nabla\vv_h}\nn \ds = \int_K \nabla \lift{\hjmp{\ww_h}} \Fip \nabla \vv_h \dx
$
in the definition of $a_h\rb{\cdot,\cdot}$.
Our $\HDIV$-HDG method with ($\HM{1}{}$) lifting results from adding the form
\begin{equation*}
	\nu \sigma_\ell \int_\Omega \nabla_h \lift{\hjmp{\uu_h}} \Fip \nabla_h \lift{\hjmp{\vv_h}} \dx
\end{equation*}
to the left-hand side of \eqref{eq:HdivHDGMethod}.
One easily checks that this additional term ensures non-negativity of the bilinear form~$a_h \rb{\cdot,\cdot}$ for any $\sigma_\ell \geqslant 1$. 
Hence, we do not require the penalty parameter~$\sigma_K$ to obey a largeness constraint (usually related to the constant in discrete trace inequalities).
In fact, we simply use~$\sigma_K=k$ and~$\sigma_\ell=2$ and thus the ($\HM{1}{}$) lifting method enforces weak (tangential) continuity of the discrete~$\HDIV$ solution in a (much) weaker sense compared to the SIP method.
~\\

Fig.~\ref{fig:3DTGV-ILES-viscosity} shows the dissipation rates for the~$\HDIV$ method for~$(k,N)=(8,4)$ for our basic choice SIP with upwinding, compared against the ($\HM{1}{}$) lifting variant with upwinding, Lax--Friedrichs, and the central flux choice for the discrete convection term.
Comparing the two results with upwinding, one can observe that convection dissipation plays a larger role when the ($\HM{1}{}$) lifting method is chosen.
The ($\HM{1}{}$) lifting method is less dissipative in the viscous term, but the convection stabilisation takes over this role such that the total dissipation rate again is not different.
In agreement with the last subsection, the same observation holds true for the Lax--Friedrichs convection term.
When the ($\HM{1}{}$) lifting method is used without convection stabilisation, surprisingly, it still works and indeed gives comparably accurate results.
We conclude that the particular choice of the discretisation of the viscous term does not have a large impact on the results for the TGV problem.

\subsubsection{Variation of penalty factors for the \emph{L}$\mathbf{^2}$-DG method}

\begin{figure}[b]
\centering
 	\includegraphics[width=0.275\textwidth]
		{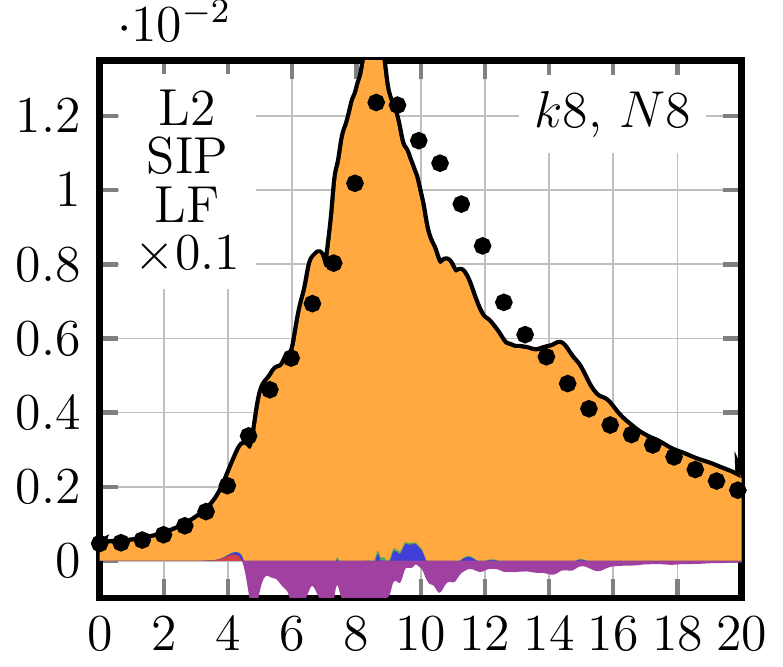}
 	\includegraphics[width=0.275\textwidth]
		{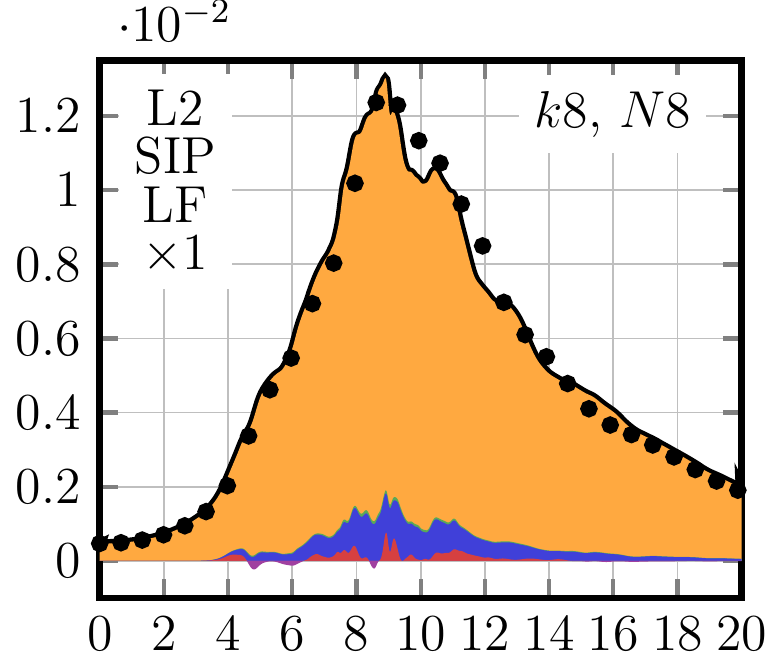}
 	\includegraphics[width=0.125\textwidth]
		{TaylorGreen/ILES-plot/tikz/3DTGV_ILES_L2_legend.pdf} \\		
 	\includegraphics[width=0.275\textwidth]
		{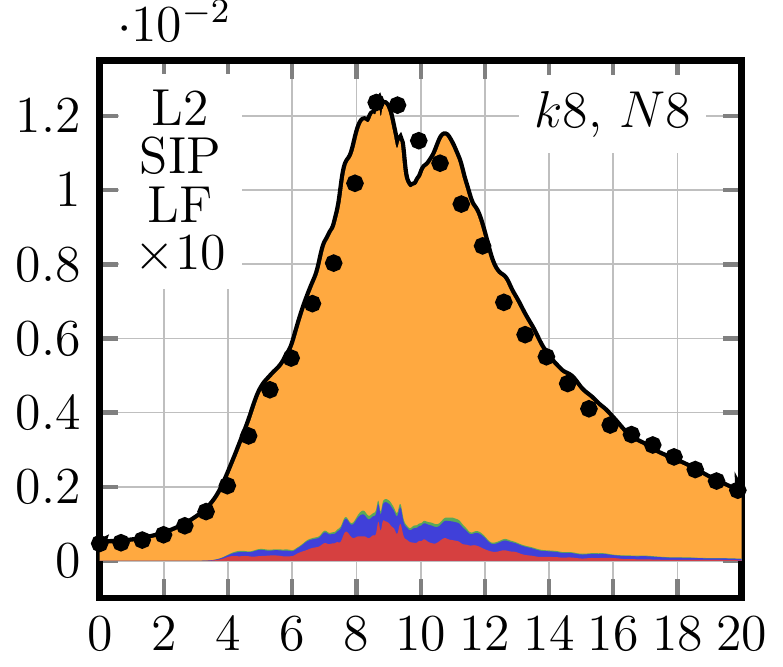}
 	\includegraphics[width=0.275\textwidth]
		{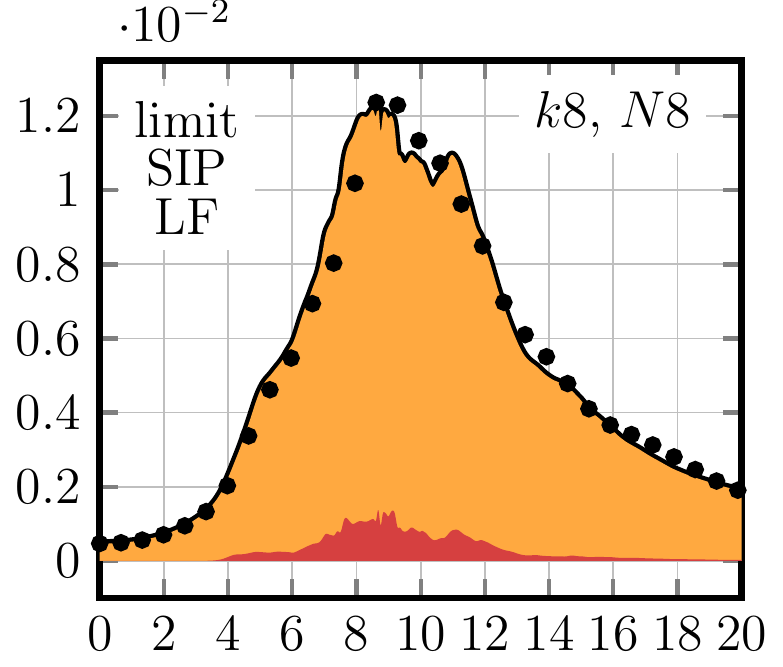}		
 	\includegraphics[width=0.125\textwidth]
		{TaylorGreen/ILES-plot/tikz/3DTGV_ILES_L2_legend.pdf}		
 	\caption{$\LTWO$ dissipation mechanisms for different penalties for $k=8$, $N=8$, where the lower right plot shows results for the limit method ($\zeta\to\infty$) described in Appendix~\ref{app:limitmethod}. The abscissa shows the time $t$.}
 	\label{fig:3DTGV-ILES-penalty}
\end{figure}

Next, we investigate the influence of the penalty factor~$\zeta$ for the~$\LTWO$-based method.
Fig.~\ref{fig:3DTGV-ILES-penalty} shows results for values of~$\zeta = 0.1, 1, 10, \infty$ (from left to right and top to bottom) considering a spatial resolution of~$k=8, N=8$. 
Note that the limit method is characterised in Appendix~\ref{app:limitmethod}.
For a penalty factor to~$\zeta=0.1$, the convective term exhibits a negative dissipation rate which is due to sign-indefinite terms in the energy balance, see equation~\eqref{eq:DiscEnergyBalanceL2}, and instabilities might occur for smaller penalty factors~$\zeta \rightarrow 0$. 
For~$\zeta=1$, only minor undershoots can be observed where the dissipation of the convective term takes a negative value.
For a penalty factor of~$\zeta=10$, the numerical dissipation of the convective term is non-negative for all times and the numerical dissipation of the convective term is larger than that of the penalty terms. 
Finally, the results for the limit method~$\zeta=\infty$ show that, by construction, the solution will be normal-continuous and exactly divergence-free.
As a consequence, the only dissipation that is not related to viscosity stems from the convection stabilisation (Lax--Friedrichs).
Moreover, note that for this problem, the total amount of numerical dissipation is similar for the~$\LTWO$-method with~$\zeta=1,10$ compared to the limit method with~$\zeta\to\infty$.
~\\

This parameter study nicely demonstrates the impact of the penalty terms on the dissipation rate of the convective term. 
However, let us explicitly emphasise that a potential conclusion like ``the convective term stabilises the discretisation scheme by providing the required numerical dissipation for under-resolved problems'' drawn from the~$\zeta=10$ results is misleading. 
In fact, the convective term would not be able to stabilise the scheme without further action. 
As demonstrated in~\cite{Fehn2018a}, the~$\LTWO$-based discretisation scheme without additional penalty terms does not lead to a robust discretisation in the under-resolved regime potentially leading to a blow-up of the solution. 
Similarly, the fact that the numerical dissipation of the continuity penalty term is small compared to the overall dissipation rate for the example considered here does not imply that the continuity penalty term is not required to obtain a robust discretisation scheme. 
As shown in~\cite{Fehn2018a}, the continuity penalty term is indeed an essential ingredient for the robustness of the~$\LTWO$-based method. 
Finally, we mention that the penalty factor is not intended to be a parameter that can be used to adjust the discretisation scheme. 
Instead, our goal is a parameter-free turbulent flow solver and designing the penalty parameter in a way to obtain a robust discretisation scheme for a constant penalty factor that is chosen once and for all (the default value is~$\zeta =1$ unless specified otherwise).

\section{Turbulent channel flow}
\label{sec:Channel}

The 3D turbulent channel flow problem is a frequently used benchmark problem for assessing the ability of flow solvers to deal with wall-bounded turbulence, see Refs.~\cite{MoinKim82,KimMoinMoser87,MoserKimMansour99}.
As the domain for all channel flows we consider the rectangular cuboid~$\Omega=\rb{0,L_x}\times\rb{0,L_y}\times\rb{0,L_z}$ with~$L_x=2\pi\delta_c$,~$L_y=2\delta_c$,~$L_z=\pi\delta_c$ and channel half-width~$\delta_c=1$.
In~$x$-direction (streamwise) and~$z$-direction (spanwise) periodic boundary conditions are prescribed whereas for~$y\in\set{0,L_y}$ the no-slip condition~$\uu=\zero$ is imposed.
Due to sharp velocity gradients, it is common practice to stretch the mesh in $y$-direction (wall normal direction).
We choose the stretching function~$\Phi \colon\sqb{0,1}\to\sqb{0,L_y}$,
\begin{equation} \label{eq:ChannelMapping}
	y \mapsto \delta_c \frac{\tanh\rb{C\sqb{2y-1}}}{\tanh\rb{C}} + \delta_c,
\end{equation}
with a constant~$C=1.8$ for all our simulations.
Therefore, given a mesh with~$N$ elements in each direction, the resulting meshes for the channel flow problem consist of~$N^3$ hexahedra which will be equidistant in~$x$- and~$z$-directions and stretched in~$y$-direction.
Usually, the whole motion is driven solely by a constant pressure gradient source term~$\ff=\rb{f_p,0,0}^\dag$ acting in the streamwise direction; cf., for example, Ref.~\cite{ChaconLewandowski14} (Sec.~13.4).
~\\

In order to distinguish different turbulent channel flow situations, the friction Reynolds number~$\Rey_\tau$ is considered most frequently.
It is defined as~$\Rey_\tau=u_\tau \delta_c/\nu$ where~$u_\tau$ denotes the so-called wall friction velocity which, in turn, depends on the wall shear stress~$\tau_w$ as~$u_\tau^2 = \tau_w/\rho$.
Under the assumption of a statistically steady state flow and with~$\rho=1$, one obtains~$\tau_w=f_p\delta_c$; see Ref.~\cite{DuboisJauberteauTemam99} (Sec.~2.3).
Then, by choosing~$f_p=1$, one obtains~$u_\tau=1$ and hence, friction Reynolds number and viscosity are connected by the simple relation~$\nu=1/\Rey_\tau$.
~\\

The relevant quantities of interest for the channel flow involve averaging in time~$\bra{\cdot}_t$ and in spatial directions of homogeneity~$\bra{\cdot}_s$, where the abbreviation~$\bra{\cdot}=\bra{\bra{\cdot}_s}_t$ is used.
Then, we consider the following normalised quantities: mean velocities~$\bra{u_i}^+=\bra{u_i}/u_\tau$, Reynolds stresses~$\bra{u_i^\prime u_j^\prime}^+=\bra{u_i u_j}/{u_\tau^2} - \bra{u_i}\bra{u_j}/{u_\tau^2}$, and normalised root mean square (rms) velocities~$u_{i,\rms}^+= \abs{\bra{u_i^\prime u_i^\prime}^+}^\half$.
Furthermore, so-called wall units~$y^+= y u_\tau/\nu$ for~$y\in\sqb{0,\delta_c}$ are used.
When normalising the numerical results,~$u_\tau$ is the friction velocity obtained as a result of the numerical simulation and not the theoretical value~$u_\tau=1$. 

\subsection{Affine vs.\ isoparametric finite element meshes}
\label{sec:AffineVsIso}
In Sec.~\ref{sec:ReTau395} below we will compare results based on finite elements meshes using either \emph{affine} or \emph{isoparametric} mappings. We use this section to explain the possible impact of the choice of finite element meshes.
~\\

In the case of affine linear mappings, the stretching function $\Phi$ in \eqref{eq:ChannelMapping} is approximated by a continuous piecewise linear approximation $\Phi_h^1$ on an equidistant mesh for $[0,1]$ with $N$ elements while in the isoparametric case the approximation is done with $\Phi_h^k$, a continuous piecewise polynomial of degree $k$.
\begin{figure}[t]
\centering
 	\includegraphics[width=0.4\textwidth]{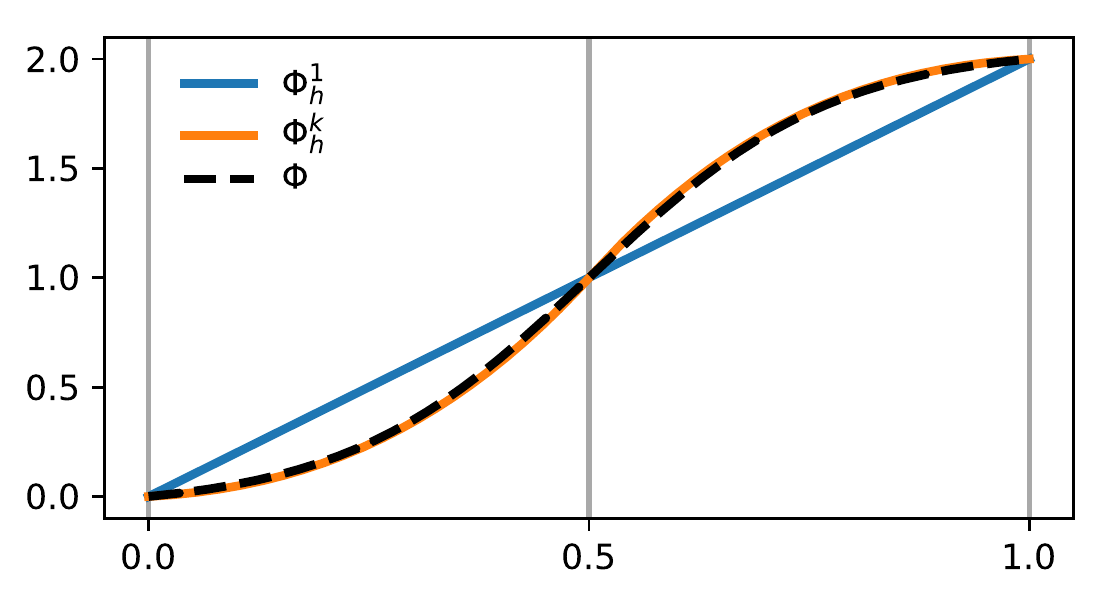} 
	 	\goodgap \goodgap \goodgap
 	\includegraphics[width=0.4\textwidth]{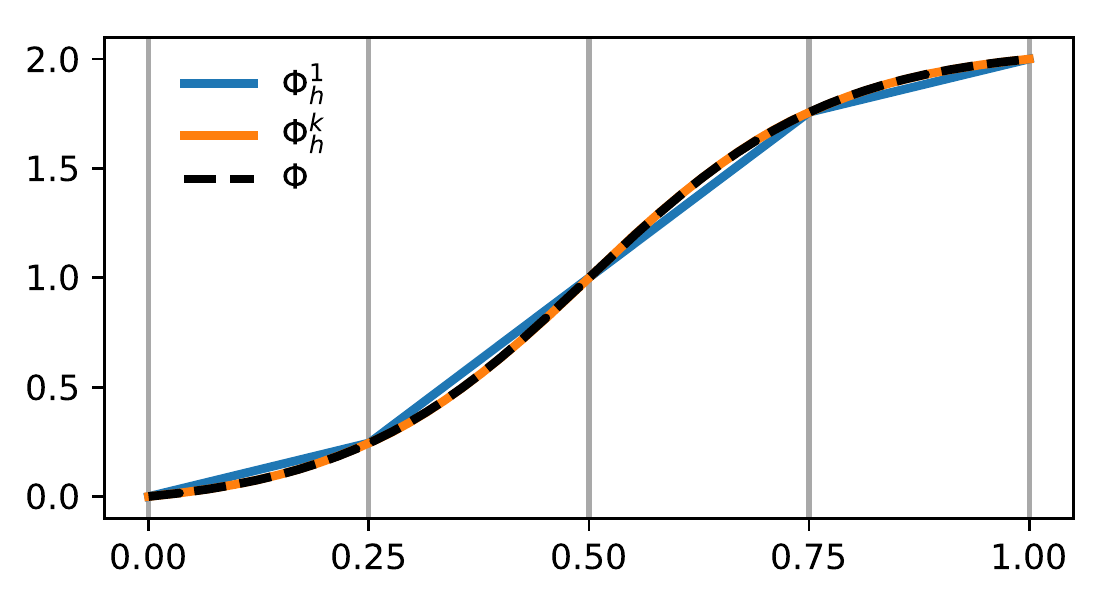} 
 	\caption{Comparison of approximations of the desired mesh stretching \eqref{eq:ChannelMapping} with affine linear and isoparametric ($k=3$) mappings for $N=2$ (left) and $N=4$ (right). }
 	\label{fig:affine-iso-sketch}
\end{figure}
In Fig.~\ref{fig:affine-iso-sketch} both approximations and the stretching function \eqref{eq:ChannelMapping} for $N=2$ and $N=4$ are sketched (for $k=3$) for illustration purposes. 
One can observe that for $N=2$ there is no stretching in the affine case and also for $N=4$ the difference between an affine stretching and the isoparametric stretching is significant. 
From this simple observation, we realise that while the analytical stretching function is used for the purpose of accumulating resolution close to the boundary, the realisation of the corresponding stretching function by the finite element mesh is essential. 
Note that this aspect is particularly relevant for high-order finite element discretisations considered here as opposed to low-order discretisations for which the number of elements is much larger.
~\\

The choice of affine or isoparametric finite element meshes also influences the choice of finite element spaces.
In Section~\ref{sec:DG-solvers} we introduced the finite element spaces based on the assumption of affine linear mappings. If this assumption is no longer fulfilled and the element mappings~$\Phi_K: \hK \to K$, where~$\hK = [0,1]^3$ is the reference hexahedron, are not affine linear the local finite element spaces are adapted accordingly. Let $\hQk{k}{\rb{\hK}}$ be the tensor-product polynomial space as introduced (without the hat notation) in Section \ref{sec:DG-solvers} but w.r.t.~$\hK = [0,1]^3$ and~$\hRTk{k}{\rb{\hK}}$ accordingly. Then, we define the mapped polynomial spaces as
$$
\Qk{k}{\rb{K}} \coloneqq \hQk{k}{\rb{\hK}} \circ \Phi_K^{-1},\quad\QQk{k}{\rb{K}} \coloneqq \hQQk{k}{\rb{\hK}} \circ \Phi_K^{-1}\quad\text{and}\quad \RTk{k}{\rb{K}} \coloneqq (\det(\nabla \Phi_K))^{-1} (\nabla \Phi_K) (\hRTk{k}{\rb{\hK}} \circ \Phi_K^{-1})
$$ where the latter is the well-known Piola transformation. Let us stress that the resulting spaces contain non-polynomial functions.
From here on, we want to focus on the practical impact that a decision between affine linear and isoparametric finite element meshes can have.
~\\

Therefore, let us consider a simple 1D configuration with the underlying question of how the approximation properties of both approaches differ for approximating a typical (averaged) turbulent velocity profile.
The following simple (explicit) formulation of a corresponding law of the wall has been derived in Reichardt~\cite{Reichardt51}:
\begin{equation}\label{eq:reichardt}
	\bra{u}^+ = 2.5 \ln\rb{1+0.4y^+} 
		+ 7.8 \sqb{ 1 - \exp\rb{ -\frac{y^+}{11}} - \frac{y^+}{11} \exp\rb{-0.33 y^+} }.
\end{equation}
The accuracy of this formula shall suffice for the intended purpose of showing the different abilities to approximate functions with sharp gradients in a boundary layer as~$y\to 0$.
~\\

\begin{figure}[t]
\centering
 	\includegraphics[width=0.4\textwidth]{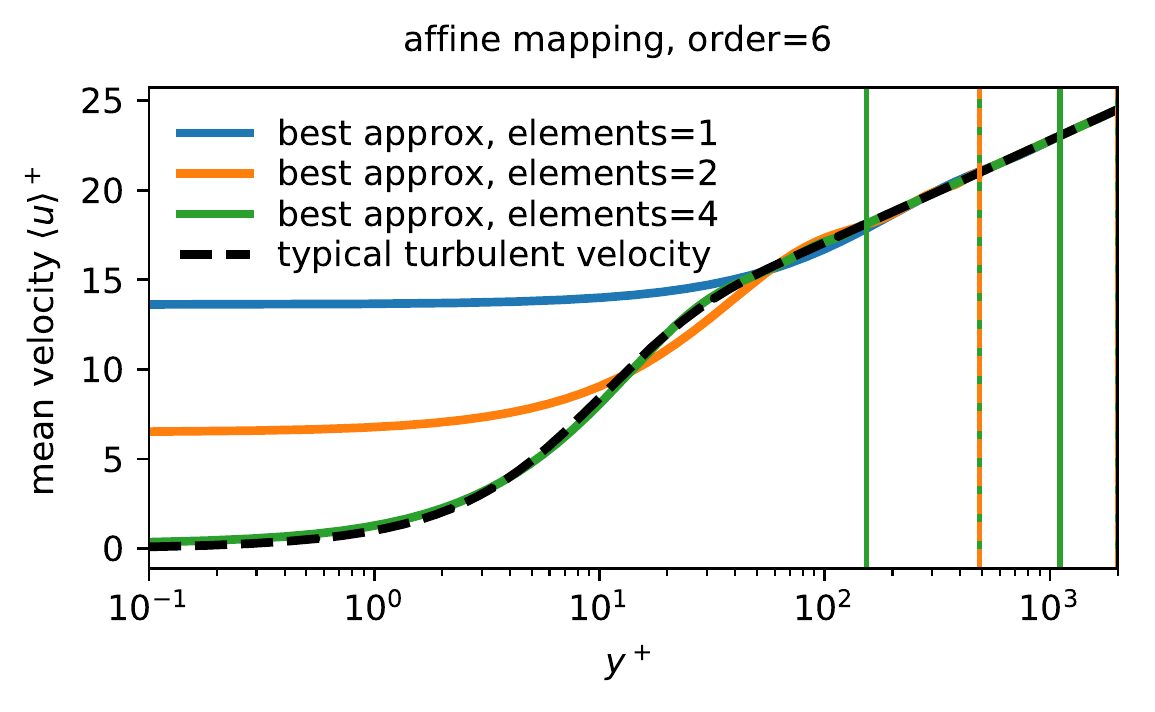} 
 		\goodgap \goodgap \goodgap
 	\includegraphics[width=0.4\textwidth]{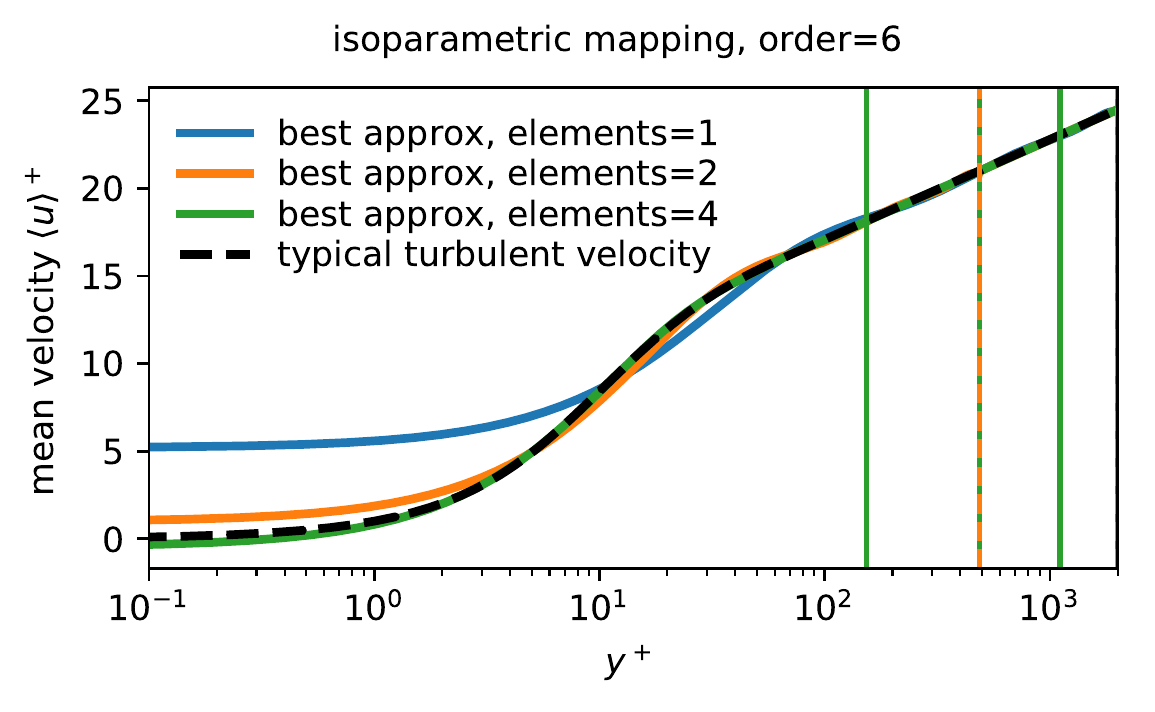} 	
 	\caption{Approximation properties resulting from affine (left) and isoparametric (right) finite elements. Simplified 1D example where the $\Ltwo$-best approximation of a typical turbulent velocity profile is shown on different meshes for $k=6$. The vertical lines correspond to the location of vertices in the corresponding mesh.}
 	\label{fig:channel-affine-iso}
\end{figure}

In Fig.~\ref{fig:channel-affine-iso}, a simple 1D example demonstrates the different approaches by showing the $\Ltwo$-best approximation of $\bra{u}$ on the inverval $y^+ \in (0,2000)$.
Here, the mesh stretching \eqref{eq:ChannelMapping} is approximated with an affine (left) and an isoparametric mapping (right) of order $k=6$ where globally discontinuous polynomials of element-wise order $k=6$ are used to approximate $\bra{u}^+$.
The resulting element boundaries are indicated by vertical lines, at least when there are more than one.
One can observe that the isoparametric approach roughly allows to use a once less refined mesh in this setting.
In the following section, the different approaches will be compared based on the actual 3D turbulent channel flow problem.
Note that the corresponding combination of polynomial order and meshes will be identical to the ones chosen in Fig.~\ref{fig:channel-affine-iso}.

\subsection{Mesh convergence study for $\Rey_\tau=$ 395}
\label{sec:ReTau395}
\begin{figure}[t]
\centering
 	\includegraphics[width=0.33\textwidth]
		{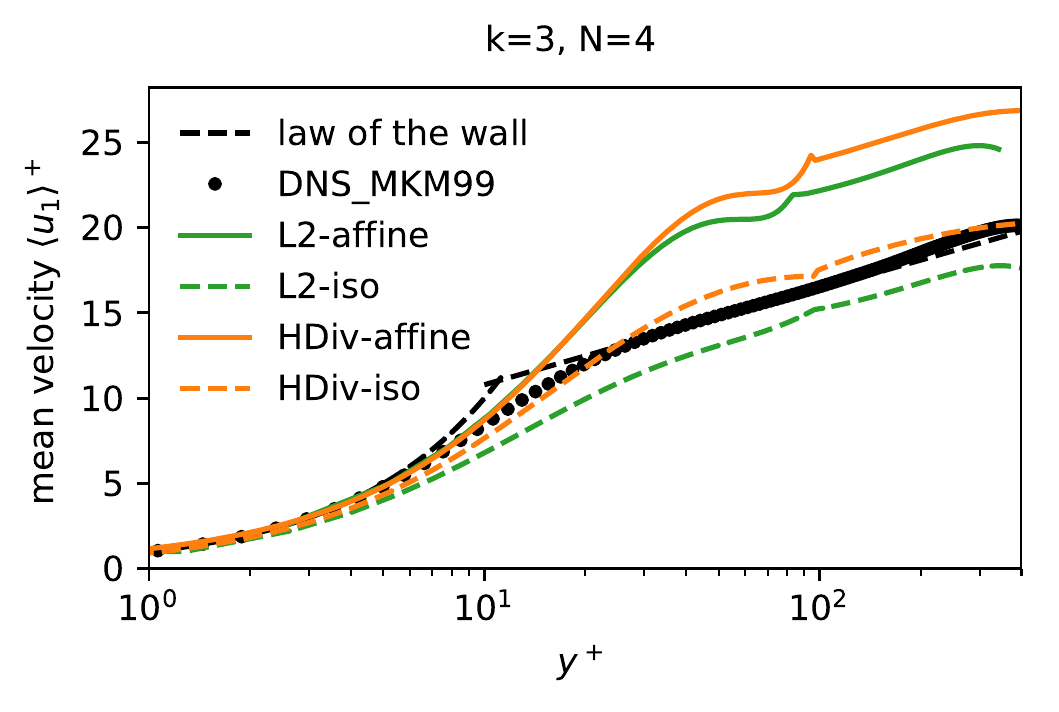}
 	\includegraphics[width=0.33\textwidth]
		{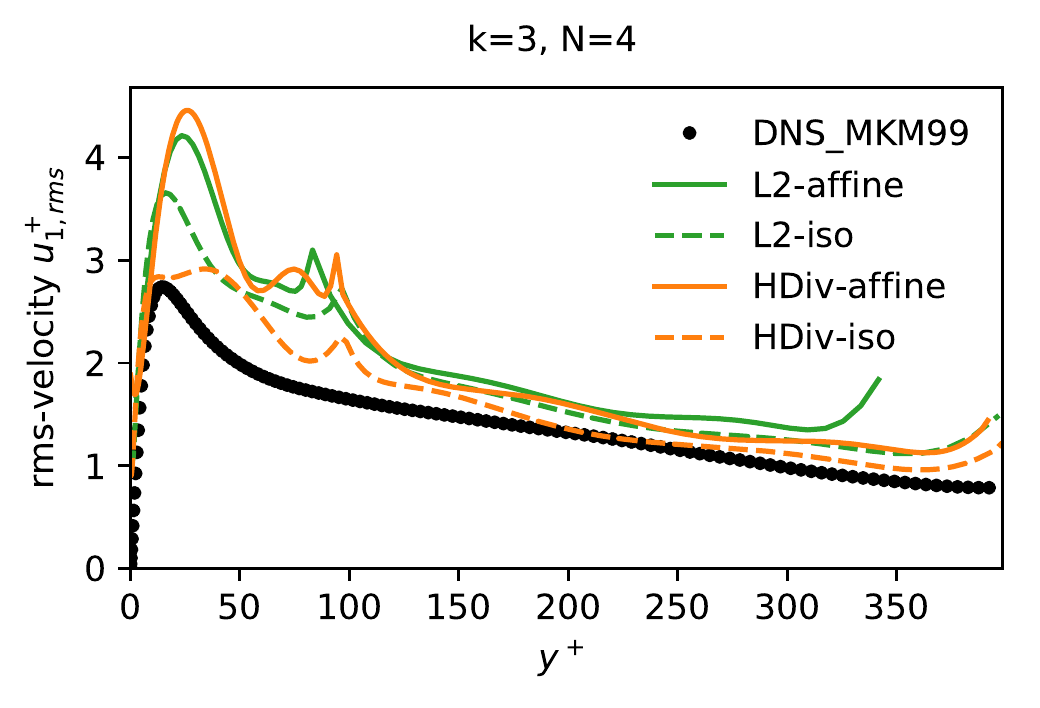}
 	\includegraphics[width=0.33\textwidth]
		{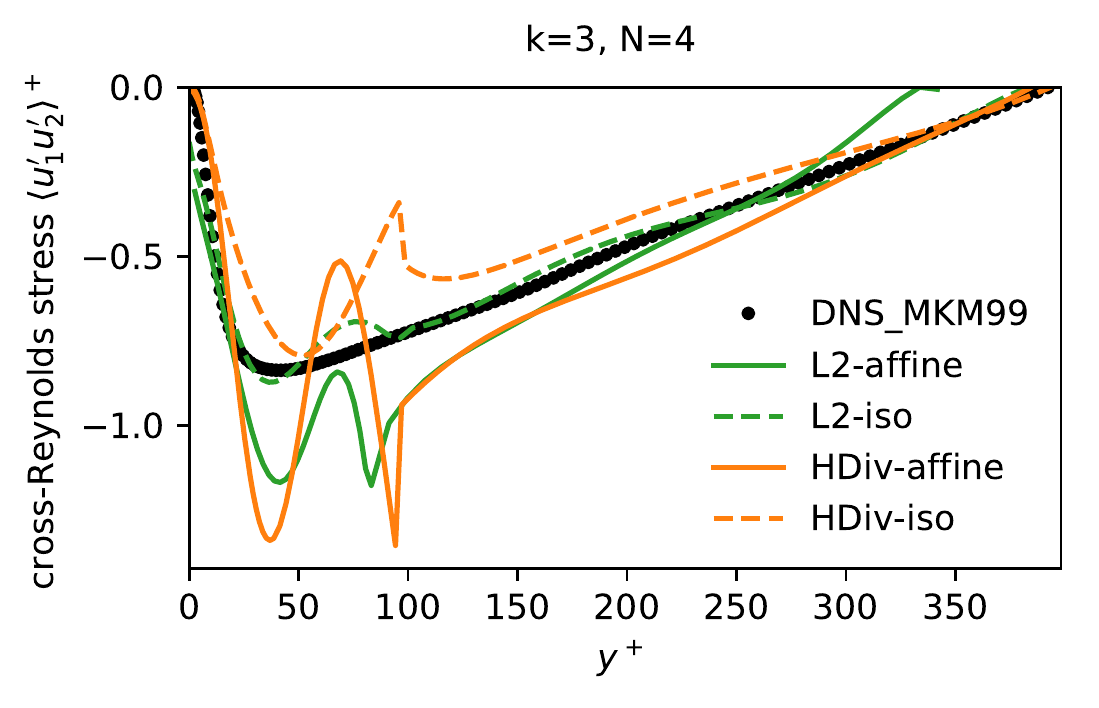} \\
 	\includegraphics[width=0.33\textwidth]
		{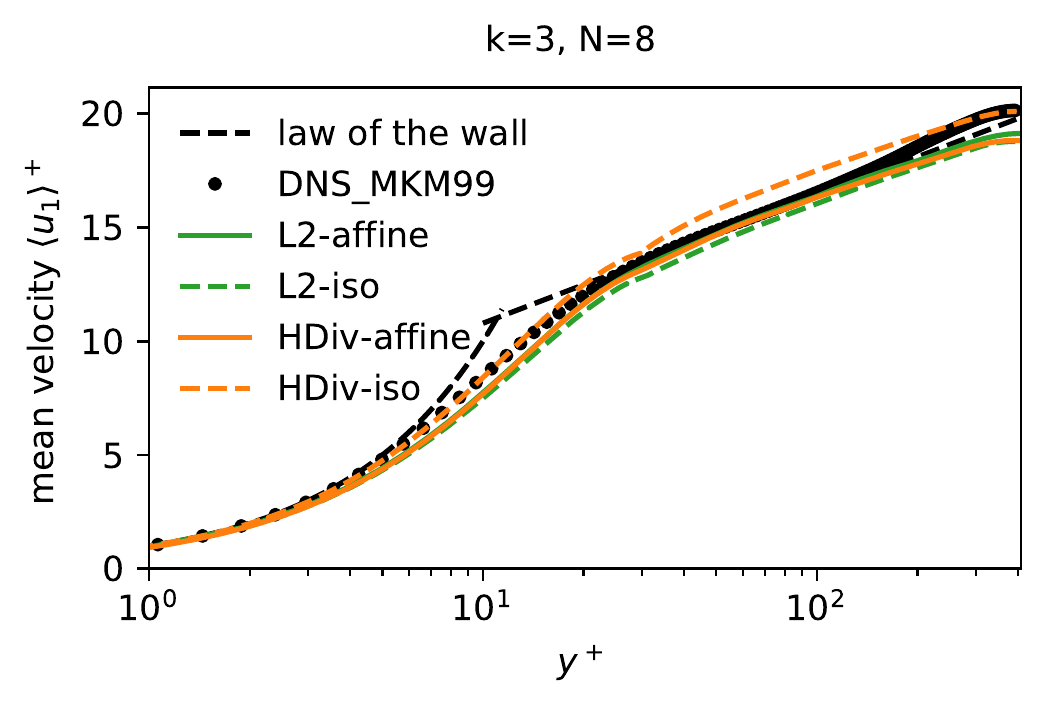}
 	\includegraphics[width=0.33\textwidth]
		{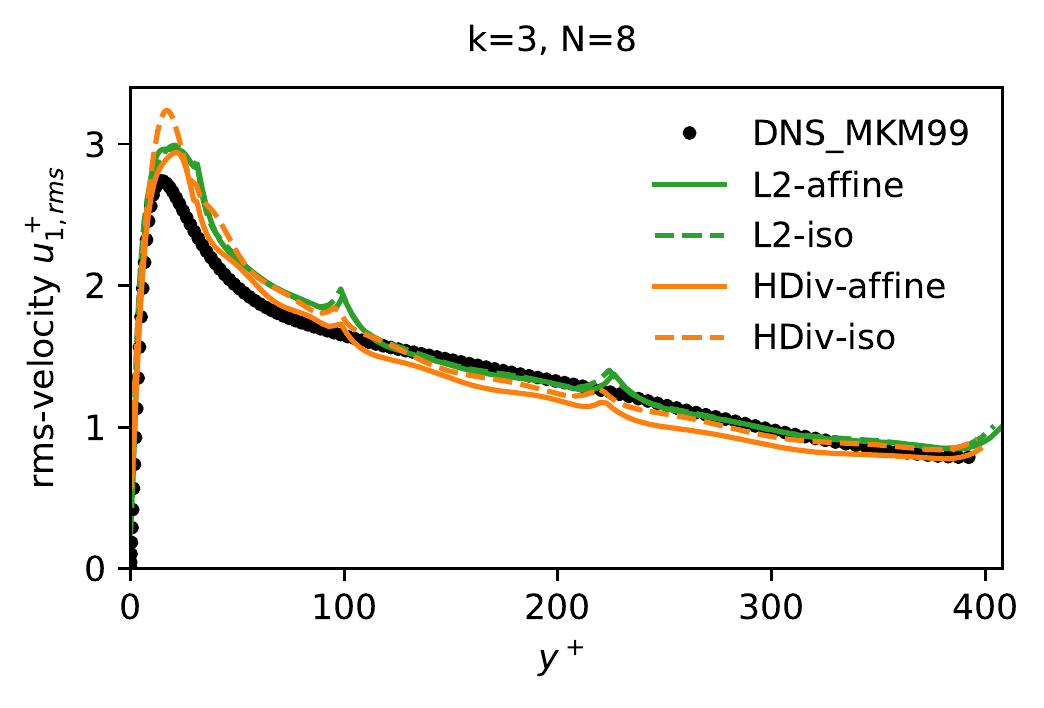}
 	\includegraphics[width=0.33\textwidth]
		{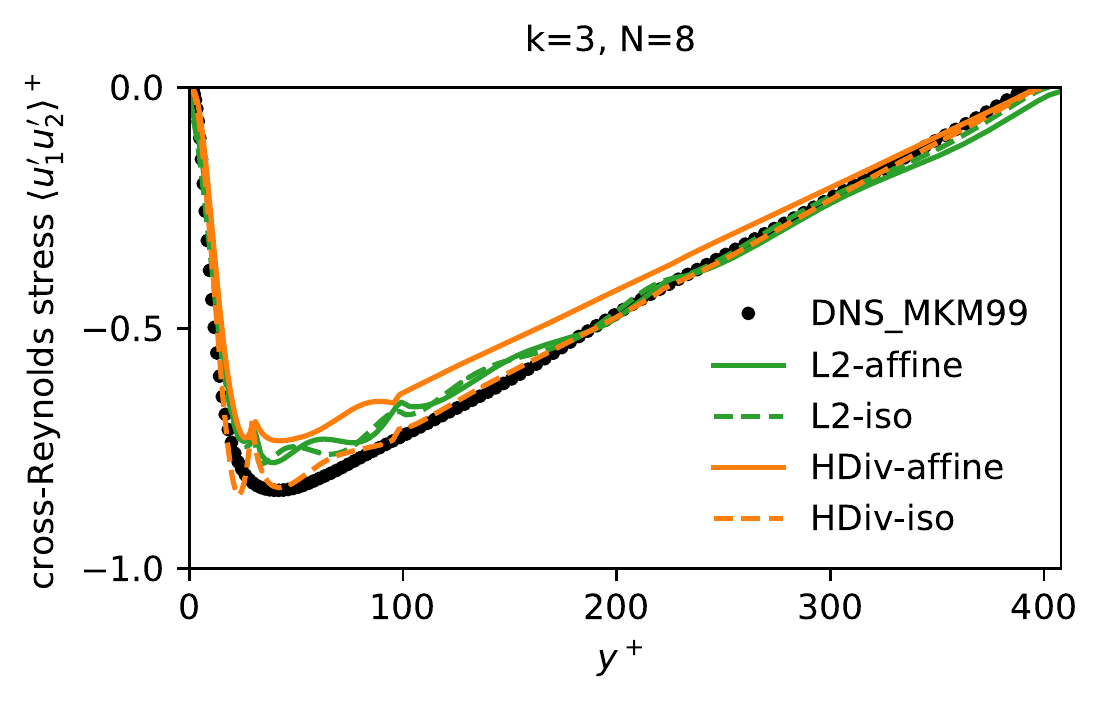} \\
 	\includegraphics[width=0.33\textwidth]
		{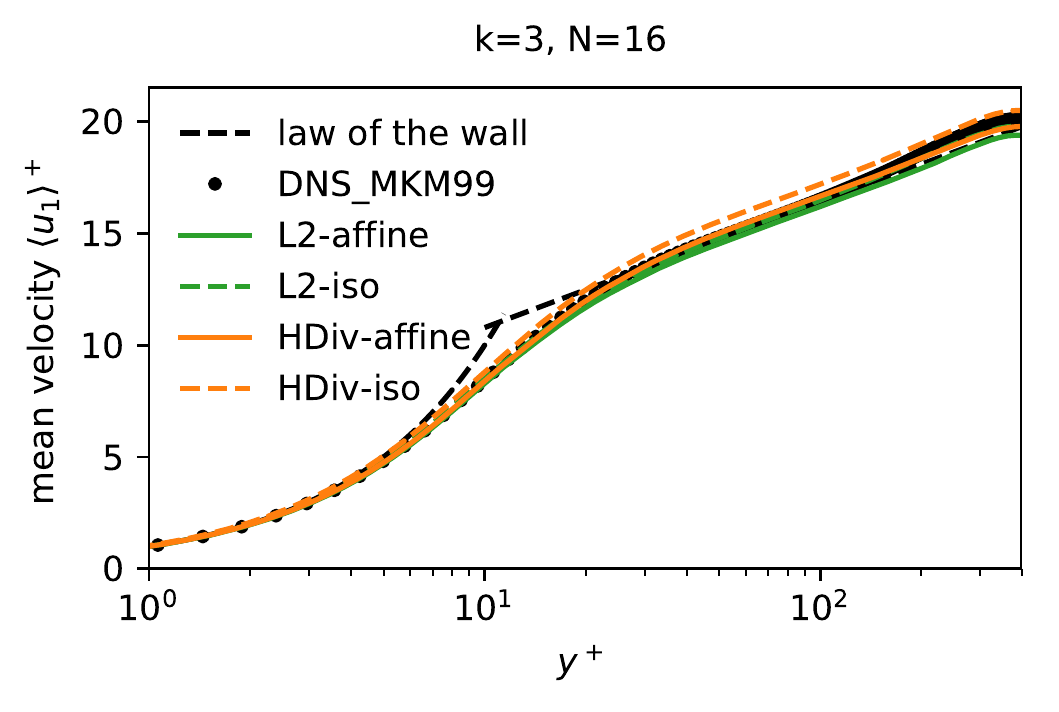}
 	\includegraphics[width=0.33\textwidth]
		{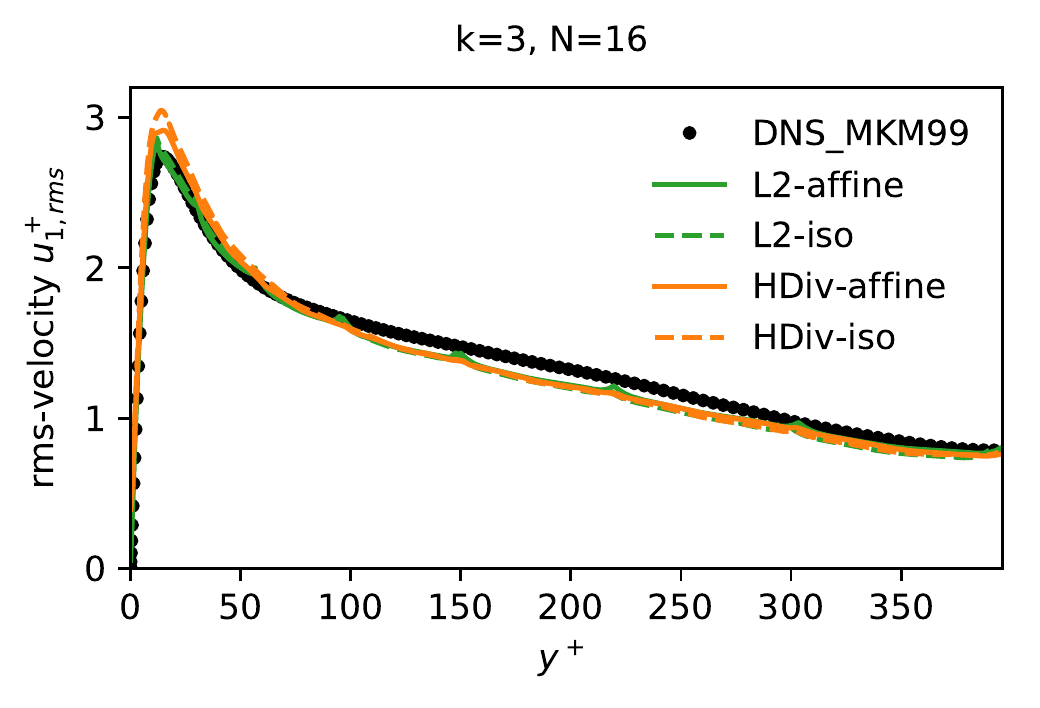}
 	\includegraphics[width=0.33\textwidth]
		{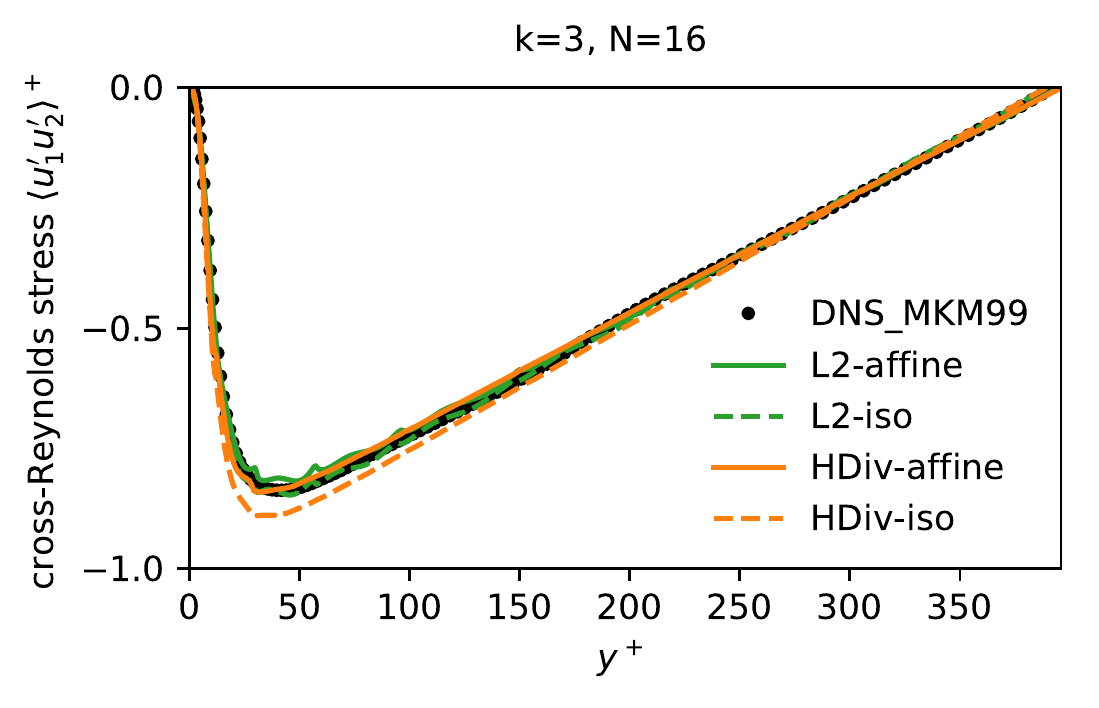}
 	\caption{Averaged channel profiles for $k=3$ at $\Rey_\tau=395$. Both $\LTWO$ and $\HDIV$ results are shown for $N=4,8,16$ with affine and isoparametric mapping of the finite element spaces.}
 	\label{fig:3DCh-k3-comparison}
\end{figure}

\begin{figure}[t]
\centering
 	\includegraphics[width=0.33\textwidth]
		{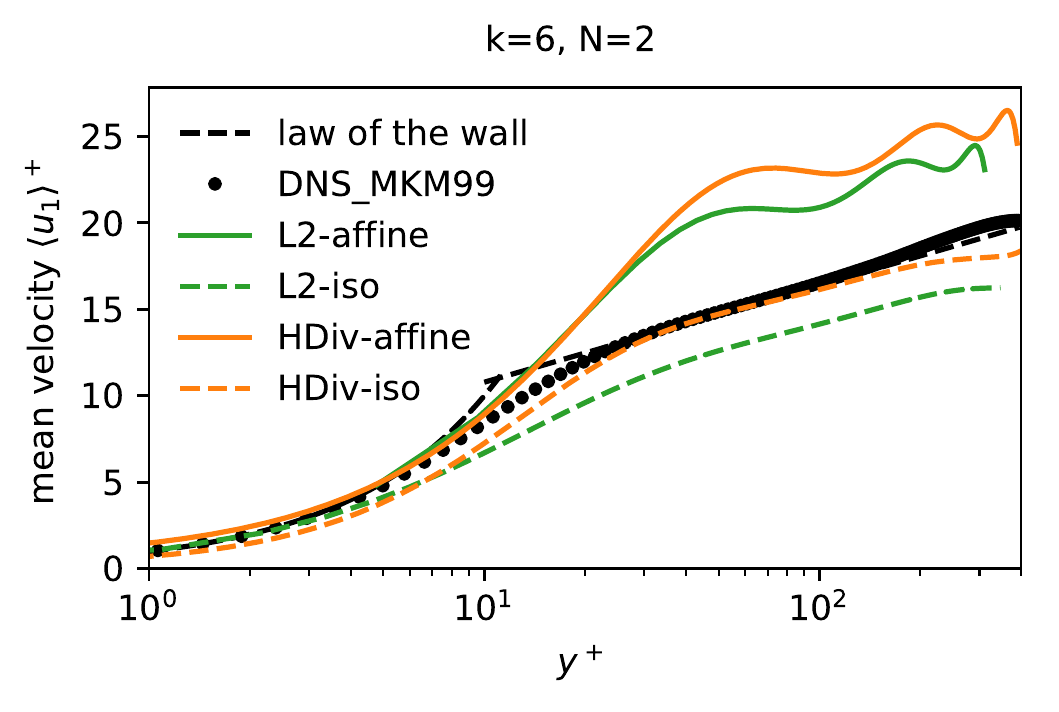}
 	\includegraphics[width=0.33\textwidth]
		{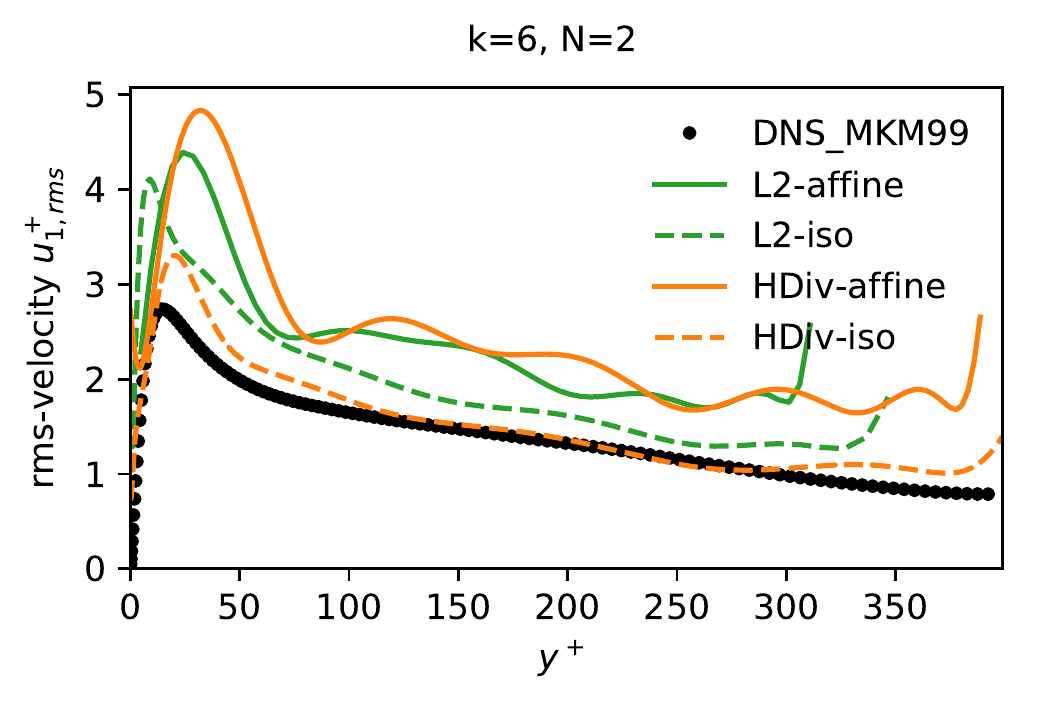}
 	\includegraphics[width=0.33\textwidth]
		{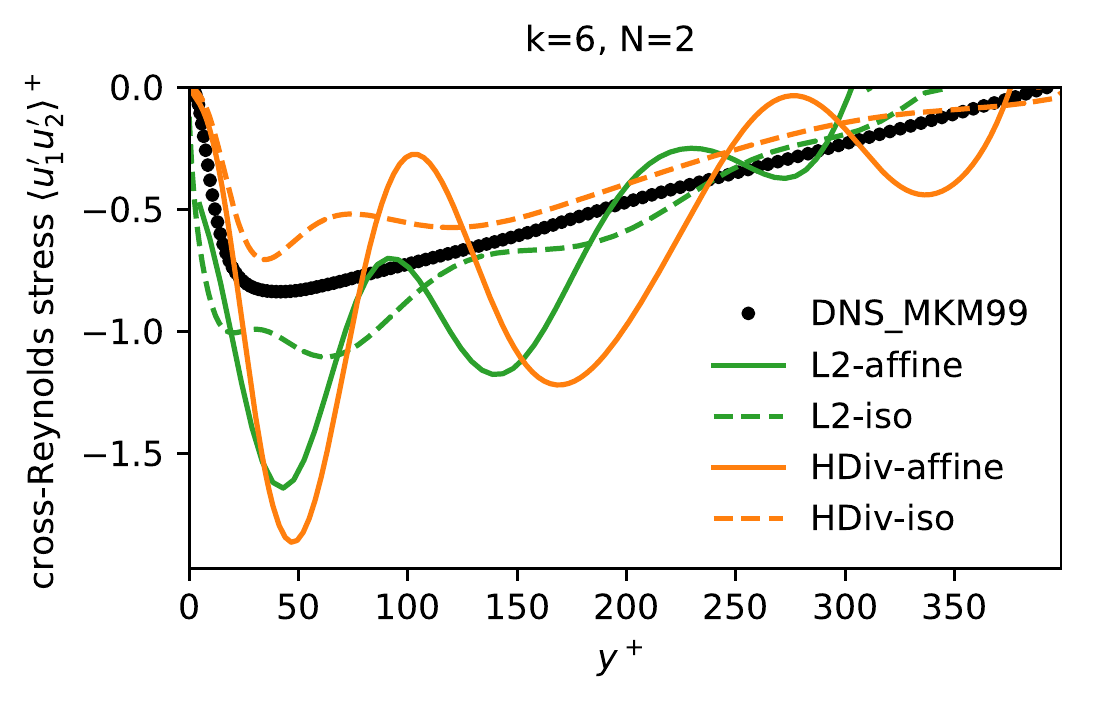} \\
 	\includegraphics[width=0.33\textwidth]
		{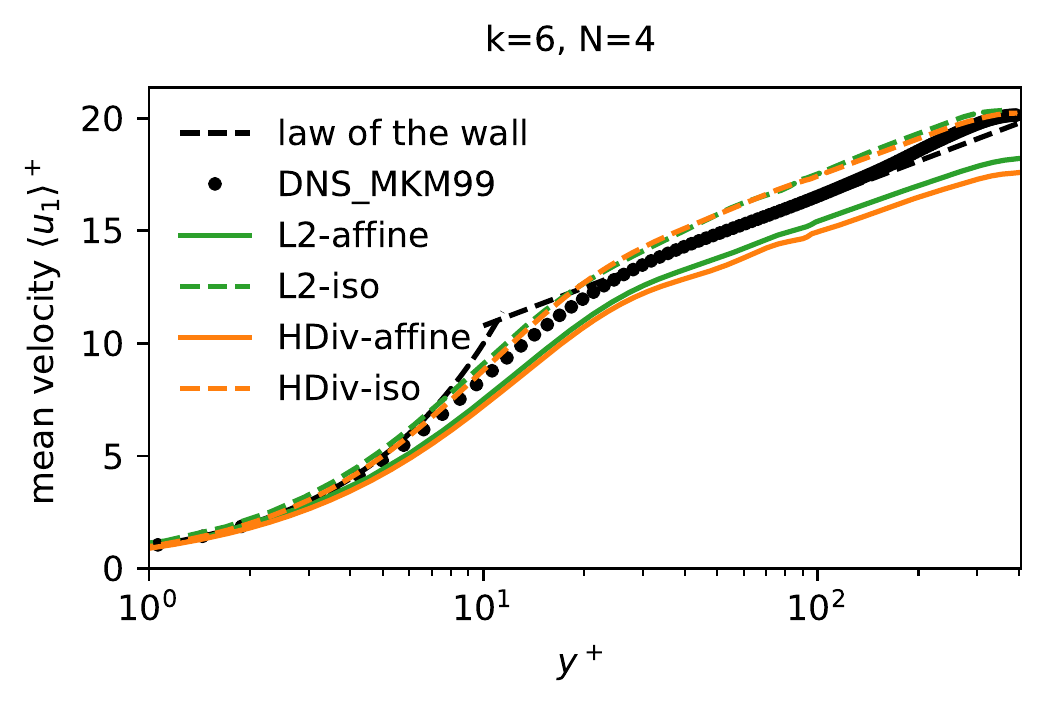}
 	\includegraphics[width=0.33\textwidth]
		{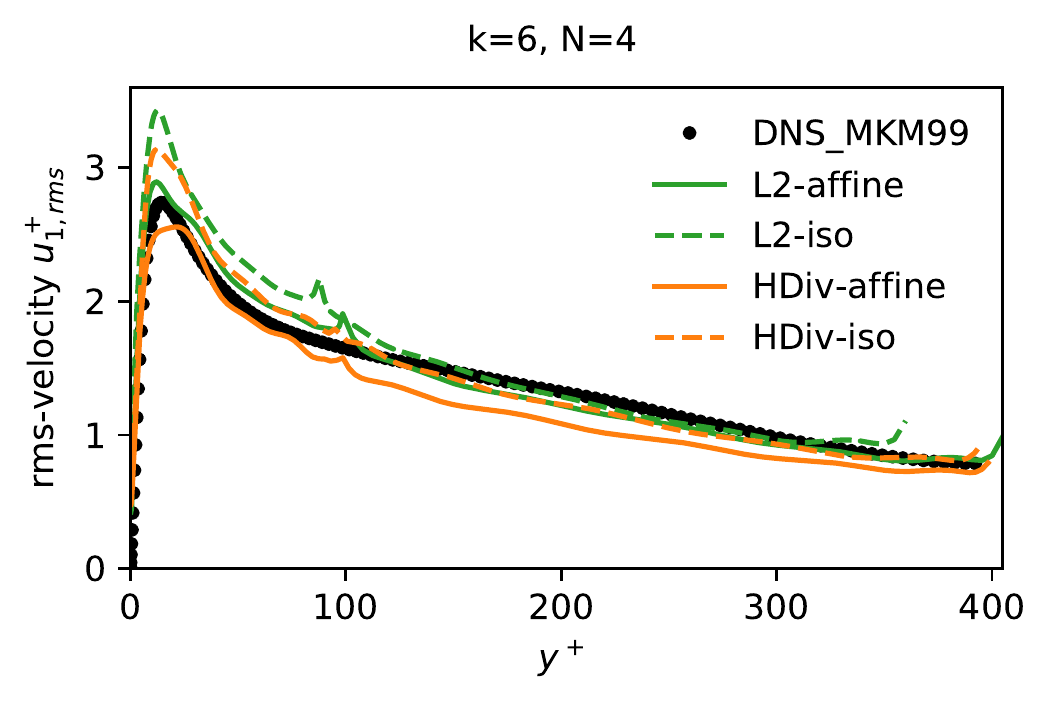}
 	\includegraphics[width=0.33\textwidth]
		{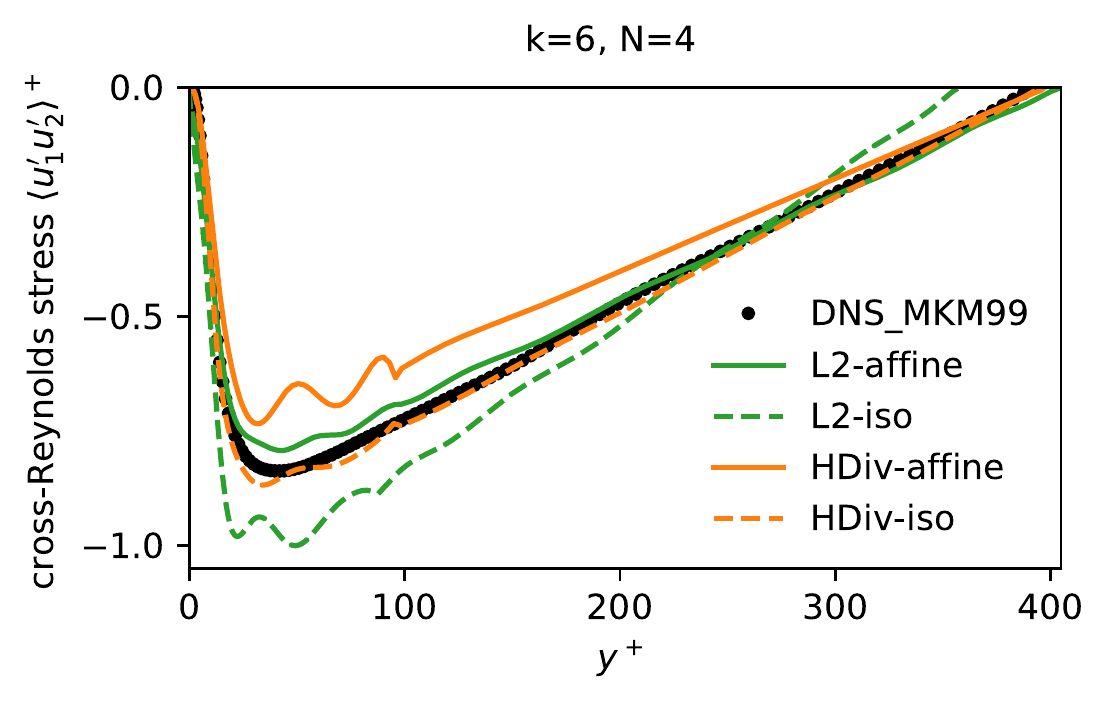} \\
 	\includegraphics[width=0.33\textwidth]
		{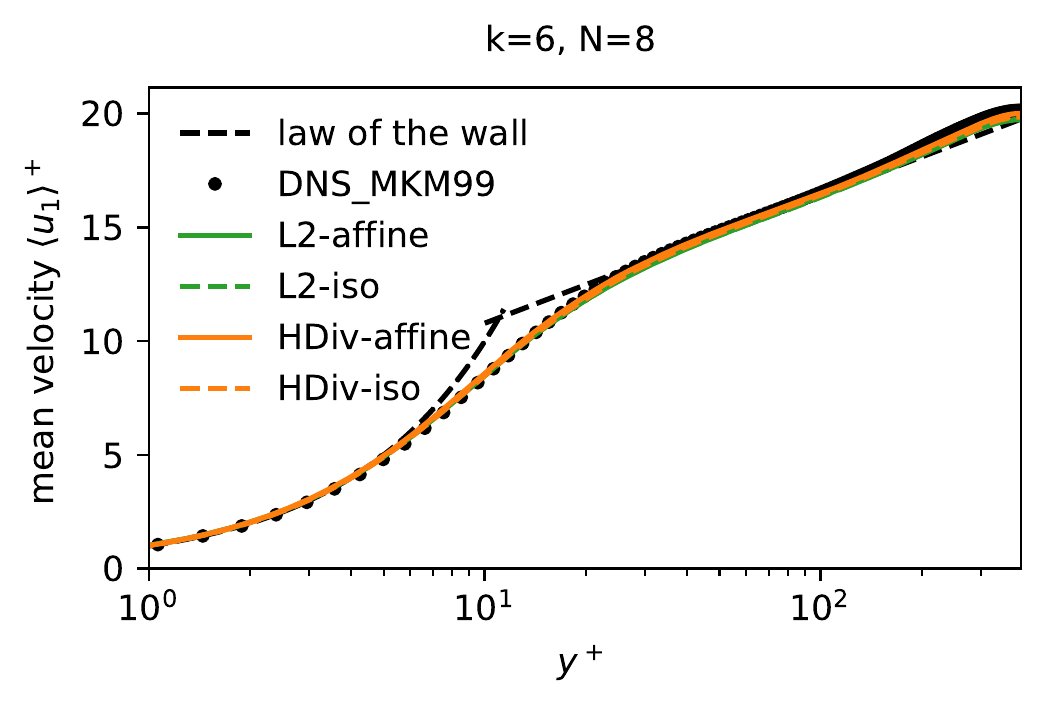}
 	\includegraphics[width=0.33\textwidth]
		{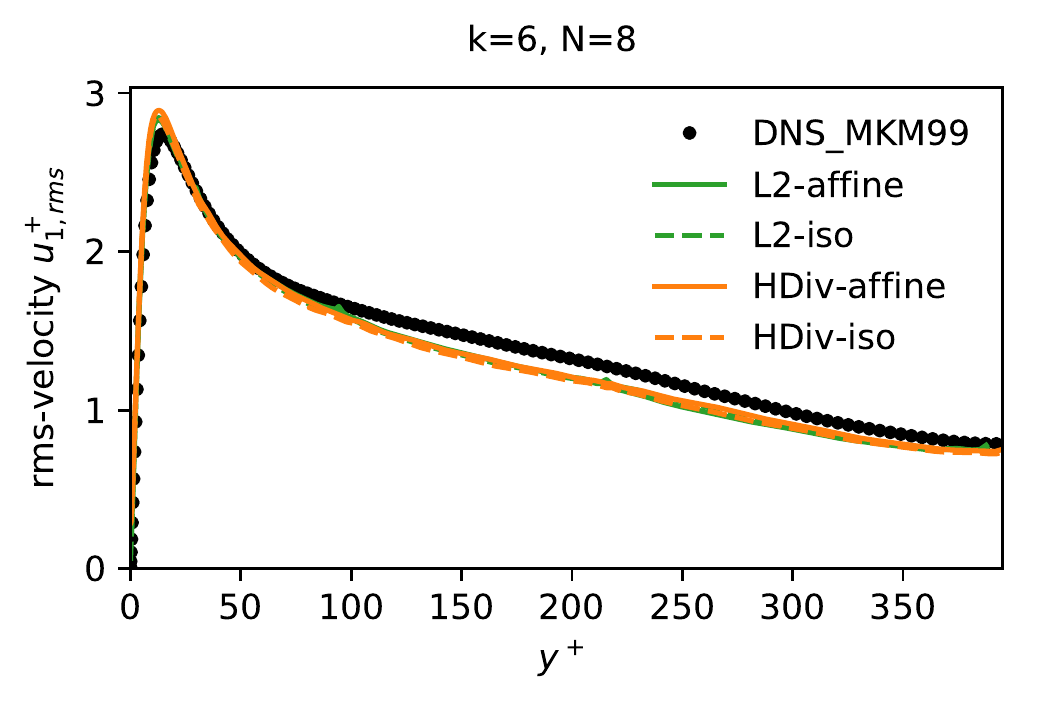}
 	\includegraphics[width=0.33\textwidth]
		{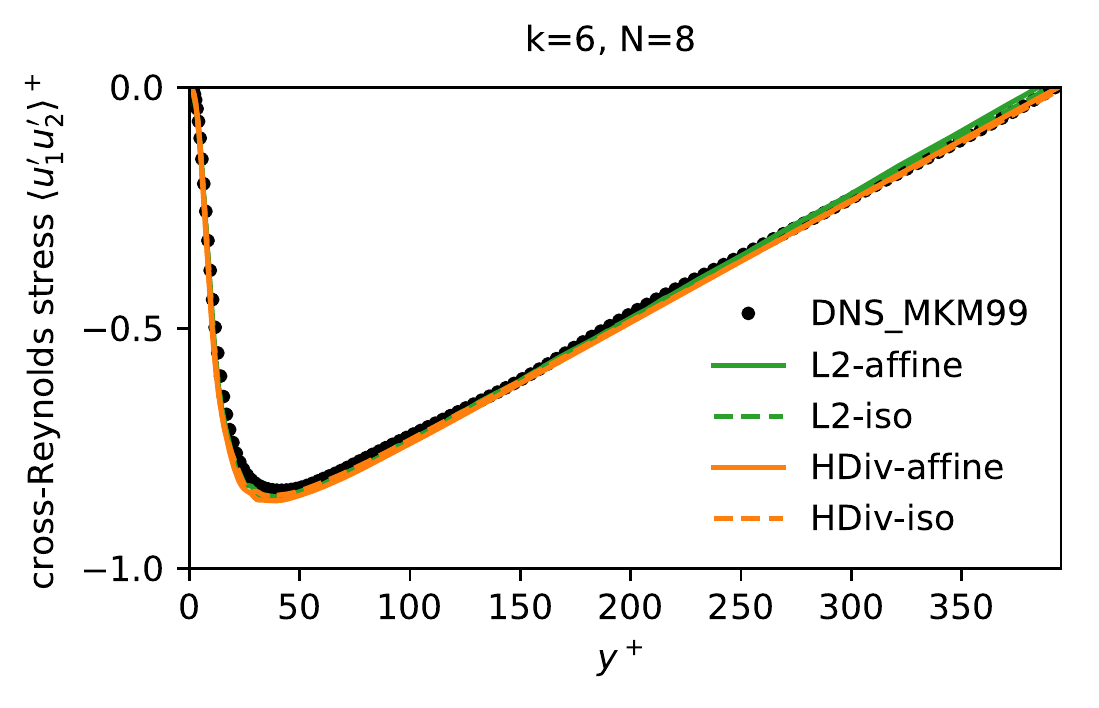}
 	\caption{Averaged channel profiles for $k=6$ at $\Rey_\tau=395$. Both $\LTWO$ and $\HDIV$ results are shown for $N=2,4,8$ with affine and isoparametric mapping of the finite element spaces.}
 	\label{fig:3DCh-k6-comparison}
\end{figure}

We investigate the accuracy of the two DG methods by comparing statistical quantities of the turbulent channel flow to accurate DNS reference data from Ref.~\cite{MoserKimMansour99}. 
Two different polynomial degrees,~$k=3$ and~$k=6$, are studied for three different meshes with~$N=4,8,16$ elements for~$k=3$ and~$N=2,4,8$ elements for~$k=6$. 
The results are shown in Fig.~\ref{fig:3DCh-k3-comparison} for~$k=3$ and in Fig.~\ref{fig:3DCh-k6-comparison} for~$k=6$, where the~$\LTWO$-based discretisation and the~$\HDIV$-based discretisation are compared, considering both affine mappings (solid lines) and isoparametric mapping (dashed lines) for each discretisation.
~\\ 

For both polynomial degrees, the numerical results converge towards the DNS reference data under mesh refinement with accurate predictions of the mean velocity profile and Reynolds stresses on the finest meshes. 
Comparing the~$k=3$ results to those for~$k=6$, the solution quality is comparable for the coarse mesh and the intermediate mesh. 
Note that on the coarsest mesh, the domain is discretised by only two elements in wall normal direction for each half of the channel for~$k=3$, and by only one element for~$k=6$. 
Remarkably, even this strongly under-resolved simulations deliver an at least meaningful prediction of the involved statistical quantities, e.g., the Reynolds stress tensor.
On the fine mesh, the different methods analysed here show less variations in the results for~$k=6$ than for~$k=3$, which can be seen as an indication of an improved accuracy of the high-order simulations with~$k=6$ compared to those with~$k=3$ once the flow becomes resolved, i.e., the results are less sensitive to changes in the discretisation scheme for the higher order method (note also that the number of degrees of freedom is slightly lower for the~$k=6$ simulations). 
In contrast, no clear improvement of the~$k=6$ simulations over the~$k=3$ simulations can be observed in the limit of under-resolution.
~\\

Comparing the results for the~$\LTWO$-based discretisation to the~$\HDIV$-based discretisation, no noticeable differences between the two methods can be observed in terms of accuracy. 
Furthermore, the results demonstrate rigorously that the spatial resolution described by~$k$ and~$h$ (or~$N$) is the relevant parameter that drives the accuracy of the results and that shifts the results towards the DNS data, having a much higher influence on the solution quality than the particular DG discretisation method itself.
~\\

Regarding the polynomial degree used for the mapping of the geometry, a rather clear trend can be observed on the coarsest meshes: The isoparametric mapping shows more accurate results than the affine mapping. 
The results for the affine mapping on the coarse mesh appear too inaccurate with the profile for the mean velocity being significantly higher than the DNS reference results and the Reynolds stresses showing larger oscillations compared to the results with isoparametric mapping. 
Given the under-resolution of the flow field on the coarsest mesh, the mean velocity profile is predicted very accurately in case of the isoparametric mapping. 
However, let us point out that a seemingly exact prediction of the mean velocity profile in terms of the mean velocity in the middle of the channel for these coarse meshes might be coincidence and it cannot be guaranteed that this result is robust under variations of the DG discretisation scheme, e.g., when changing the numerical fluxes or the discretisation parameters for the convective and viscous terms. 
In Sec.~\ref{sec:ParameterVariationsTurbulentChannel} below, we explicitly demonstrate the impact of the DG discretisation parameters on the accuracy of the results for the coarsest mesh. 
With increasing resolution, the differences between the affine and isoparametric mappings diminish. 
This is in agreement with Fig.~\ref{fig:channel-affine-iso} showing that the affine mapping is of course also able to resolve sharp gradients if the mesh is fine enough.
~\\

Due to the above observations, in the following we restrict ourselves to showing results only for the isoparametric setting.
Furthermore, as~$k=6$ does not show a clear advantage compared to~$k=3$ for the coarsest spatial resolutions, all following investigations are performed only for~$k=3$.
However, we do not expect different results (qualitatively) for different polynomial orders.

\subsection{Influence of the SIP parameter and convection stabilisation}
\label{sec:ParameterVariationsTurbulentChannel}

\begin{figure}[!ht]
\centering
 	\includegraphics[width=0.33\textwidth]
		{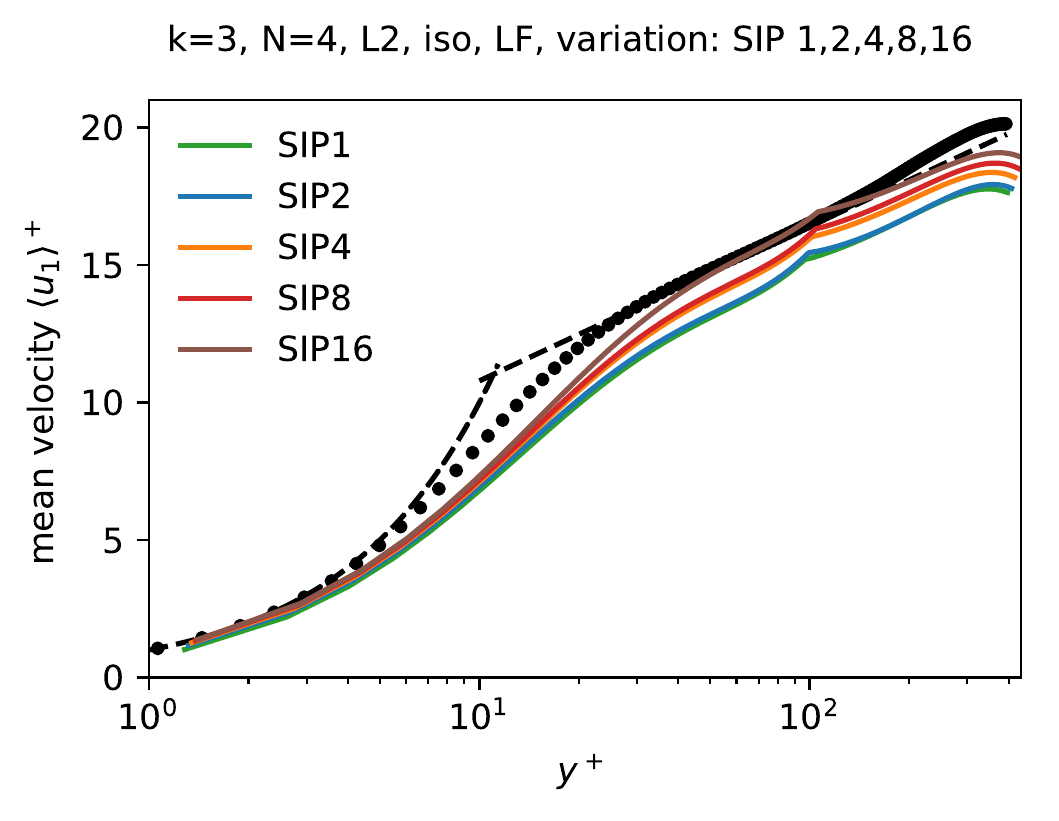}
 	\includegraphics[width=0.33\textwidth]
		{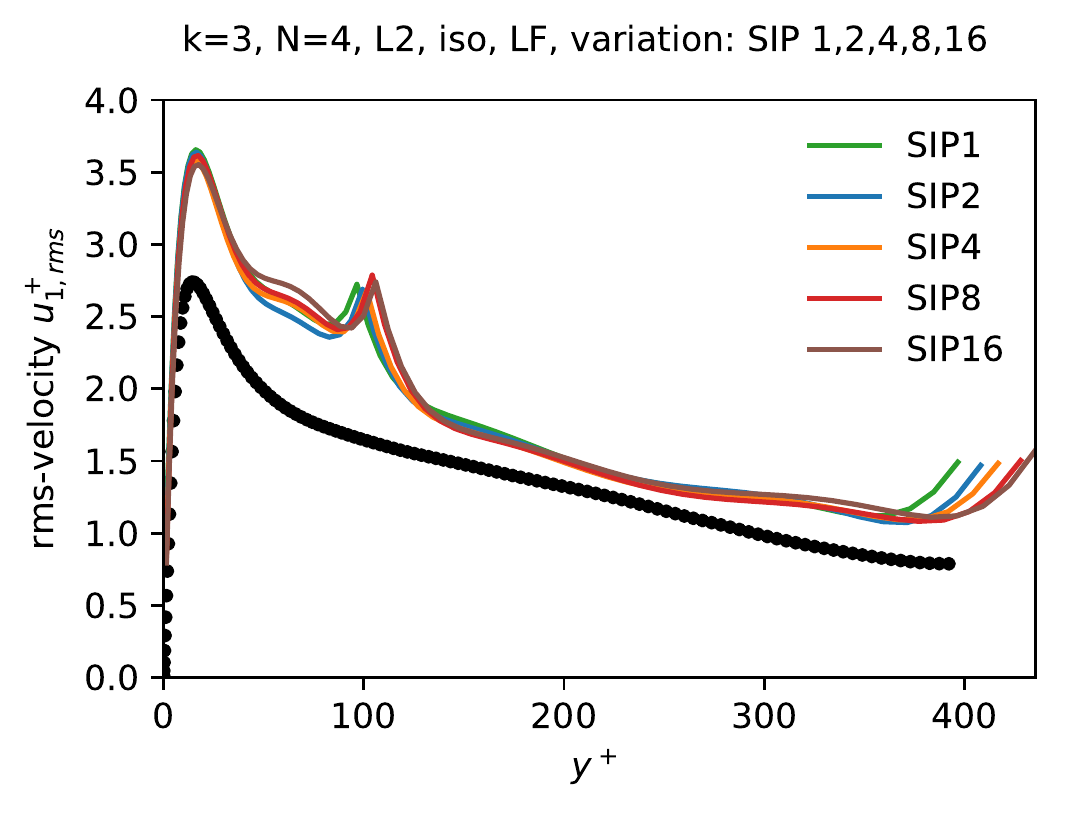}
 	\includegraphics[width=0.33\textwidth]
		{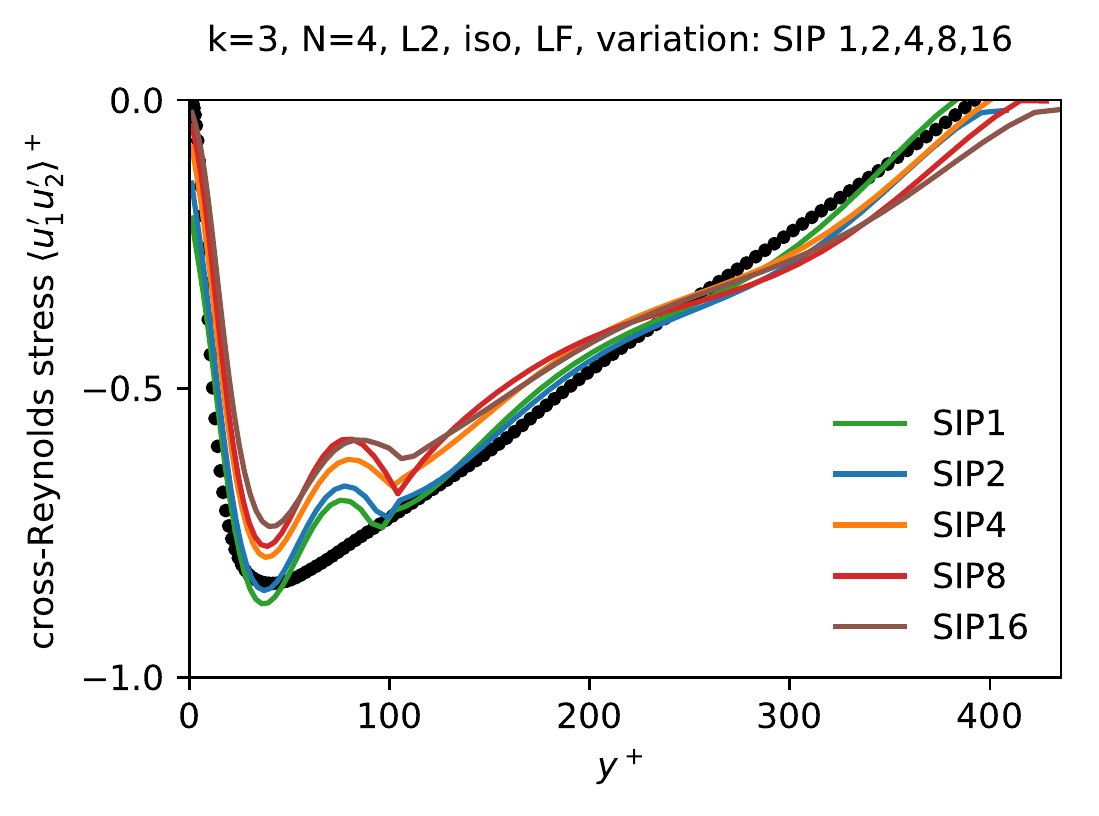} \\
 	\includegraphics[width=0.33\textwidth]
		{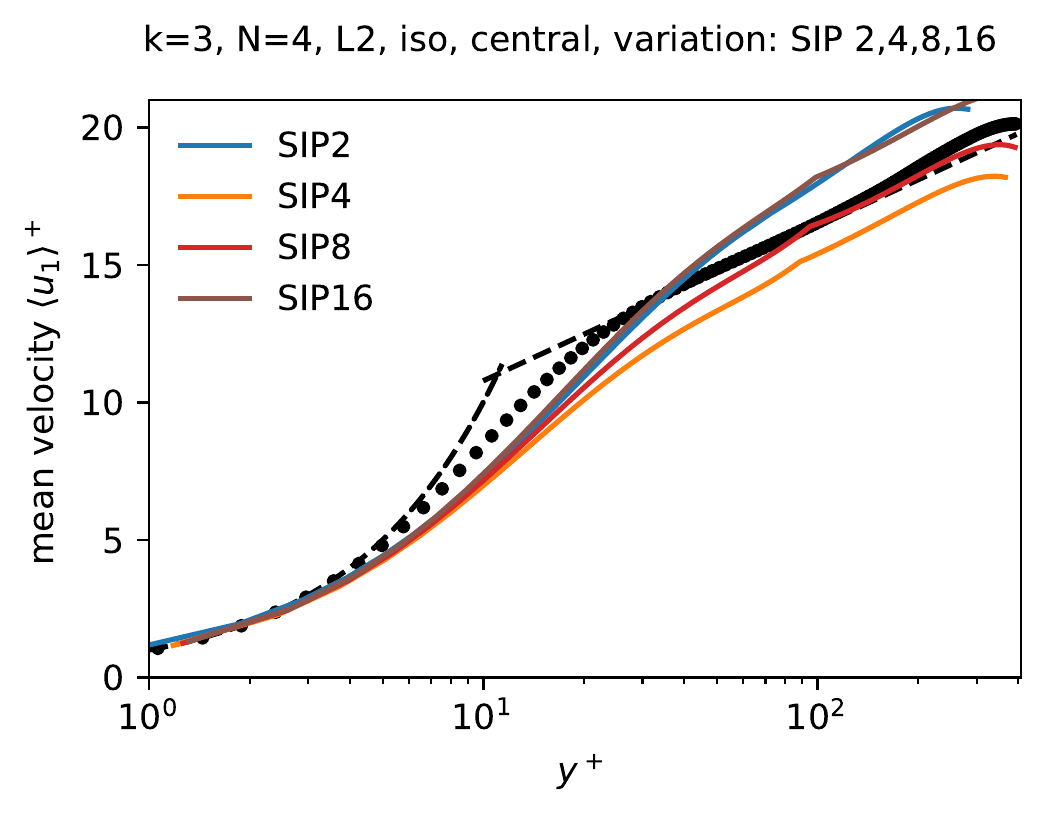}
 	\includegraphics[width=0.33\textwidth]
		{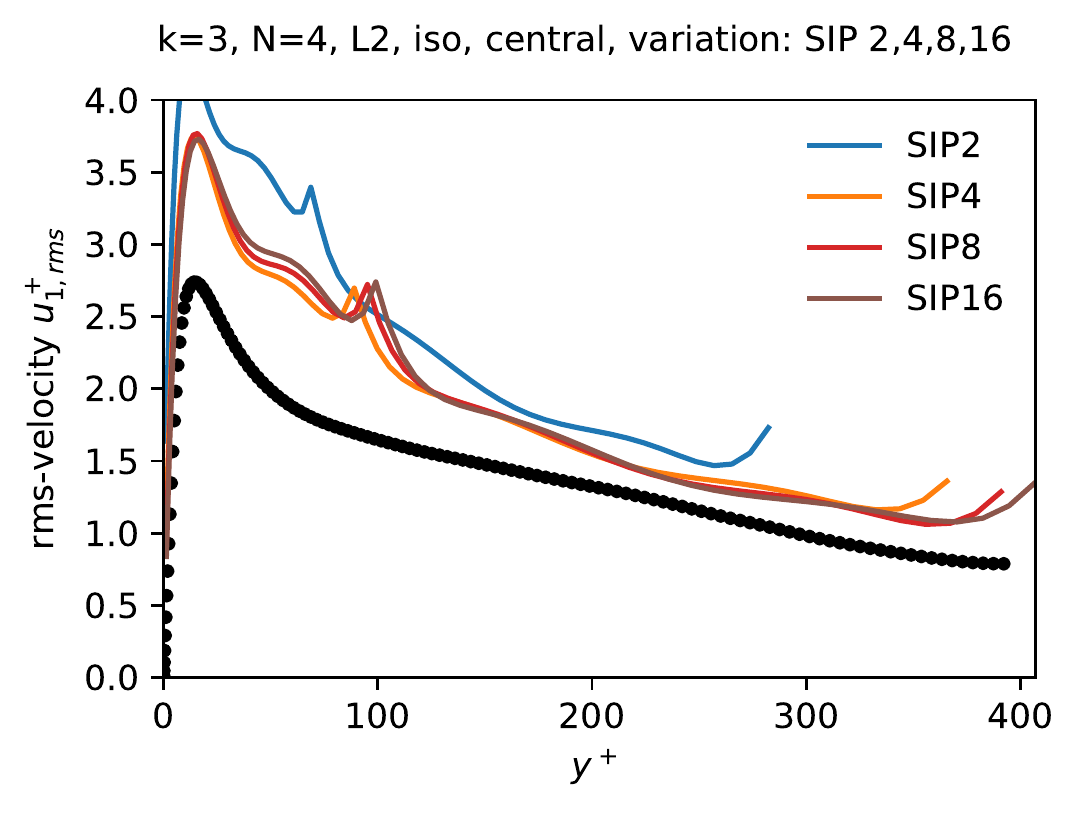}
 	\includegraphics[width=0.33\textwidth]
		{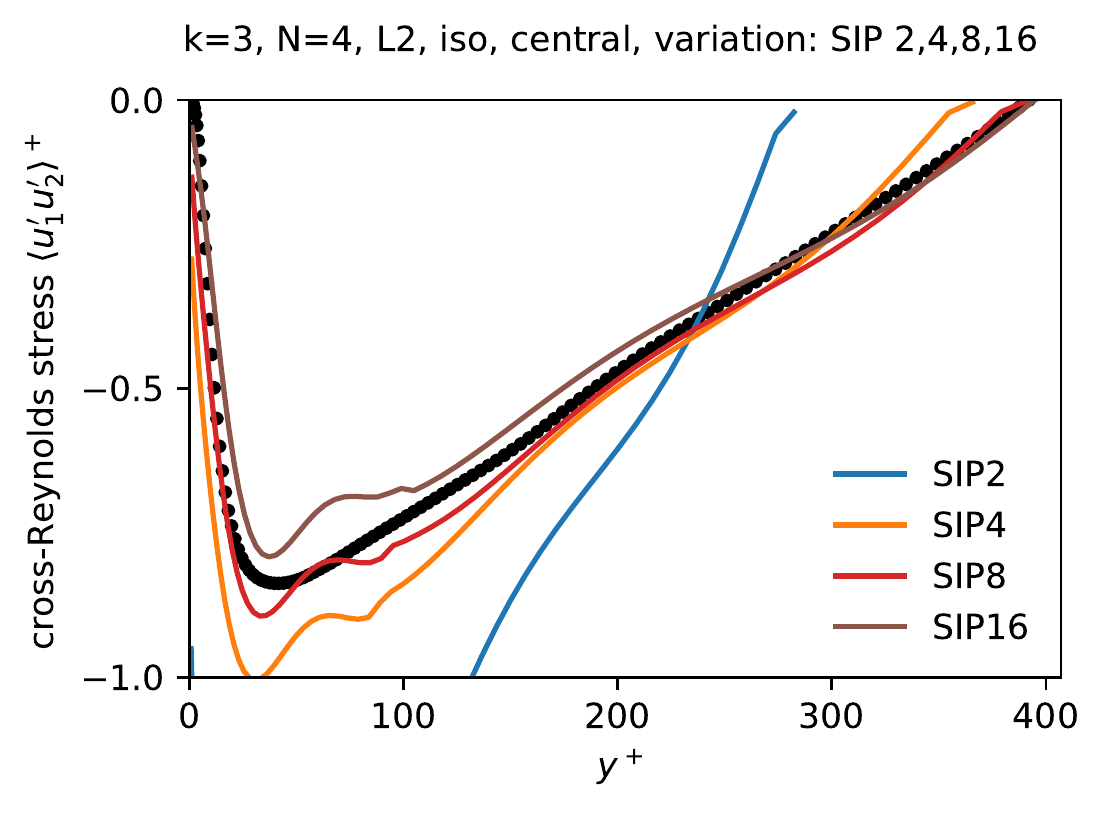}
 	\caption{$\LTWO$: Averaged channel profiles for the most under-resolved setting $k=3$, $N=4$ with isoparametric mapping. The basic SIP parameter is multiplied by $1,2,4,8,16$. Top: computations with Lax--Friedrichs stabilisation; bottom: convective formulation with central flux for convection term.}
 	\label{fig:3DCh-k3-L2}
\end{figure}

\begin{figure}[!ht]
\centering
 	\includegraphics[width=0.33\textwidth]
		{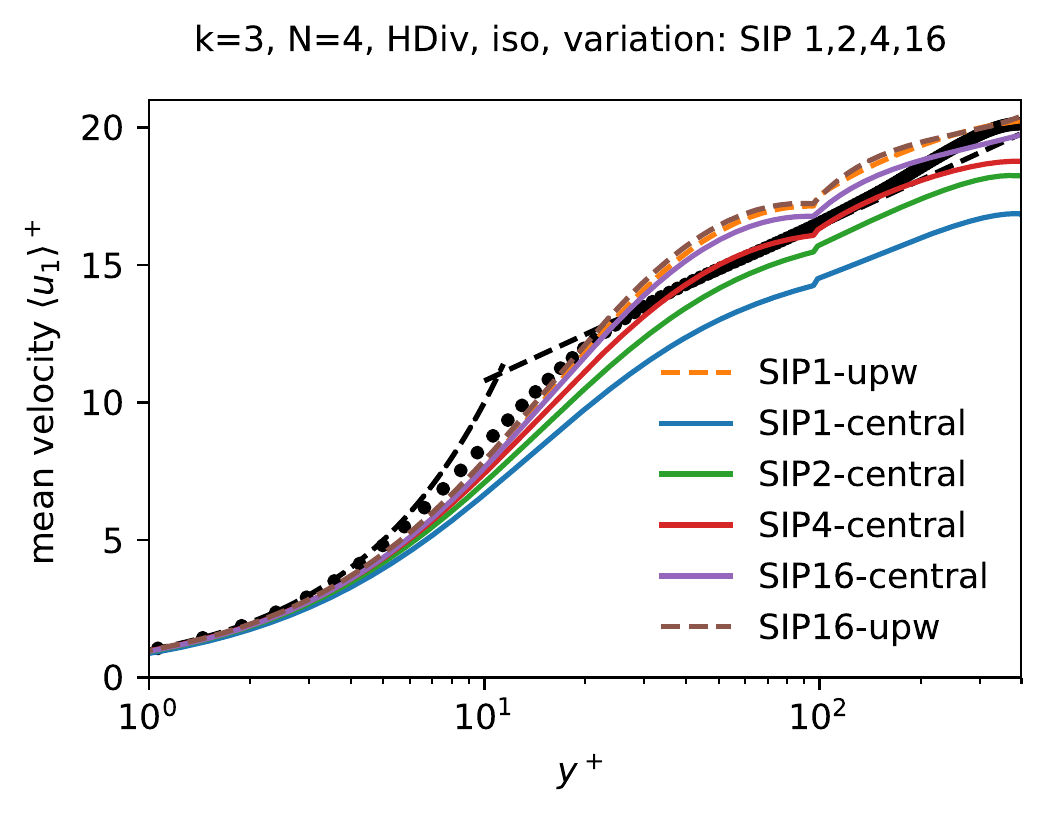}
 	\includegraphics[width=0.33\textwidth]
		{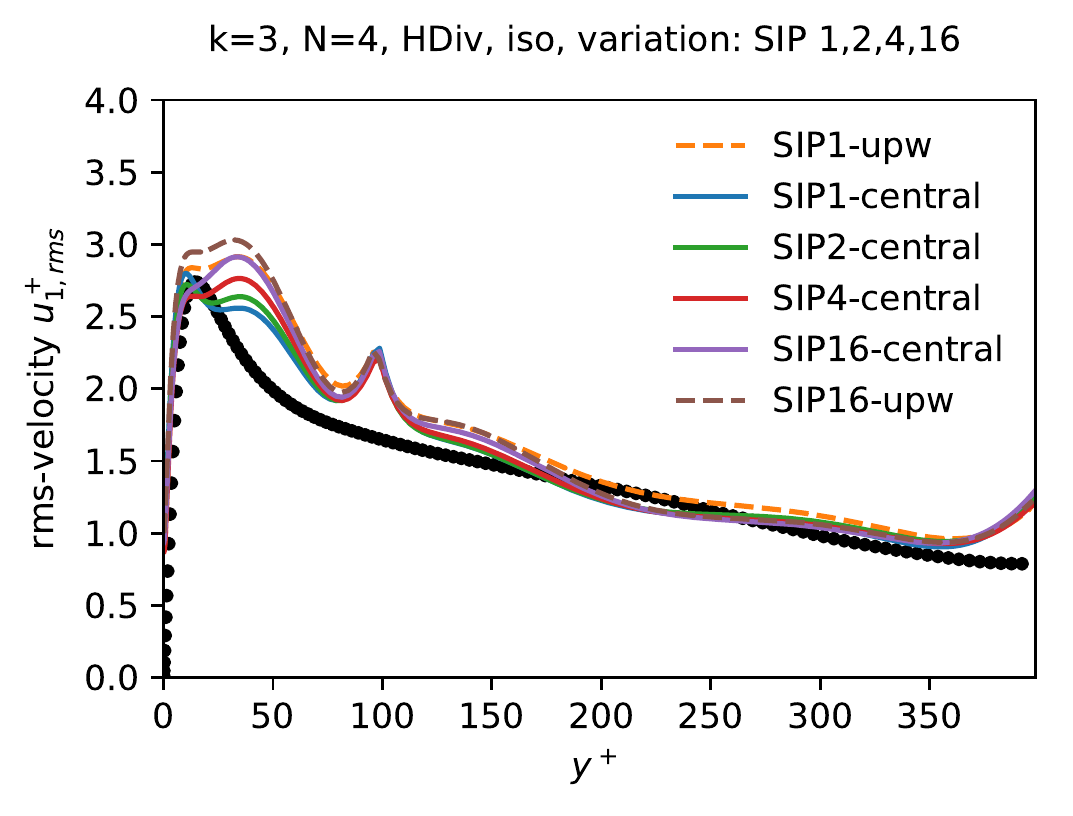}
 	\includegraphics[width=0.33\textwidth]
		{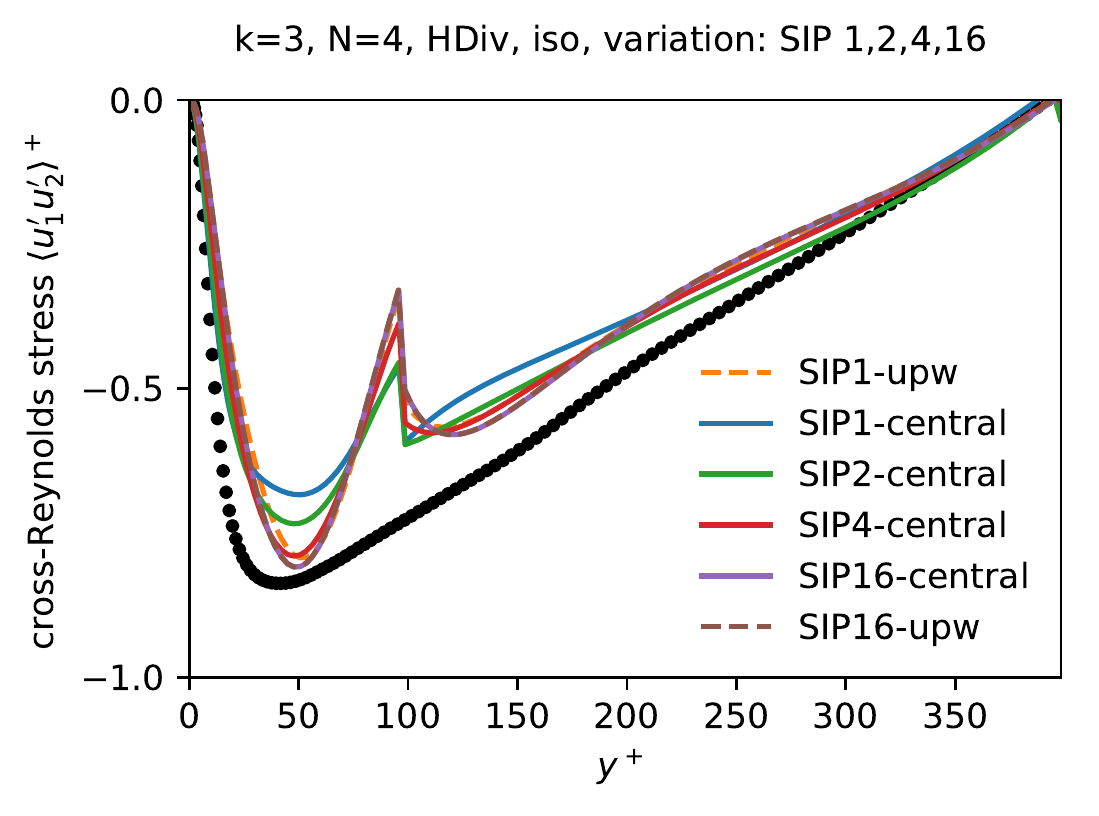}
 	\caption{$\HDIV$: Averaged channel profiles for the most under-resolved setting $k=3$, $N=4$ with isoparametric mapping. The basic SIP parameter is multiplied by $1,2,4,16$ and the two limit cases with and without upwinding are investigated.}
 	\label{fig:3DCh-k3-HDiv}
\end{figure}

Having analysed the convergence behaviour of the considered discretisation schemes towards the reference data under mesh refinement, it remains to demonstrate the sensitivity of the results for a given spatial resolution with respect to the parameters of the discretisation scheme. 
Instead of tuning the parameters to improve the solution quality for a given mesh, we want to demonstrate the ``guaranteed'' accuracy of the discretisation schemes, defined by the least accurate results achieved for ``unsuitable'' or ``non-optimised'' parameters. 
For these investigations, we focus on the coarsest mesh for degree~$k=3$ since the variations in the results (i.e., the discretisation error) can be expected to be largest for the lowest spatial resolution. 
On the one hand, we compare results for ``upwind'' fluxes to central fluxes for the convective term to investigate its impact for wall-bounded turbulent flows. 
On the other hand, we study the impact of the SIP penalty factor, which can be expected to be the second main parameter that determines the dissipation properties of the methods.
Results for these method variations are shown in Fig.~\ref{fig:3DCh-k3-L2} for the~$\LTWO$-method and in Fig.~\ref{fig:3DCh-k3-HDiv} for the~$\HDIV$-method. 
~\\

For the~$\LTWO$-method, we multiply our basic SIP penalty parameter described above with factors of~$1,2,4,8,16$ and investigate both the ``standard'' Lax--Friedrichs (LF) flux formulation (top row in Fig.~\ref{fig:3DCh-k3-L2}) and the convective formulation of the convective term with central flux as described in Sec.~\ref{sec:TGVConvetiveTerm} (bottom row in Fig.~\ref{fig:3DCh-k3-L2}). 
For the LF formulation, the SIP penalty factor has a rather small impact on the results. 
The values of the mean velocity profile increase for increasing penalty factors and the curves move closer to the DNS reference data. 
Moreover, the wall shear stress increases slightly with increasing penalty factor. 
For the central flux formulation, the SIP penalty factor has a larger impact. 
Without upwind stabilisation in the convective term, a SIP penalty factor~$\geq 2$ was required to render the method stable, which can be explained by the fact that the upwind stabilisation term and the SIP penalty term are the only terms in the discretisation scheme that weakly enforce tangential continuity in the velocity field and that the ``portion'' of the convective term has to be taken over by the SIP term if the upwind stabilisation is switched off. 
For a small SIP penalty factor of~$2$, the prediction of the wall shear stress is still inaccurate, but for increasing penalty factors of~$4,8,16$ the accuracy achieved with the LF formulation is retained.
Analogously, switching off the upwind-like stabilisation term in the LF formulation leads to qualitatively similar results as those shown in Fig.~\ref{fig:3DCh-k3-L2} for the convective formulation with central flux.
Since the LF formulation produces better results with the SIP penalty factor as defined in Sec.~\ref{sec:L2DG}, the LF formulation is the preferred choice here, but we mention that a purely central convective flux can be used as well even though it remains unclear whether this extends to more complex problems. 
Since we want to avoid parameter adjustments for the present discretisation schemes, the LF formulation with minimal SIP penalty factor is advantageous given the fact that the iteration counts of iterative solvers will deteriorate with increasing penalty factors and given the fact that parameter tuning is not a reliable or preferable tool in under-resolved turbulence simulations in our opinion.
~\\

For the $\HDIV$ method, our basic SIP penalty parameter described above is multiplied with factors $1,2,4,16$, where $8$ is omitted only in order to improve the visibility of the results.
Fig.~\ref{fig:3DCh-k3-HDiv} shows the resulting channel profiles with different SIP parameters and with and without upwinding (convection stabilisation).
The first conclusion can be drawn from comparing SIP1-upw with SIP16-upw (dashed lines), which show that once upwinding is used, the particular parameter for the SIP mechanism does not play a major role anymore.
On the other hand, SIP1 without upwinding (central) obviously leads to worse (but still stable) results.
Thus, we conclude also for the $\HDIV$ method that a certain amount of numerical dissipation is beneficial for channel flow problems.
Interestingly, when the SIP parameter is increased for the central flux situation, the curves approach the SIP1-upw curves in a continuous fashion up to the point that SIP16-central is very similar to SIP16-upw.
From this behaviour we conclude that for the $\HDIV$ method and for this problem, the particular source for numerical dissipation, be it from SIP or from upwinding, is not decisive.
A possible explanation for this phenomenon can be derived from regarding how upwinding works; cf.\ \eqref{eq:Hdiv-convection}: In principle, upwinding is an anisotropic mechanism acting only on facets where in- or outflow occurs.
However, in turbulent channel flow simulations, the turbulent mixing properties of the instantaneous flow basically result in the fact that upwinding acts isotropically.
Thus, SIP and upwinding are penalising the tangential discontinuity in a very similar way here.

\section{Summary and conclusions}
\label{sec:Conclusions}

Two conceptually different discontinuous Galerkin discretisation methods for the numerical simulation of incompressible turbulent flow problems have been analysed and compared in this work. 
Fulfilling the incompressibility constraint as well as inter-element continuity of the velocity field in the direction normal to element interfaces are of particular importance regarding the numerical robustness and accuracy of the discretisation schemes. 
Remarkably, the robustness of the considered methods allows to obtain meaningful results even in significantly under-resolved settings.
~\\

The present work shows that these aspects can be addressed successfully by the use of either specialised function spaces in an~$\HDIV$-context fulfilling these properties exactly, or by plain~$\LTWO$-conforming spaces equipped with additional stabilisation terms enforcing the above requirements in a weak sense. 
To assess the properties of the numerical discretisation schemes, the three-dimensional Taylor--Green vortex problem has been investigated as a representative of transitional flows and decaying turbulence as well as the turbulent channel flow problem as a representative of wall-bounded turbulent flows.  
The comparison of the~$\LTWO$-based and~$\HDIV$-based methods, as well as studying the impact of different discretisation variants for the convective and viscous terms and the influence of certain discretisation parameters has several important implications. 
Overall, variations of the discretisation scheme were found to have only a small or moderate influence on the accuracy of the results. 
This insensitivity can be seen as an advantage of the methods analysed here in the sense that accurate results are not only achieved for a certain set of parameters but that this accuracy can be expected in general.
At the same time and as shown in this work, this implies that the possibilities to optimise high-order DG discretisations by choosing one or the other flux formulation appear to be limited as compared to the influence other parameters -- most importantly the spatial resolution parameters~$h$ and~$k$ -- have on the accuracy of the results. 
~\\

Interestingly, it is commonly believed that the discretisation of the convective term plays a crucial role in DG discretisations rendering the method stable for convection-dominated problems by the choice of suitable flux functions such as upwind fluxes. 
Let us emphasise that even in under-resolved settings, basically independent of the chosen convective flux, we still have been able to obtain accurate results with respect to e.g.\ predictions of statistical quantities such as the Reynolds stress tensor.
In this context, a key message of the present work is that -- for DG discretisations of the incompressible Navier--Stokes equations -- measures enforcing mass conservation and energy stability in a weak sense or in a local (exact) sense are more important to achieve a robust discretisation scheme. 
Indeed, the present work shows that convection stabilisation (upwind or Lax--Friedrichs) is neither necessary to ensure energy-stability of the overall method nor crucial for the robustness of the simulations.
Whereas the~$\LTWO$ method relies on divergence and normal-continuity stabilisation for energy-stability, the divergence-free $\HDIV$ method is designed to be energy stable without any additional stabilisation.
Thus, concerning robustness for incompressible flows, the crucial ingredient turns out to be a meaningful fulfilment of the divergence-free constraint and normal-continuity of the velocity, realised weakly~($\LTWO$) or strongly~($\HDIV$) in this work.
Nonetheless, the use of upwind-like fluxes for the convective term was found advantageous, especially for the wall-bounded turbulent channel flow problem, since alternative central flux formulations have shown a larger sensitivity with respect to the discretisation parameters of the viscous term.
~\\
 
We conclude that both~$\LTWO$-based and~$\HDIV$-based discretisations show convincing results in highly under-resolved scenarios and are promising candidates as generic turbulent flow solvers. 
Preconceptions like ``high-order methods are not robust'' or ``no-model LES is not physical'' are widespread and the results shown here -- albeit promising -- are certainly not enough to conclusively demonstrate the opposite.
Therefore, given the maturity of the results shown here, the applicability of these novel high-order DG discretisations to complex engineering applications should be considered as part of future work.
  

\appendix

\section{Characterisation of a limit method}
\label{app:limitmethod}

The $\LTWO$-DG method is defined by the velocity and pressure spaces
\begin{equation*}
	\VV_h = \QQdk{k}{}, \quad
	 \Q_h = \Qdk{k-1}{} \cap \Lpz{2}{\OMEGA} .
\end{equation*}
and the variational formulation \eqref{eq:L2DGMethod} which depends on the stabilisation parameters $\tau_D$ and $\tau_C$, cf.\ \eqref{DivergencePenalty} and \eqref{ContinuityPenalty}. 
~\\

One may ask the question if the limit $\tau_D, \tau_C \to \infty$ can be characterised as a meaningful discretisation. 
To this end, we first realise that in the limits $\tau_D, \tau_C \to \infty$, the solution $\uu_h$ has to fulfil $j_{\dvg,h}\rb{\uu_h,\vv_h} = 0$ and $j_{\mathrm{conti},h}\rb{\uu_h,\vv_h} = 0$ for all $\vv_h \in \VV_h$, as both bilinear forms are positive semi-definite. 
The subspace of functions in $\VV_h$ that fulfil these constraints are pointwise divergence-free and normal-continuous across element interfaces. 
In this sense, the limit shares the properties of the $\HDIV$-based method. 
However, the finite element space $\VV_h$ and hence the pointwise divergence-free and normal-continuous subspace
\begin{equation*}
\VV_h^0 = \set{ \uu_h \in \VV_h\colon
 j_{\mathrm{div},h}\rb{\uu_h,\vv_h} = j_{\mathrm{conti},h}\rb{\uu_h,\vv_h} = 0 \text{ for all } \vv_h \in \VV_h }
\end{equation*}
are different from the $\HDIV$-based method (on non-simplex meshes).
~\\

Let us -- for ease of presentation -- assume that the mesh consists of straight, affine linearly mapped hexahedral elements. 
In this case, if $\uu_h \in \VV_h^0$, the weak divergence-constraint $d_h\rb{\uu_h,q_h} = 0$ for all $q_h \in \Q_h$, cf.\ \eqref{L2VelocityDivergenceWeakForm}, can be rewritten as $d_h(\uu_h,q_h) = g_h(q_h,\uu_h) = 0$ for all $q_h \in \Q_h$.
~\\

To characterise $\VV_h^0$ through constraints by Lagrange multiplier functions we rewrite
\begin{subequations}
\begin{align}
 j_{\mathrm{div},h}\rb{\uu_h,\vv_h} &= 0 \quad \forall \vv_h \in \VV_h
\quad\text{as} & d_{\mathrm{div},h} ( \uu_h, q_h) =
	\int_\Omega \rb{\DIVh \uu_h} q_h \dx & = 0 \quad \forall q_h \in Q_h^{\text{vol}} = \rb{\DIVh \VV_h} \\
 \text{and} \quad j_{\mathrm{conti},h}\rb{\uu_h,\vv_h} &= 0 \quad \forall \vv_h \in \VV_h
 \quad\text{as} & d_{\mathrm{conti},h} ( \uu_h, \hat{q}_h) =
\sum_{F\in\Fi}  \int_F \rb{\jmp{\uu_h}\ip\nn} \hat{q}_h \ds  & = 0 \quad \forall \hat{q}_h \in Q_h^{\text{skel}} = \text{tr}^n|_{\Fi} \VV_h.
\end{align}
\end{subequations}
Here $Q_h^{\text{skel}}$ is the space of functions defined on the (interior) skeleton of the mesh that is obtained by taking the normal trace of functions in  $\VV_h$. 
We can explicitly characterise this space as the following space of facet unknowns:
\begin{equation*}
Q_h^{\text{skel}} = \set{\hat{q}_h\in\Lp{2}{\rb{\Fi}}\colon \restr{q_h}{F}\in\Qk{k}{\rb{F}}, ~\forall\, F\in\Fi}.
\end{equation*}
This space is well-known in the community of hybrid mixed and hybrid DG methods. 
Less established is the space $Q_h^{\text{vol}}$ which, however, can also be characterised explicitly as follows:
\begin{equation*}
Q_h^{\text{vol}} = \set{ q_h \in \Qdk{k}{} : q_h|_T \in \Qk{k}{\rb{K}} \setminus \set{\prod_{i=1}^d \xi_i^k }, \forall\, K\in\T} \cap \Lpz{2}{\OMEGA},
\end{equation*}
where $\xi_i,~i=1,2,3$ are the spatial coordinates. We added the zero-mean constraint for uniqueness of solutions in the later given variational formulation. 
This means $Q_h^{\text{vol}}$ is $\Qdk{k}{}$ except for the element-wise highest-order bubble. 
Moreover, $\Q_h^{\text{vol}}$ is a superset of $Q_h$ (since the pressure is approximated by polynomials of degree~$k-1$) so that the weak divergence-constraint $d_h\rb{\uu_h,q_h} = 0$ holds for all $q_h \in \Q_h$, cf.\ \eqref{L2VelocityDivergenceWeakForm}, and the corresponding Lagrange-multiplier $p_h \in Q_h$ become superfluous.
~\\

With these characterisations, one can define the limit method as
\begin{subequations} \label{eq:LimitL2DGMethod}
	\begin{empheq}[left=\empheqlbrace]{align} 
	&\text{Find }\rb{\uu_h,p_h, \hat{p}_h}\colon \rsb{0,\tend} \to \VV_h \times \Q_h^{\text{vol}} \times \Q_h^{\text{skel}}
	\text{ with }\uu_h\rb{0} =\uu_{0h}\text{ s.t., }\forall\,\rb{\vv_h,q_h,\hat{q}_h}\in\VV_h \times \Q_h^{\text{vol}} \times \Q_h^{\text{skel}},
	\nonumber \\ \nonumber
		&\rb{\partial_t\uu_h,\vv_h} 
		+ \nu a_h\rb{\uu_h,\vv_h}
		+ c_h\rb{\uu_h;\uu_h,\vv_h}
		+ d_{\text{div},h}\rb{\vv_h,p_h}
		+ d_{\text{conti},h}\rb{\vv_h,\hat{p}_h}
		+ d_{\text{div},h}\rb{\uu_h,q_h}
		+ d_{\text{conti},h}\rb{\uu_h,\hat{q}_h}
		=\rb{\ff,\vv_h}.	
	\end{empheq} 
\end{subequations}
~\\

Let us note that the solution $\uu_h$ will be in $\VV_h^0$ which has dimension $\dim(\VV_h)-\dim(Q_h^{\text{vol}})-\dim(Q_h^{\text{skel}})$.
\begin{thmRem}
In this section we assumed affine linear element geometries. 
This assumption can be dropped if the (discontinuous) finite element space $\VV_h$ is defined through Piola transformations from the reference element to the physical element, cf.\ Section \ref{sec:AffineVsIso}.
Thereby, the normal traces and the divergence of $\VV_h$ can be characterised through polynomials again. 
\end{thmRem}

\bibliography{DG-turbulence-BibTeX.bib}
\end{document}